\documentclass[preprint,review,12pt]{elsarticle}

\usepackage{amssymb}
\usepackage[colorlinks=true,linkcolor=black,citecolor=blue,urlcolor=blue]{hyperref}

\usepackage{amsmath}
\usepackage{array}
\usepackage{booktabs}
\usepackage{tabularx}
\usepackage{tikz}
\usepackage[edges]{forest} 
\usetikzlibrary{trees}
\usepackage{adjustbox} 
\usepackage{multirow}
\usepackage{caption}
\usepackage{natbib} 
\usepackage{tikz}
\usepackage{geometry}
\DeclareUnicodeCharacter{FF0C}{,} 

\journal{Information Fusion}
\makeatletter
\renewcommand\paragraph{\@startsection{paragraph}{4}{\parindent}%
    {1.5ex plus .5ex minus .2ex}%
    {-1em}%
    {\normalfont\normalsize\bfseries}}
\makeatother
\begin{document}

\begin{frontmatter}



\title{Integrating Artificial Intelligence into Operating Systems: A Survey on Techniques, Applications, and Future Directions}


\author[1]{Xinkui Zhao}
\author[1]{Yifan Zhang}
\author[1]{Ziying Li}
\author[1]{Guanjie Cheng\corref{cor1}}
\author[2]{Jianwei Yin}
\author[3]{Zhiquan Liu}
\author[4]{Lufei Zhang}
\author[5]{Zuoning Chen}
\cortext[cor1]{Corresponding author. Email: chengguanjie@zju.edu.cn}

\affiliation[1]{organization={School of Software Technology, Zhejiang University},
            city={Ningbo},
            state={Zhejiang},
            country={China}}

\affiliation[2]{organization={College of Computer Science and Technology, Zhejiang University},
            city={Hangzhou},
            state={Zhejiang},
            country={China}}

\affiliation[4]{organization={State Key Laboratory of Mathematical Engineering and Advanced Computing},
            city={Wuxi},
            state={Jiangsu},
            country={China}}

\affiliation[5]{organization={Chinese Academy of Engineering},
            city={Beijing},
            country={China}}
            
\affiliation[3]{organization={College of Cyber Security, Jinan University},
    city={Guangzhou},
    country={China}}
    
\begin{abstract}
Heterogeneous hardware and dynamic workloads exacerbate long‑standing operating system (OS) bottlenecks in scalability, adaptability, and manageability. At the same time, advances in machine learning (ML), large language models (LLMs), and agent-based intelligence create opportunities for OS automation and self‑optimization, yet current efforts remain fragmented without a unifying perspective. This survey provides a comprehensive overview of techniques, architectures, application scenarios, challenges, and future directions at the intersection of AI and OS. Specifically, we trace the paradigm shift from heuristic and rule based designs to AI enhanced systems, highlighting the respective strengths of ML, LLMs, and agent-based methods across the OS stack. We review representative progress in \emph{AI for OS}, covering core components and the broader OS ecosystem, and in \emph{OS for AI}, which includes component level optimizations and architecture level innovations that support short and long context inference, distributed training, and edge inference. To inform engineering practice, we synthesize common evaluation dimensions, summarize methodological pipelines, and distill actionable patterns that balance real-time constraints with predictive accuracy. We further identify key challenges, including system complexity, performance overhead, model drift, limited explainability, and privacy and safety risks, and outline practical strategies such as modular, AI-ready kernel interfaces; unified toolchains and benchmarks; hybrid rules plus AI decision frameworks with trustworthy guardrails; and verifiable in kernel inference. Finally, we propose a three-stage developmental roadmap, namely AI powered, AI refactored, and AI driven OSs, aimed at bridging research prototypes and production systems, thereby advancing the design of next generation intelligent operating systems and enabling scalable, reliable deployment of modern AI workloads.

\end{abstract}


\end{frontmatter}


\section{Introduction}
\label{sec:Introduction}
The evolution of operating systems (OSs) has been shaped by the ongoing interplay between technological advancement and the evolving demands of computing environments~\cite{peterson1985operating,tanenbaum2015modern,stallings2011operating}. From the manually configured systems of the early computing era to today’s highly complex platforms, OS development has consistently aimed to optimize resource management, enhance usability, and accommodate emerging computing paradigms~\cite{mckusick1996design,ceruzzi2003history}. Foundational milestones, such as the development of Unix~\cite{ritchie1974unix} and the emergence of graphical user interfaces, laid the groundwork for modern operating systems by enabling multitasking, modularity, and user accessibility. As computing platforms have diversified—from desktops and mobile devices to embedded and edge systems~\cite{satyanarayanan2017emergence,heiser2016l4}—OS design has continued to evolve, adapting to the specialized demands of heterogeneous hardware and increasingly sophisticated applications.

As computing environments have become increasingly heterogeneous and interconnected, the complexity of managing modern operating systems has risen substantially~\cite{ekmecic1996survey,shelepov2009hass}. The vast scale and diversity of hardware platforms~\cite{wentzlaff2009factored,barak2005design}, workloads~\cite{beckman2006influence}, and user expectations~\cite{matias2013operating,wong2000operating} present significant challenges for conventional, manually managed OS architectures. Human-driven approaches and static rule-based mechanisms are often inadequate for handling the dynamic and unpredictable characteristics of contemporary systems~\cite{mao2016resource,casas2021drsir}, resulting in inefficiencies, scalability constraints, and heavy maintenance overhead. At the same time, rapid advances in artificial intelligence (AI) have transformed diverse domains, including healthcare~\cite{rajpurkar2022ai,hamet2017artificial}, finance~\cite{cao2022ai,zheng2019finbrain,milana2021artificial}, autonomous vehicles~\cite{parekh2022review,bathla2022autonomous}, and natural language processing~\cite{wu2023brief,achiam2023gpt,liu2024deepseek}. AI techniques such as expert systems, machine learning (ML), and large language models (LLMs) have demonstrated strong capabilities in learning from data, adapting to evolving conditions, and automating complex decision-making processes~\cite{li2023towards}.

This convergence presents a dual opportunity. On one hand, integrating artificial intelligence (AI) techniques into operating systems (OSs) addresses two critical challenges: (1) the limitations of manual, rule-based policies that degrade under increasingly heterogeneous hardware and dynamic workloads; and (2) the difficulty of manually balancing multiple, often conflicting objectives such as latency, throughput, energy efficiency, fairness, and isolation. AI-enabled OSs can allocate resources more efficiently, anticipate user needs, and proactively mitigate security and performance issues. These capabilities are particularly vital in emerging domains such as the Internet of Things (IoT)~\cite{madakam2015internet}, edge computing, and large-scale cloud infrastructures~\cite{filelis2018framework,polze2012trends}, where scalable and intelligent coordination across vast device and service ecosystems is essential. 

On the other hand, innovations in OS design can accelerate AI development. Modern AI workloads are increasingly complex, data-intensive, and heterogeneous, requiring high throughput, low latency, and scalable support for CPUs, GPUs, and specialized accelerators. Traditional OS infrastructures—originally designed for general-purpose computing—often become performance bottlenecks in AI training and inference pipelines. Recent advances in kernel-bypass networking, library OSs, and accelerator-aware scheduling demonstrate that rethinking OS abstractions and components can yield substantial performance gains for AI workloads.

Taken together, the relationship between AI and OSs is no longer unidirectional but mutually reinforcing. AI introduces new opportunities for intelligent automation in OS management, while recent advances in OS design establish the foundational support required to scale and deploy next-generation AI systems.\begin{figure}[t]
    \centering
    \includegraphics[width=1\linewidth]{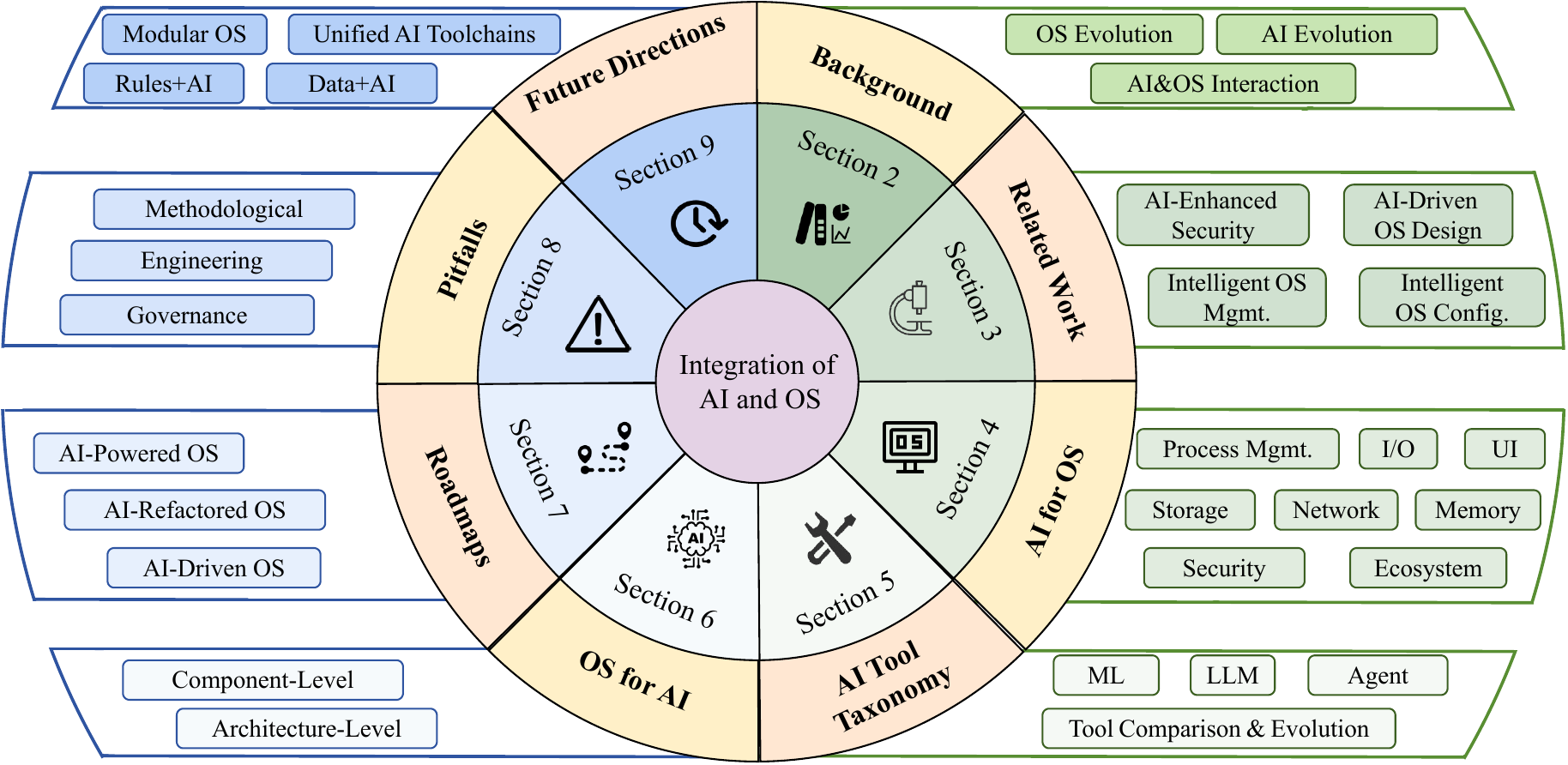}
    \caption{Overall structure of the survey. Section~\ref{sec:background} introduces the background; Section~\ref{sec:related} reviews related work; Sections~\ref{sec:module},~\ref{sec:tool}, and~\ref{sec:structure} discuss \textit{AI for OS}, the taxonomy of AI tools, and \textit{OS for AI}, respectively. Section~\ref{sec:stages} presents the developmental roadmap; Section~\ref{sec:pitfalls} analyzes common pitfalls; and Section~\ref{sec:future} outlines future research directions.}
    \label{fig:overview}
\end{figure}

This survey provides a comprehensive overview of the emerging intersection between artificial intelligence and operating systems, as illustrated in Figure~\ref{fig:overview}. We categorize existing and ongoing research into two primary directions: (\textit{i}) \textbf{AI for OS}, which applies AI techniques to enhance OS design, operation, and usability; and (\textit{ii}) \textbf{OS for AI}, which focuses on advances in OS architecture that enable the efficient execution of AI workloads. Based on an extensive review of recent studies, we analyze the benefits, limitations, and open challenges in each direction and identify promising directions for future research.

This paper is organized as follows. Section~\ref{sec:related} reviews related research at the intersection of artificial intelligence (AI) and operating systems (OSs). Section~\ref{sec:module} examines the integration of AI techniques into core OS components, including scheduling, memory management, I/O, storage, networking, and security. Section~\ref{sec:tool} classifies and compares AI tools applied in this domain based on their efficiency, versatility, interpretability, and overhead. Section~\ref{sec:structure} investigates the complementary perspective of \textit{OS for AI}, analyzing component-level optimizations and architectural innovations that improve the performance of AI workloads. Section~\ref{sec:stages} introduces a developmental roadmap tracing the evolution from \textit{AI-powered} to \textit{AI-refactored} and ultimately \textit{AI-driven} operating systems. Section~\ref{sec:pitfalls} discusses common limitations identified in existing studies, while Section~\ref{sec:future} highlights promising directions for future research. Finally, Section~\ref{sec:conclusion} concludes the paper. To help readers navigate the survey, Figure~\ref{fig:overview} illustrates the overall structure and organization of the work.

Here are the highlights of this paper:
\begin{itemize}
\item We conduct a comprehensive survey at the intersection of artificial intelligence (AI) and operating systems (OSs), unifying perspectives from both \textbf{AI for OS} and \textbf{OS for AI} paradigms.

\item We examine how a wide spectrum of AI methodologies—from traditional learning models to generative and agent-based intelligence—are reshaping the design, optimization, and management of modern operating systems.

\item We analyze the mutual influence between AI techniques and OS architecture, outlining a progressive roadmap that captures the evolutionary path toward fully intelligent operating systems.

\item We highlight overarching research challenges and opportunities shaping this emerging field, emphasizing the need for systematic integration, reliability, and interpretable intelligence across the OS lifecycle.
\end{itemize}

\section{Background}
\label{sec:background}

\subsection{Operating Systems: Fundamentals and Evolution}

\begin{figure}[htbp]
    \centering
    \includegraphics[width=0.55\linewidth]{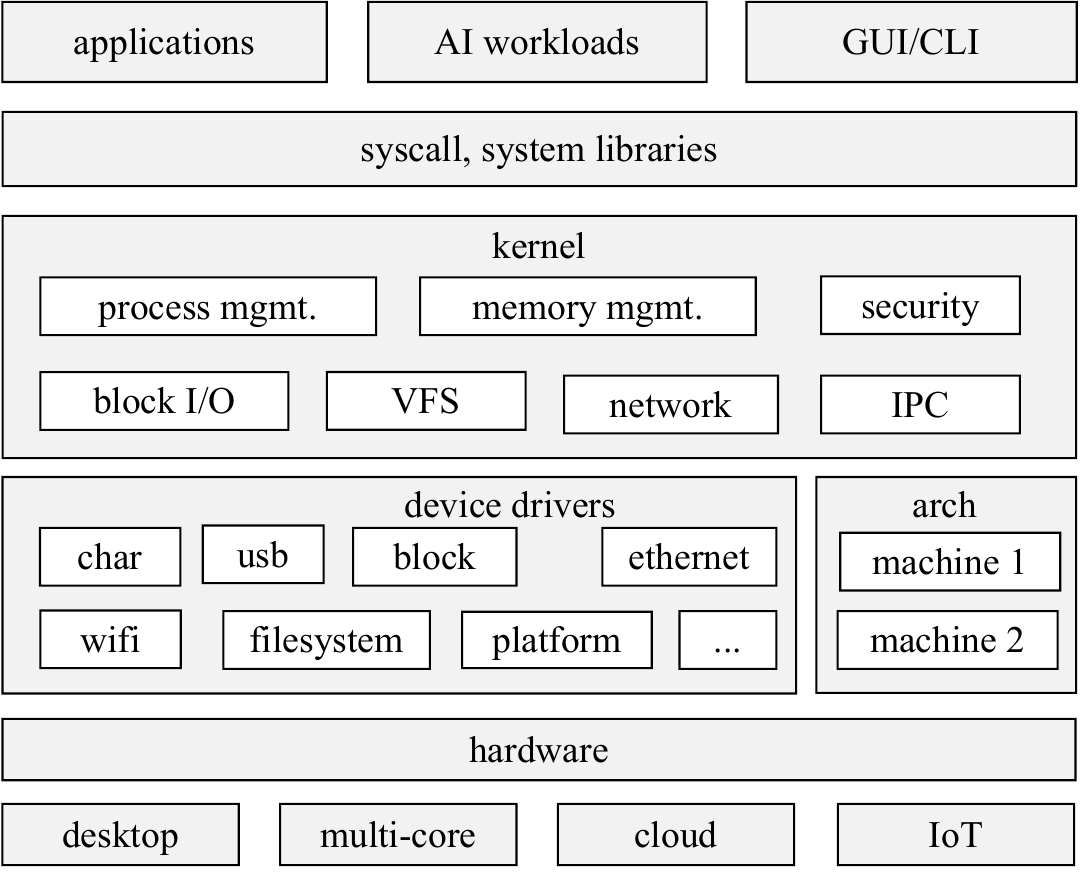}
    \caption{Fundamental structure of an operating system. The diagram spans from hardware and device drivers at the base, through core kernel subsystems (process, memory, I/O, networking, inter-process communication, and security), to system libraries, user interfaces, and applications. It illustrates how OSs abstract heterogeneous hardware into uniform services for diverse applications.}
    \label{fig:os_struct}
\end{figure}

OSs form the foundational software layer that mediates between hardware resources and user applications. They manage processes, memory, storage, input/output (I/O) devices, and security, while providing abstractions and services that facilitate application development, as illustrated in Figure~\ref{fig:os_struct}. Landmark developments such as time-sharing systems, Unix, and graphical user interfaces established the groundwork for today’s sophisticated computing platforms~\cite{peterson1985operating,hansen1973operating,tanenbaum1997operating}. Over time, the role of OSs has expanded beyond single-user desktop environments to encompass multi-core servers~\cite{li2007efficient,li2010operating}, mobile systems~\cite{okediran2014mobile,hall2009operating}, Internet of Things (IoT) devices~\cite{zikria2019internet,javed2018internet}, and large-scale cloud infrastructures~\cite{musse2016cloud,pianese2010toward}.

This expansion has substantially increased system heterogeneity and complexity. Modern OSs must meet diverse and often competing requirements for efficiency, fairness, reliability, energy conservation, and security. Traditional heuristic-based methods—such as static scheduling and rule-driven memory management—are increasingly inadequate for dynamic and unpredictable workloads. These challenges underscore the growing demand for adaptive, data-driven approaches to OS design and optimization.

\subsection{Classification Framework: AI for OS and OS for AI}

A useful framework for examining this interdisciplinary field is the \emph{two-way relationship} between AI and OSs. Existing research can be broadly categorized into two complementary directions:

\begin{itemize}
    \item \textbf{AI for OS}: AI techniques—ranging from classical machine learning (ML) models to large language model (LLM)-based agents—can be integrated into operating systems to enhance performance, security, and usability. Representative applications include intelligent scheduling, predictive memory management, anomaly detection, malware prevention, and natural language–based user interaction. This research direction focuses on making operating systems more adaptive, efficient, and intelligent.
    
    \item \textbf{OS for AI}: Advances in operating system design and architecture can, in turn, accelerate AI workloads. Modern AI applications are inherently heterogeneous, spanning CPUs, GPUs, FPGAs, and specialized accelerators, while traditional OS abstractions often introduce throughput and latency bottlenecks. This direction explores how rearchitecting OS mechanisms and abstractions can deliver substantial performance gains for AI systems.
\end{itemize}








\section{Related Work}
\label{sec:related}
Previous research has made substantial progress in integrating AI with operating systems~\cite{zhang2019learned,shankar2025machine,Vishwakarma2021AIBO}. For example, Agarwal et al.\cite{agarwal2023artificial} investigate the application of quantum computing in cybersecurity, highlighting how AI and qubit-based operating systems can significantly enhance security protocols. Their findings suggest that quantum-enhanced systems hold great promise for addressing complex security challenges. Similarly, Wohlrab et al.\cite{wohlrabapplication} explore the adoption of object-oriented programming and AI-driven methodologies to structure and optimize operating systems, advocating for more compact and customizable designs that can effectively respond to the evolving and diverse requirements of modern computing environments. This work underscores the necessity of adaptability and efficiency in OS architecture. Korshun et al.\cite{korshun2023automation} focus on the role of AI and machine learning in OS management, proposing a comprehensive framework for their integration to improve resource allocation, security, performance, and overall system adaptability. Their research highlights the importance of intelligent systems capable of self-optimization and dynamic adaptation to changing conditions. Furthermore, Ranasinghe et al.\cite{ranasinghe2009artificial} assess the impact of various AI techniques on critical aspects of distributed operating systems, such as resource management, fault tolerance, and security. Safarzadeh et al.\cite{safarzadeh2021artificial} specifically examine the improvements that machine learning can bring to OS resource management, while works like~\cite{sibai2023},~\cite{qiu2020survey}, and~\cite{sk2022literature} address security concerns in operating systems. Additionally, Pereira et al.\cite{pereira2021learning} explores the use of AI approaches in software configuration management.

While these works have provided valuable insights and advancements, they often concentrate on specific types of operating systems or may now be considered somewhat outdated, particularly given the recent emergence and rapid development of large language models~\cite{zhao2023survey}. As we move forward, it will be crucial to explore how AI technologies can be integrated into operating systems to drive the next wave of innovation in computing.

\definecolor{connect-line}{RGB}{0,0,0}
\definecolor{middle-color}{RGB}{255,255,255}
\definecolor{leaf-color}{RGB}{255,255,255}
\definecolor{line-color}{RGB}{25,25,112}

\definecolor{black}{RGB}{0,0,0}


\definecolor{pure}{RGB}{112,25,25}
\definecolor{node}{RGB}{25,25,112}
\definecolor{graph}{RGB}{25,112,25}

\tikzstyle{pure-leaf}=[draw=pure,
    rounded corners,minimum height=1em,
    fill=leaf-color!40,text opacity=1, align=center,
    fill opacity=.5,  text=black,align=left,font=\scriptsize,
    inner xsep=3pt,
    inner ysep=1pt,
]
\tikzstyle{pure-middle}=[draw=pure,
    rounded corners,minimum height=1em,
    fill=middle-color!40,text opacity=1, align=center,
    fill opacity=.5,  text=black,align=left,font=\scriptsize,
    inner xsep=3pt,
    inner ysep=1pt,
]
    
\tikzstyle{node-leaf}=[draw=node,
    rounded corners,minimum height=1em,
    fill=leaf-color!40,text opacity=1, align=center,
    fill opacity=.5,  text=black,align=left,font=\scriptsize,
    inner xsep=3pt,
    inner ysep=1pt,
]
\tikzstyle{node-middle}=[draw=node,
    rounded corners,minimum height=1em,
    fill=middle-color!40,text opacity=1, align=center,
    fill opacity=.5,  text=black,align=left,font=\scriptsize,
    inner xsep=3pt,
    inner ysep=1pt,
]

\tikzstyle{graph-leaf}=[draw=graph,
    rounded corners,minimum height=1em,
    fill=leaf-color!40,text opacity=1, align=center,
    fill opacity=.5,  text=black,align=left,font=\scriptsize,
    inner xsep=3pt,
    inner ysep=1pt,
]
\tikzstyle{graph-middle}=[draw=graph,
    rounded corners,minimum height=1em,
    fill=middle-color!40,text opacity=1, align=center,
    fill opacity=.5,  text=black,align=left,font=\scriptsize,
    inner xsep=3pt,
    inner ysep=1pt,
]

\tikzstyle{leaf}=[draw=line-color,
    rounded corners,minimum height=1em,
    fill=leaf-color!40,text opacity=1, align=center,
    fill opacity=.5,  text=black,align=left,font=\scriptsize,
    inner xsep=3pt,
    inner ysep=1pt,
    ]
\tikzstyle{middle}=[draw=line-color,
    rounded corners,minimum height=1em,
    fill=middle-color!40,text opacity=1, align=center,
    fill opacity=.5,  text=black,align=left,font=\scriptsize,
    inner xsep=3pt,
    inner ysep=1pt,
    ]
\begin{figure*}[t]
\centering
\resizebox{\linewidth}{!}{%
\begin{forest}
  for tree={
    forked edges,
    grow=east,
    reversed=true,
    anchor=base west,
    parent anchor=east,
    child anchor=west,
    base=middle,
    font=\scriptsize,
    rectangle,
    line width=0.9pt,
    draw=connect-line,
    rounded corners,align=left,
    minimum width=2em,
    s sep=5pt,
    inner xsep=3pt,
    inner ysep=1pt,
  },
  where level=1{text width=4.5em}{},
  where level=2{text width=6em,font=\scriptsize}{},
  where level=3{font=\scriptsize}{},
  where level=4{font=\scriptsize}{},
  where level=5{font=\scriptsize}{},
[Categorization of Integrating AI into OS, black,rotate=90,anchor=north,edge=pure
    [OS Core, pure-middle, edge=pure, text width=4.5em
        [Process Mgmt., pure-middle, edge=pure, text width=6em
            [{%
                Fingler et al.~\cite{fingler2023towards};Yang et al.~\cite{yang2022improvement}; Mohammad et al.~\cite{hedayati2019multi}; Blocher et al.~\cite{blocher2021switches};\\
                Choi et al.~\cite{choi2022serving};
                Kim et al.~\cite{kim2023dream}; Hu et al.~\cite{hu2022primo}; Subedi et al.~\cite{subedi2019leveraging}
                Qiu et al.~\cite{qiu2020firm};\\
                Aaen Springborg et al.~\cite{aaen2023automatic}; Horstmann et al.~\cite{8869545};
                Balkanski et al.~\cite{balkanski2024energyefficientschedulingpredictions};\\
                Gupta et al.~\cite{gupta2024relief};
                Goel et al.~\cite{10.1145/3697835};
                Biswas et al.~\cite{biswas2023machine};
                Augustine et al.~\cite{augustine2021applying};\\
                Chen et al.~\cite{chen2020machine};
                Goodarzy et al.~\cite{goodarzy2021smartos};
                Tetzlaff et al.~\cite{tetzlaff2010intelligent}
                }, pure-leaf, text width=26em, edge=node]
        ]
        [I/O, pure-middle, edge=pure, text width=6em
            [{%
                Fingler et al.~\cite{fingler2023towards};
                Sun et al.~\cite{sun2021linux};
                Bateni et al.~\cite{bateni2020neuos};
                Hanel et al.~\cite{hanel2020vortex};\\
                Hao et al.~\cite{hao2020linnos};
                Arnold et al.~\cite{arnold2019enablingfddmassivemimo};
                Kurniawan et al.~\cite{kurniawan2025heimdall}
                }, pure-leaf, text width=26em, edge=node]
        ]
        [Storage Mgmt. , pure-middle, edge=pure, text width=6em
            [{%
                Akgun et al.~\cite{10.1145/3568429};
                Fingler et al.~\cite{fingler2023towards};
                Al-Saleh et al.~\cite{10184243};
                Liang et al.~\cite{liang2019cognitive};\\
                Hanel et al.~\cite{hanel2020vortex};
                Hildebrand et al.~\cite{hildebrand2020autotm};
                Hao et al.~\cite{hao2020linnos};
                Wu et al.~\cite{wu2024mitigating};\\
                Wang et al.~\cite{wang2024learnedftl};
                Gupta et al.~\cite{gupta2024relief};
                Balkanski et al.~\cite{balkanski2024energyefficientschedulingpredictions};
                Yuan et al.~\cite{yuan2022adaptive}
                }, pure-leaf, text width=26em, edge=node]
        ]
        [Memory Mgmt., pure-middle, edge=pure, text width=6em
            [{%
                Zhang et al.~\cite{zhang2022software};
                Lagar et al.~\cite{lagar2019software};
                Gupta et al.~\cite{gupta2024relief};
                Chang et al.~\cite{10.1145/3625549.3658659};\\
                De et al.~\cite{de2022using};
                Wu et al.~\cite{wu2024mitigating};
                Hildebrand et al.~\cite{hildebrand2020autotm};
                Miura et al.~\cite{9355915}
                }, pure-leaf, text width=26em, edge=node]
        ]
        [Network, pure-middle, edge=pure, text width=6em
            [{%
                Kwon et al.~\cite{9289257};
                Li et al.~\cite{9123561};
                Coronado et al.~\cite{9110260};
                Zhong et al.~\cite{10649557};\\
                Lee et al.~\cite{youngmin2023deeplearningbasedmodeling};
                Kavitha et al.~\cite{kavitha2019operating};
                Sedláček et al.~\cite{10575154};
                Veliyath et al.~\cite{10677020};
                }, pure-leaf, text width=26em, edge=node]
        ]
        [Security, pure-middle, edge=pure, text width=6em
            [{%
                Benabderrahmane et al.~\cite{benabderrahmane2024hack};
                Qin et al.~\cite{qin2019msndroid};
                Cheng et al.~\cite{cheng2024lightweight};
                Perez et al.~\cite{perez2022development};\\ 
                Maasmi et al.~\cite{9665597};
                Wihar et al.~\cite{wihar2024novel};
                De Wit et al.~\cite{panman2022dynamic};
                Jurečková et al.~\cite{jurevckova2024online};\\
                Cheng et al.~\cite{cheng2023conditional};
                Alzaylaee et al.~\cite{Alzaylaee_2020};
                Su et al.~\cite{su2025secure}
            }, pure-leaf, text width=26em, edge=node]
        ]
        [OS GUI/CLI, pure-middle, edge=pure,text width=6em
            [{%
                Mei et al.~\cite{mei2024aios};
                Wu et al.~\cite{wu2024copilot};
                Xu et al.~\cite{xu2024osagent};
                Shi et al.~\cite{shi2024commandspromptsllmbasedsemantic};
                He et al.~\cite{he2024pc};\\
                Christianos et al.~\cite{christianos2024lightweight};
                Kamath et al.\cite{kamath2024herding};
                Zhang et al.~\cite{zhang2024enhanced};
                Ye et al.~\cite{ye2025mobile};\\
                Zhao et al.~\cite{zhao2025stackpilot};
                Wang et al.~\cite{wang2025mobilea3gent}
                Zhang et al.~\cite{zhang2025agentcpm}; 
                Rivard et al.~\cite{rivard2025neuralos};
                }, pure-leaf, text width=26em, edge=node]
        ]
    ]
    [OS Ecosystem, node-middle, edge=node, text width=4.5em
        [Coding, node-middle, edge=node, text width=6em
            [{%
                Del Vecchio et al.~\cite{delvecchio2024dynamiccodeorchestrationharnessing};
                Zheng  et al.~\cite{zheng2023kenkernelextensionsusing};
                Zheng et al.~\cite{zheng2024kgent};
                Islam et al.~\cite{islam2024llmpoweredcodevulnerabilityrepair};\\
                Zhou et al.~\cite{zhou2025benchmarking}
                }, node-leaf, text width=26em, edge=node]
        ]
        [Verification, node-middle, edge=node,text width=6em
            [{%
                 Yang  et al.~\cite{yang2023kernelgptenhancedkernelfuzzing};
                Zhang et al.~\cite{zhang2024ecg};
                Zhang et al.~\cite{zhang2024selene}
                }, node-leaf, text width=26em, edge=node]
        ]
        [Ops\&Maintenance, node-middle, edge=node,text width=6em
            [{%
                Dusane  et al.~\cite{9453065};
                Akram et al.~\cite{10.1145/3208040.3208051};
                Zhang et al.~\cite{8758626};
                Kim et al.~\cite{kim2025logs};\\
                Zhong et al.~\cite{zhong2024logparser};
                Huang et al.~\cite{huang2025logrules};
                Guan et al.~\cite{guan2024wattscope}
                }, node-leaf, text width=26em, edge=node]
        ]
        [OS Tuning, node-middle, edge=node,text width=6em
            [{%
                Cui  et al.~\cite{cui2022linux};
                Wu et al.~\cite{9078880};
                Herzog et al.~\cite{10.1145/3447555.3466566};
                Chen et al.~\cite{chen2024autoos};\\
                Lin et al.~\cite{lin2025byos};
                Cortez et al.~\cite{cortez2017resource}
                }, node-leaf, text width=26em, edge=node]
        ]
    ]
]
\end{forest}
}
\caption{A categorization of integration of AI and OS.}
\label{fig:taxonomy-techniques}
\end{figure*}

\section{What OS Modules Are Often Enhanced?}
\label{sec:module}


Modern operating systems sit at the intersection of increasingly diverse hardware and complex application demands. They must allocate resources, enforce isolation, and deliver consistent performance across highly dynamic environments. While traditional kernels rely on fixed heuristics, such static approaches fail to adapt to evolving workloads and heterogeneous contexts~\cite{schupbach2012tackling}. These limitations motivate the integration of intelligent mechanisms that enable operating systems to learn, adapt, and optimize continuously.

Figure~\ref{fig:taxonomy-techniques} provides an overview of how AI techniques are being integrated into modern operating systems. The taxonomy differentiates between core OS modules—including scheduling, inter-process communication (IPC), I/O, storage, memory management, networking, and security—and the broader OS ecosystem, which covers interfaces, coding, verification, operations and maintenance, and system tuning. Representative research efforts are identified in each category, illustrating the wide range of AI applications across both low-level kernel functionality and higher-level system services. This categorization demonstrates that intelligence is no longer confined to isolated optimizations but increasingly permeates the entire operating system stack and its surrounding ecosystem.

\subsection{Core OS}
\subsubsection{Process Management}

Process management is a core subsystem of the operating system, encompassing process creation and termination, scheduling, synchronization, IPC, and resource allocation. Together, these functions determine how effectively an OS shares CPU time, manages memory, and coordinates tasks across heterogeneous hardware. They ensure fair and efficient use of resources, directly influencing task latency, utilization, and overall user experience~\cite{li2023batch,liu2022multi}. However, as hardware architectures and application workloads grow increasingly complex, traditional schedulers based on fixed, heuristic-driven rules often struggle to adapt. Such approaches are inflexible and perform poorly in dynamic or heterogeneous environments, leading to inefficiencies in managing variable workloads and resource contention. Recent research demonstrates that AI techniques are being integrated into these modules. In scheduling, for example, AI-augmented policies improve fairness and responsiveness by dynamically adapting to workload characteristics, surpassing the limitations of static heuristics employed in conventional schedulers.

On individual hosts, ML techniques can refine ticket distribution, improve adaptability, and adjust scheduling variables with greater precision. The goal is to ensure fair allocation of CPU time and memory while maintaining system performance and minimizing task completion time. Numerous studies have sought to optimize existing process scheduling algorithms, such as the Completely Fair Scheduler (CFS), which is widely used in the Linux kernel to allocate CPU time equitably among processes. However, the vanilla CFS exhibits significant limitations in multicore environments, where its complexity grows rapidly~\cite{lozi2016linux}. Chen et al.\cite{chen2020machine} observed that while CFS achieves high CPU utilization, it overlooks contention for lower-level hardware resources. To address this, they proposed an ML-based, resource-aware load balancer for Linux that employs a lightweight data collection method for training. Their approach uses a multi-layer perceptron (MLP) model trained to replicate CFS load-balancing decisions, combined with an in-kernel inference mechanism for real-time scheduling. Leveraging eBPF for dynamic tracing, the MLP provides accurate approximations of CFS decisions with low overhead and minimal additional latency, thereby enabling efficient load management. Building on this work, Fingler et al.\cite{fingler2023towards} ported the model and algorithm to CUDA and integrated them into a kernel module to further reduce inference cost and improve efficiency.

Goodarzy et al.\cite{goodarzy2021smartos} also challenge the vanilla CFS for its limitations in allocating CPU, memory, I/O, and network bandwidth. To address this, they propose SmartOS, an operating system that automatically learns which tasks the user considers most important at a given time. Based on these learned preferences, SmartOS dynamically adjusts the allocation of system resources, prioritizing CPU, memory, I/O, and network bandwidth for the tasks the user is actively focused on. The authors demonstrate a Linux-based implementation and show that a reinforcement learning approach can quickly learn and adapt resource allocation to meet user demands. Similarly, Kim et al.\cite{kim2020memory} identify CFS’s difficulties in managing excessive memory traffic generated by deep learning services. They propose a memory-aware fair-share scheduling algorithm aimed at reducing the vulnerability of quality-of-service (QoS) applications to memory-related interference.

Beyond process scheduling on individual hosts, machine learning techniques have also been applied to improve resource allocation in computer clusters, such as supercomputers. Springborg et al.~\cite{aaen2023automatic} present Chronus, a Python application designed to work alongside the widely used Simple Linux Utility for Resource Management (SLURM) scheduler. Chronus performs extensive benchmark tests on high-performance computing (HPC) clusters, systematically varying parameters such as core counts, processor speeds, and hyperthreading. During these tests, it records both performance metrics and energy consumption, producing data used to train an ML model that identifies patterns and predicts energy-efficient configurations for specific jobs and systems. Benchmark results demonstrate substantial energy savings, suggesting that this approach can be scaled to broader HPC deployments, enabling both ecological benefits and cost efficiency without compromising computational performance.

In parallel architectures, scheduling and task allocation must account for both interprocessor communication overhead and power consumption, which are often the primary performance bottlenecks. Conventional heuristics rely on conservative assumptions and are unable to capture dynamic runtime behavior. Tetzlaff et al.~\cite{tetzlaff2010intelligent} address this limitation by using machine learning to automatically generate architecture- and application-specific heuristics. Their method predicts runtime behavior during an offline training phase, enabling task mapping that is both power- and communication-aware without introducing additional compile-time costs. In doing so, scheduling is extended beyond fairness and load balancing to encompass energy-efficient, IPC-aware optimization.

IPC mechanisms are critical for coordinating processes and enabling data exchange, but they also introduce both performance overheads and potential security vulnerabilities. Recent research has shown how AI can enhance IPC along these two dimensions. On the security side, host-based intrusion detection has leveraged IPC traces in memory to uncover malicious behaviors. By capturing memory-resident IPC exchanges, constructing graphs of process communication, and applying graph neural networks, researchers have demonstrated accurate classification of malicious nodes within process graphs~\cite{augustine2021applying}. This transforms IPC from a passive communication channel into an active sensor for intrusion detection. On the efficiency side, large-scale scientific workflows face costly data staging and exchange overheads. DESTINY~\cite{subedi2019leveraging} addresses this by learning data access patterns with machine learning, proactively delivering data, and, when possible, bypassing traditional IPC through shared memory exposure. This approach reduces communication costs and achieves up to 75\% faster read response times. Together, these works illustrate how AI-enhanced IPC can simultaneously improve system security and data movement efficiency, expanding the role of IPC beyond traditional synchronization into a performance- and security-critical subsystem of the OS.


\paragraph{Insights and Challenges} The rapid evolution of workloads and the increasing complexity of runtime environments have driven the development of new scheduling strategies beyond traditional heuristic-based approaches. AI tools are central to this paradigm shift, enabling operating system schedulers to make more intelligent and dynamic resource allocation decisions. While conventional schedulers rely on static rules and heuristics, AI-centric approaches leverage data-driven models to adapt to complex workload patterns, resource contention, and diverse system constraints. These advancements are evident in both individual hosts and large-scale high-performance computing clusters, where AI techniques have led to improvements in fairness, efficiency, and adaptability. AI schedulers can optimize system performance in real-time, predict application behaviors more accurately, and balance competing objectives such as throughput, latency, and energy consumption.

However, integrating AI into OS scheduling introduces new challenges. Running models within the OS kernel or cluster managers increases computational overhead, making it essential to balance inference accuracy with low latency. ML schedulers also depend on large volumes of high-quality training data, which can be difficult to obtain and maintain in dynamic, heterogeneous environments. Ensuring scalability across multicore systems and generalizability to unseen scenarios remains an open issue. Additionally, embedding ML models into existing system architectures increases the complexity of design and implementation.

\subsubsection{I/O Layer}

The I/O subsystem plays a crucial role in ensuring predictable and efficient performance under diverse workloads, particularly in high-concurrency environments. Traditional heuristic-based scheduling often fail to capture the nonlinear interplay between workloads, device internals, and performance outcomes, leading to suboptimal throughput and unpredictable latency behaviors. To address these limitations, recent research has applied AI mechanism to enhance I/O performance like scheduling, parameter tuning, and admission control.

Predictability is a fundamental requirement of the operating system I/O subsystem, as it directly impacts the ability to deliver consistent and reliable performance across varying workloads and operational conditions. In modern computing environments, the I/O layer must support diverse applications ranging from latency-sensitive services to high-throughput data processing. Ensuring predictable I/O behavior is therefore essential for maintaining Quality of Service (QoS) guarantees and meeting Service Level Agreements (SLAs)~\cite{dean2013tail,hauser2018predictability}. Building on the theme of predictability, LinnOS~\cite{hao2020linnos} integrates a lightweight neural network directly into the kernel I/O path. Treating SSDs as black-box devices, it performs fine-grained per-I/O inference to predict device behavior in real-time, allowing applications to anticipate whether their performance expectations will be met. Without requiring hardware modifications or filesystem changes, LinnOS reduces I/O latency by up to 40\% and improves throughput by up to 3$\times$ under contention-heavy workloads, showing how ML can enhance I/O predictability at runtime.

More recently, Kurniawan et al.~\cite{kurniawan2025heimdall} advances ML-powered I/O admission control for flash storage. Unlike heuristics or earlier ML solutions, Heimdall executes a full machine learning pipeline—including accurate period-based labeling, three-stage noise filtering, in-depth feature engineering, and hyperparameter tuning—achieving decision accuracy of up to 93\%. It introduces deployment optimizations (e.g., Python-to-C conversion, quantization, and joint inference) to reach sub-µs inference latency with only 28KB memory overhead. Evaluations on production traces from Microsoft, Alibaba, and Tencent show that Heimdall reduces average I/O latency by 15–35\% compared to state-of-the-art heuristics and up to 2$\times$ compared to baseline approaches. Furthermore, Heimdall has been successfully deployed at the Linux kernel block layer and integrated with Ceph, proving its practicality in both single-node and distributed environments.


\paragraph{Insights and Challenges}
The adoption of AI in the I/O layer shows strong potential to transform scheduling and admission control from static, heuristic-based policies into adaptive, data-driven mechanisms. Systems such as LinnOS demonstrate how lightweight AI models embedded in the kernel can provide per-I/O inference of device behavior, enabling predictable performance. More recent work such as Heimdall illustrates how carefully engineered AI pipelines can deliver robust admission control with sub-microsecond inference overhead, substantially reducing both average and tail latencies. Together, these advances suggest that the I/O layer can evolve into a self-optimizing subsystem capable of balancing high throughput and low latency even under highly variable workloads.

Despite these advances, key challenges remain. First, inference overhead is critical: AI models must remain lightweight enough to meet microsecond-level latency constraints while maintaining sufficient accuracy to guide admission and scheduling. Second, the dynamic variability of device internals makes it difficult for static models to sustain performance over time, pointing to the need for adaptive or continuously updated model pipelines. Third, integrating AI-driven I/O control into production systems such as the Linux kernel or distributed storage frameworks raises concerns about compatibility, deployment complexity, and long-term maintainability. Addressing these challenges is essential to ensure that AI-enhanced I/O systems are both practical and reliable in real-world environments.

\subsubsection{Storage Management}

At the storage management level, the operating system bridges logical data requests with the underlying hardware. This layer handles critical functions such as logical-to-physical address translation, wear management, and error handling—factors that directly influence both the performance and endurance of flash-based and emerging non-volatile memory (NVM) devices. Unlike the I/O subsystem, which primarily governs request scheduling and admission, storage management operates closer to the hardware, determining how data is placed, accessed, and maintained.

Wang et al.~\cite{wang2024learnedftl} propose LearnedFTL, a machine learning–enhanced Flash Translation Layer (FTL) design at the page level. By modeling the quasi-linear relationship between logical and physical page numbers, LearnedFTL mitigates irregularities in flash address mapping that frequently cause double reads during random access. Training of learned indexes is integrated into the garbage collection process, keeping models current with minimal overhead. To guard against mispredictions, a lightweight bitmap filter ensures correctness in address translation. Compared with state-of-the-art FTL designs, LearnedFTL reduces translation-induced double reads by 55.5\% and achieves improvements of up to 12.2× in P99 tail latency and 8.2× in average throughput under random workloads.

Reliability is another central concern in storage management, particularly for NVM technologies that exhibit distinctive failure modes. Wu et al.~\cite{wu2024mitigating} present LearnWD, a machine learning–based approach to mitigating the write-disturbance (WD) problem in NVMs such as phase-change memory (PCM), ReRAM, and STT-RAM. Using clustering techniques, LearnWD classifies stale data blocks by susceptibility to errors and employs out-of-place updates to remap writes onto more resilient regions. This speculative remapping lowers the frequency of WD errors while maintaining efficiency. Evaluations across 15 real-world datasets show that LearnWD reduces WD errors by 20.1\%, decreases write latency by 11.0\%, and extends device endurance by 21.9\%.

Collectively, these works demonstrate how machine learning can advance storage management by improving both address translation efficiency and memory reliability. Whereas I/O-layer innovations primarily optimize request-level policies, approaches such as LearnedFTL and LearnWD embed intelligence deeper in the data path, enabling storage systems to deliver higher performance and longer lifetimes across diverse workloads and hardware technologies.

\paragraph{Insights and Challenges}  
\begin{figure}[h]
    \centering
    \includegraphics[width=1\linewidth]{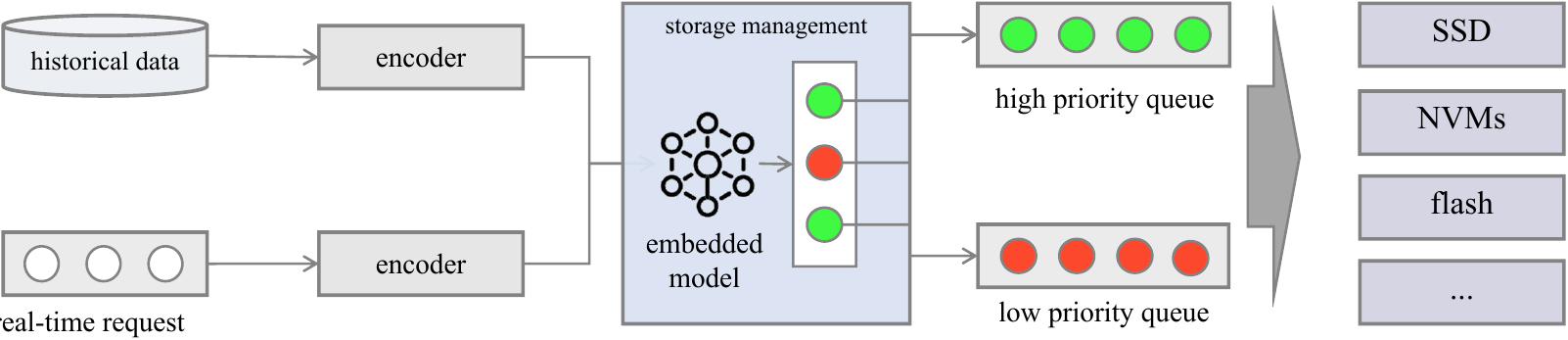}
    \caption{General ML for Block Layer Enhancement Workflow}
    \label{fig:storage}
\end{figure}

Recent advances demonstrate that integrating machine learning into the block layer can substantially improve both performance and reliability. As illustrated in Figure~\ref{fig:storage}, a common design pattern is to treat ML models as black boxes that predict the future state of read/write requests. Based on these predictions, the system selects and executes the operation most likely to meet performance goals. These studies reveal a key insight: the block layer, traditionally governed by rigid heuristics, can be reimagined as a self-optimizing component that adapts in real-time to workload diversity and device variability.

Nonetheless, several challenges remain. First, the stringent latency requirements of block-level operations place tighter constraints on inference overhead than at higher layers, necessitating models that are both lightweight and accurate. Second, the heterogeneity of modern storage devices—spanning NAND flash, PCM, ReRAM, and STT-RAM—complicates the development of models that can generalize across technologies. Third, block-layer ML systems must remain robust in the face of device-specific failures, workload bursts, and long-term degradation, ensuring that learned models do not compromise system reliability. Overcoming these challenges will be essential for deploying ML-powered block layers at scale in production environments.

\subsubsection{Memory management}
Memory management and storage management are fundamental components of an operating system, each responsible for optimizing resource utilization and providing abstraction for different hardware resources. Memory management focuses on volatile primary memory (RAM), abstracting physical memory into virtual memory for fast, temporary storage. 

With the emergence of advanced hardware technologies such as 3D-stacked memory, significant improvements in memory bandwidth and latency have been realized. However, traditional heuristic-based memory address mapping algorithms often fail to fully leverage these advancements. To bridge this gap, Zhang et al.~\cite{zhang2022software} propose Software-Defined Address Mapping (SDAM), a hardware-software co-design mechanism that dynamically adjusts address mapping strategies based on program access patterns, thereby maximizing hardware utilization in 3D memory. SDAM further integrates machine learning techniques to automatically identify and cluster access patterns of key program variables, enabling efficient and adaptive address mapping while minimizing overhead. This approach not only alleviates the burden on programmers but also allows the system to dynamically optimize memory performance for diverse workloads.

\begin{figure}[h]
    \centering
    \includegraphics[width=0.64\linewidth]{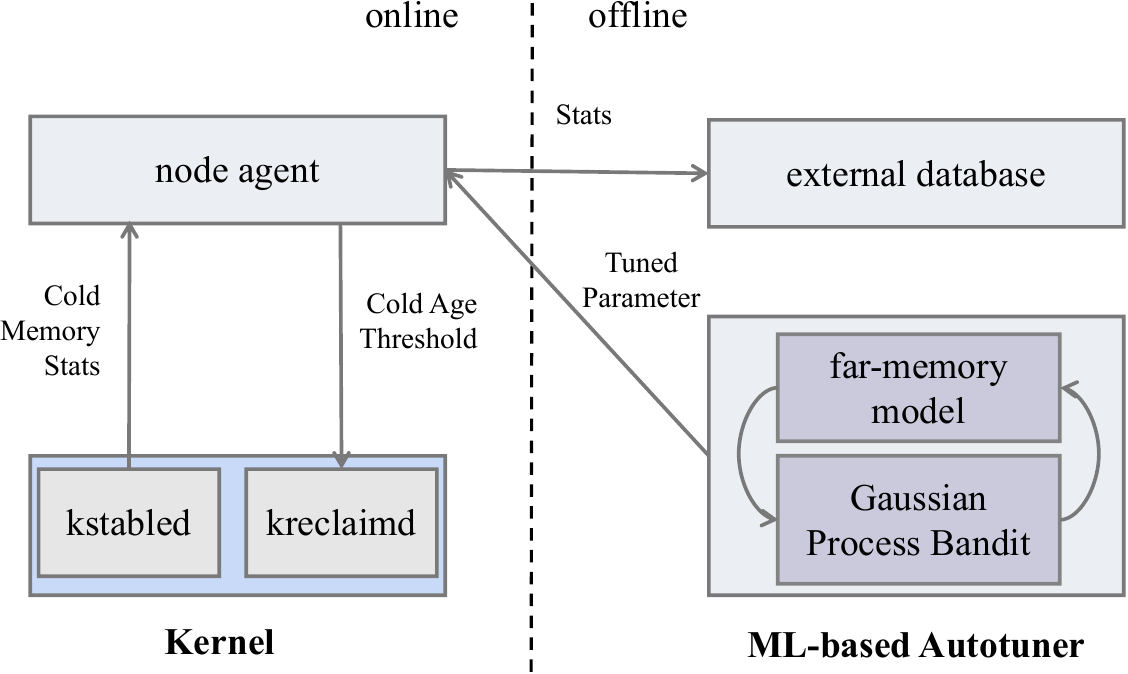}
    \caption{Workflow for the software-defined far memory system for warehouse-scale computers}
    \label{fig:sdam}
\end{figure}

In parallel with the advancements in high-performance memory technologies, new memory paradigms such as far memory are reshaping how memory capacity is scaled in modern systems~\cite{10.1145/3317550.3321433}. Addressing similar challenges of dynamic workloads and cost efficiency, Lagar-Cavilla et al.~\cite{lagar2019software} introduce a software-defined far memory system for warehouse-scale computers (Figure.~\ref{fig:sdam}) to address increasing memory demands and cost challenges. By leveraging zswap, a Linux kernel mechanism, the system proactively compresses and migrates cold memory pages to create a flexible, low-cost far memory tier in software. It employs ML-based autotuning to optimize performance under dynamic workloads while meeting strict service-level objectives (SLOs). In the online section, the kernel manages memory with two key modules: kstaled and kreclaimd. kstaled tracks page ages by scanning page table entries, while kreclaimd reclaims cold pages exceeding the Cold Age Threshold, moving them to zswap to free DRAM. The Node Agent bridges the kernel and offline components by collecting statistics, adjusting the Cold Age Threshold dynamically, and exporting data for further analysis. In the offline section, an ML-based Autotuner uses far memory data and a Gaussian Process Bandit algorithm to optimize parameter configurations for memory savings and performance. The tuned parameters are fed back to the Node Agent, enabling continuous system optimization. Successfully deployed at Google since 2016, the system achieves significant memory cost savings (4–5\% TCO reduction) with minimal performance impact, demonstrating practicality and scalability for large-scale production environments.

Building on the theme of leveraging machine learning to optimize memory systems and workload performance, Rocha et al.~\cite{de2022using} introduce PredG, a machine learning framework specifically designed to improve graph processing efficiency on Non-Uniform Memory Access (NUMA) systems. Unlike previous examples that focus on general-purpose optimization, PredG targets a specific application domain—graph processing. It identifies the optimal thread and data mappings for NUMA machines by utilizing graph features to train an artificial neural network for precise predictions. Notably, this framework eliminates the need for application execution after training, making it a powerful and efficient tool for optimizing graph workloads on NUMA architectures. This work aligns with the broader trend of automated graph machine learning, which aims to discover optimal hyper-parameters and neural architectures for diverse graph tasks and datasets without manual intervention.

New deployment paradigms are also emerging in memory management, reflecting the evolving landscape of modern computing systems. Virtualization, a critical component of contemporary computing environments, introduces unique challenges and opportunities for optimizing memory and resource management. Continuing the exploration of applying machine learning techniques to these areas, Miura et al.~\cite{9355915} tackle the issue of cache management in virtualized environments, which involve both guest and host operating systems. In such setups, the second cache (managed by the host OS) often experiences poor locality of reference, making traditional algorithms such as LRU less effective. To address this, the authors propose leveraging LSTM, a recurrent neural network model designed for time-series data, to enhance cache replacement decisions in the second cache. Through simulations and real-world workload evaluations, their approach demonstrates significant improvements in cache hit ratios and I/O throughput. This work highlights the potential of machine learning to address inefficiencies in traditional memory and storage systems, showcasing how advanced models can adapt to complex, dynamic environments. It aligns with the broader trend of using ML to optimize memory management, offering innovative solutions for improving performance in virtualized systems.

\paragraph{Insights and Challenges} A significant trend in recent years is the emergence of new hardware and deployment paradigms, which have opened up exciting opportunities for researchers to explore innovative methods. Unlike traditional heuristic-based approaches, software-defined memory management frameworks, such as SDAM and software-defined far memory systems—leverage machine learning to dynamically adapt to workload behaviors. This not only reduces manual intervention but also improves resource utilization. Despite significant advancement in memory management, several key challenges remain. First, many existing approaches incur high inference overhead, which can limit their practicality, particularly in resource-constrained or latency-sensitive environments. Second, current methods often focus on specific aspects of optimization and rely heavily on domain expertise for design and tuning. This lack of generality highlights the need for more versatile and intelligent solutions that can adapt to a wide range of hardware and workload configurations. Future research should prioritize developing low-overhead, general-purpose methods that balance performance optimization with reduced reliance on specialized knowledge, enabling more efficient and scalable memory management frameworks.

\subsubsection{Network}
The rapid advancement of the Internet and mobile communication industries has greatly increased the complexity of modern network systems. This evolution entails more sophisticated infrastructures, a wider diversity of connected devices and resources, and increasingly dynamic network architectures~\cite{7185326}. To meet the demands of this shifting environment, network systems must evolve to become more intelligent and agile, capable of adapting to heterogeneous device requirements and unpredictable operating conditions.

Artificial intelligence techniques, particularly deep neural networks (DNNs), have shown substantial promise for improving network traffic management. Kwon et al.~\cite{9289257} investigated the use of DNNs for classifying real network traffic collected from the ONOS (Open Network Operating System) platform. Experiments on simple network topologies demonstrated that DNNs can effectively support network packet classification. The study also emphasized that deploying DNNs in real-world scenarios must account for both user data packets and control packets required for network maintenance, as classification performance is highly sensitive to the composition of traffic data.

Another critical challenge in modern networking is sustaining optimal data transmission rates under dynamic conditions. Li et al.~\cite{9123561} address this issue with MLRA, a machine learning–based approach that identifies correlations among transmission rate, throughput performance, and link quality to enhance adaptive efficiency for the IEEE 802.11ac standard. MLRA integrates a two-level rate search procedure with a congestion detection mechanism to ensure both scalability and robustness. Experimental evaluations on commercial IEEE 802.11ac NICs demonstrate that MLRA consistently outperforms existing rate adaptation solutions, achieving performance gains of approximately 133\% to 658\% while incurring minimal overhead.

Beyond traffic and rate management, the integration of artificial intelligence (AI) with software-defined networking (SDN) is being explored to address broader challenges such as network resilience during natural disasters. Veliyath et al.~\cite{10677020} present an approach that leverages SDN, machine learning, and image processing to predict, respond to, and recover from flood-related disruptions. The method applies unsupervised K-Means clustering to satellite imagery for optimal node distribution and employs the supervised CatBoost algorithm with log transformation to forecast flood-affected areas. The framework simulates an SDN controller on ONOS, a platform noted for its self-healing capabilities, support for next-generation devices, and high performance, and further deploys a virtual network environment using Mininet. Simulation results indicate that this approach enhances self-healing and improves overall network performance.

\paragraph{Insights and Challenges}  
The rapid evolution of mobile network systems, driven by increasingly complex infrastructures and dynamic network structures, demands intelligent and adaptive management solutions. The integration of AI and ML into these systems offers significant transformative potential. Recent studies highlight the role of software-defined networking (SDN) combined with ML, demonstrating SDN’s ability to flexibly adapt to diverse infrastructures and policies. When augmented with ML, SDN illustrates the versatility of AI in strengthening network resilience, efficiency, and self-healing capabilities. These insights underscore the growing importance of AI-driven approaches in optimizing performance, adaptability, and disaster recovery within increasingly dynamic and complex environments.  

Despite this promise, several challenges remain. First, the performance of AI models—such as DNNs for traffic classification—depends heavily on the quality and diversity of training data. Classification accuracy varies across traffic types, necessitating attention to both delivery and maintenance packets. Second, while solutions like MLRA report substantial performance gains, their scalability in highly dynamic and heterogeneous environments with diverse devices and traffic patterns remains uncertain. Third, the computational overhead of deploying ML models, particularly in real-time scenarios such as traffic classification and disaster prediction, may overwhelm resource-constrained network devices. Fourth, many existing methods still lack sufficient generalization across platforms and problem domains. Addressing these challenges is essential to fully realize the potential of AI in modern mobile network systems.

\subsubsection{Security}
OS security is one of the most critical domains where AI techniques are increasingly employed to improve functionality and robustness.  

Advanced persistent threats (APTs) represent some of the most challenging cybersecurity risks, distinguished by their sophistication and ability to remain undetected for prolonged periods. They often circumvent traditional anomaly detection systems by exploiting vulnerabilities and employing deceptive strategies. Detecting APTs with ML is particularly difficult due to the scarcity of representative datasets and the severe imbalance between benign and malicious samples, both of which hinder accurate classification. To address this challenge, Benabderrahmane et al.~\cite{benabderrahmane2024hack} propose \textit{AE-APT}, a framework built on AutoEncoder (AE) architectures that span from conventional to Transformer-based designs. By integrating online learning–based anomaly detection, AE-APT supports real-time analysis and rapid response to emerging threats, providing a promising approach to one of the most complex problems in OS cybersecurity.

Malware detection is a key focus in AI-OS integration, particularly for Android, the most widely used mobile operating system. Qin et al.~\cite{qin2019msndroid} introduce MSNdroid, an advanced malware detector that leverages deep learning techniques to analyze native API calls, permissions, system API features, and other application-level data. By utilizing a Deep Belief Network and a comprehensive dataset of malicious and benign applications, MSNdroid achieves high detection accuracy with an impressively low false-negative rate, making it a reliable tool for combating Android malware. Similarly, Alzaylaee et al.~\cite{Alzaylaee_2020} propose DL-Droid, a deep learning-based system that employs dynamic analysis to detect malicious Android applications. By generating inputs based on application states, DL-Droid achieves a detection rate of up to 97.8\% using dynamic features alone and 99.6\% when combining dynamic and static features. This approach demonstrates the power of combining feature types to improve detection performance while enhancing code coverage. De Wit et al.~\cite{panman2022dynamic} further explore malware detection on Android by incorporating hardware-level data and classification techniques, showing the potential of app-specific metrics to boost detection rates and contribute to the growing field of AI-enhanced OS security.

Android is the most widespread smartphone operating system, and its popularity has attracted attackers who develop various malicious applications. On the defensive front, considerable research has been dedicated to identifying the types and states of Android applications based on their system-level behavior. Maasmi et al.~\cite{9665597} use a ML technique to detect whether an application is running in the foreground or not. This technique aims to enhance the accuracy of behavioral malware detection by providing informative priors or contextual metadata on the app's state to malware identification models. The results indicate that a structured ML pipeline, which first identifies the app and then detects its mode, can achieve substantially higher accuracy compared to direct mode identification.

Ransomware remains one of the most critical cybersecurity threats, targeting systems worldwide and causing substantial financial and data losses. Wihar et al.~\cite{wihar2024novel} propose a novel approach to ransomware detection using an LSTM-based model capable of analyzing system call sequences in real-time. This model identifies anomalous behavior indicative of ransomware activities with high accuracy and low false-positive rates. Integrated into the Linux kernel, this method facilitates seamless communication and classification while maintaining efficient resource utilization. Experimental evaluations show that the LSTM model provides a proactive defense mechanism against ransomware threats, significantly enhancing the security framework of Linux systems through its adaptive learning capabilities.

Extending beyond detection, Jurečková et al.~\cite{jurevckova2024online} propose a ML-based approach for the online clustering of malicious samples into malware families. Using a weighted k-nearest neighbor classifier for known families and an online k-means algorithm for emerging malware, their method achieves cluster purities ranging from 90.20\% to 93.34\%. This approach focuses on the static analysis of portable executable files for the Windows operating system, offering valuable insights for categorizing malware and understanding evolving threats.

FingerAI ~\cite{perez2022development}  is an open-source OS fingerprinting tool that uses AI models instead of traditional rule-based methods to identify operating systems from network traffic. It features modular design, supports both active and passive data collection, and applies machine learning models trained on real-world fingerprinting datasets (from tools like nmap and p0f) to improve detection accuracy, especially for new or hardened systems. Evaluations in both virtual and real environments show that fingerAI achieves effective and reliable OS classification, addressing key limitations of conventional approaches.

\paragraph{Insights and Challenges}Machine learning is transforming OS security by offering advanced capabilities in threat detection, behavior analysis, and adaptive defense. Unlike traditional rule-based systems, ML excels at identifying complex patterns in large, dynamic datasets, enabling the detection of previously unknown or evolving threats. By analyzing system behaviors, network traffic, and user activity, ML can not only identify anomalies but also predict potential attacks, allowing for proactive mitigation. Furthermore, ML’s ability to integrate static and dynamic features provides deeper insights into system vulnerabilities and attack surfaces, enhancing overall security frameworks. However, the deployment of ML in OS security reveals critical insights: robust models require diverse and high-quality training data, as well as mechanisms to adapt to the ever-changing nature of cyber threats. Real-time learning and decision-making are increasingly essential, as static models struggle to keep pace with sophisticated attackers. Moreover, the integration of ML into OS workflows must balance accuracy, speed, and resource efficiency to be practical in real-world environments. These advancements suggest that ML has the potential to shift OS security from reactive to proactive, but achieving this requires addressing challenges related to scalability, data scarcity, and adversarial robustness.

\subsubsection{OS GUI/CLI}  
As the critical bridge between hardware and software, the operating system plays a pivotal role in shaping user interaction, thereby influencing how users engage with and perceive technology. Recent research seeks to revolutionize this role, aiming to redefine user interactions within computing environments.  

Personalized services tailored to individual needs can enhance both user satisfaction and loyalty. Interactive systems, enabled by well-designed interfaces, can better interpret user intentions and provide adaptive support. Zhang et al.~\cite{zhang2024enhanced} explore the integration of large language models (LLMs), machine learning, and interaction design into recommendation and operating systems. Their findings highlight the potential of these technologies to deliver more intelligent and personalized services, thereby advancing both recommendation research and user experience design.  

\begin{figure}[h!]
    \centering
    \begin{minipage}{0.45\linewidth}
        \centering
        \includegraphics[width=\linewidth]{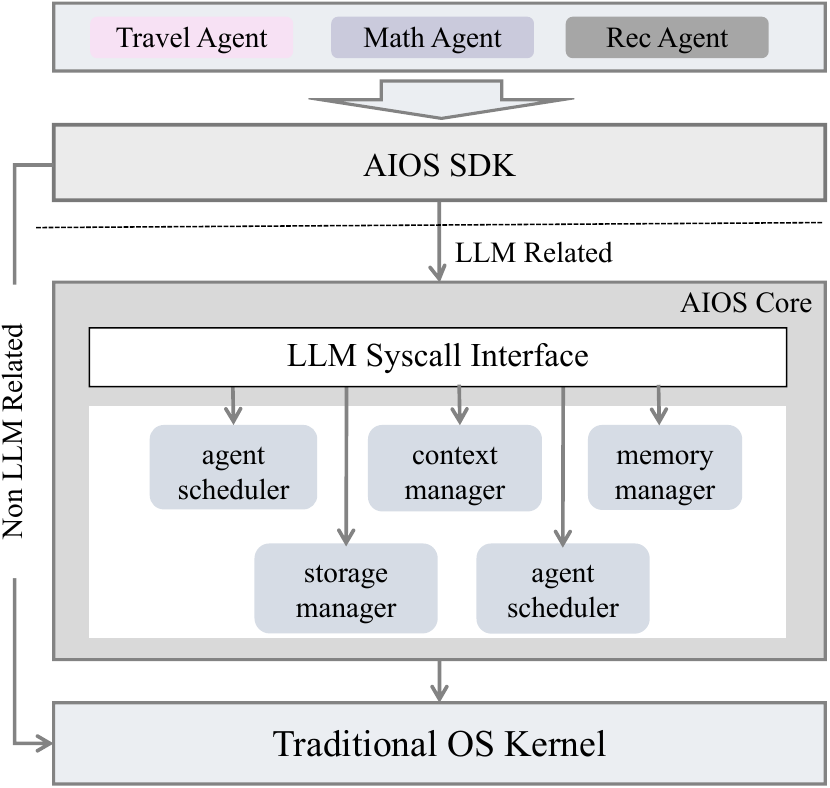}
        \caption{Architecture for AIOS}
        \label{fig:aiosarc}
    \end{minipage}
    \hfill
    \begin{minipage}{0.45\linewidth}
        \centering
        \includegraphics[width=\linewidth]{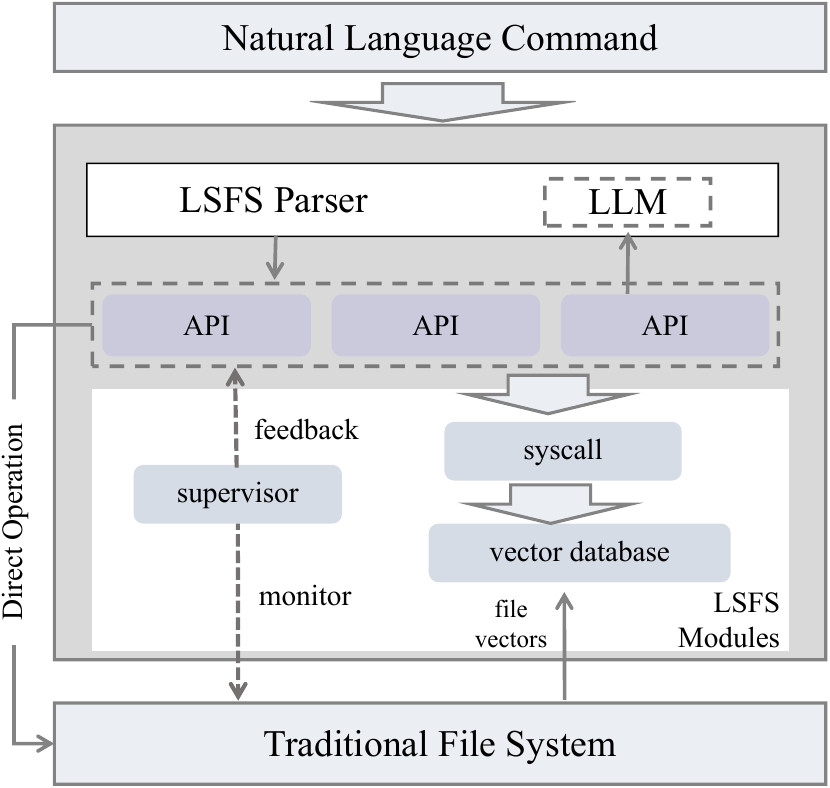}
        \caption{Architecture for LSFS}
        \label{fig:lsfs}
    \end{minipage}
\end{figure}

Beyond personalization, some studies propose deeper structural reforms to OS design through LLM integration. The AIOS-Agent ecosystem~\cite{ge2023llm,mei2024aios} positions LLMs as an “operating system with a soul,” embedding them into the OS kernel to enable intelligent decision-making and resource management, as illustrated in Figure~\ref{fig:aiosarc}. Within this framework, the LLM’s context window functions as a form of system memory, while external storage operates as a file system with advanced retrieval capabilities. Hardware and software are treated as peripherals and libraries, facilitating seamless agent–environment interactions. Natural language serves as the primary programming interface, lowering the barrier to software development and democratizing access to complex functionality.  

This ecosystem supports both single- and multi-agent applications capable of executing a wide range of tasks. Building on the AIOS concept, subsequent studies have extended the framework with additional components, enriching its functionality and broadening its applicability.

Expanding on this foundation, the CoRE (Code Representation and Execution) system~\cite{xu2024aios} introduces a unified framework for natural language programming, pseudo-code programming, and flow programming. CoRE leverages LLMs as interpreters to execute logically structured programs described in natural language, bridging the gap between human intent and machine execution. By defining a structured syntax, the CoRE language allows users to articulate program logic in a manner that is both accessible to non-programmers and precise enough for computational interpretation. Programs are decomposed into components such as step names, types, instructions, and connections, enabling clear and logical execution.

Within the AIOS ecosystem, the LLM-based Semantic File System (LSFS)~\cite{shi2024commandspromptsllmbasedsemantic} in Figure.\ref{fig:lsfs} redefines traditional file management by integrating LLMs for natural language-driven interactions and semantic understanding of file content. Unlike conventional systems that rely on metadata such as file names and timestamps, LSFS employs embedding vectors to index files based on their content, enabling advanced functionalities like semantic retrieval, grouping, and version control. Users can perform complex operations—such as file updates and change summarization—through simple natural language prompts, significantly streamlining file management. Key features include semantic indexing, natural language parsing for file operations, real-time change monitoring, and robust safety mechanisms to prevent unintended actions. Seamlessly integrated into the AIOS ecosystem, LSFS enhances intelligent multi-agent collaboration and supports capabilities such as task scheduling and knowledge integration. This innovation transforms file systems into intelligent, user-friendly components that are essential for modern AI-driven environments.

\subsection{OS Ecosystem}
Beyond the internal mechanisms of scheduling, memory management, and storage, AI tools are increasingly being applied to the broader operating system ecosystem. Rather than focusing solely on low-level kernel execution paths, this line of work explores how AI can support the development, configuration, operation, verification, and even education of operating systems.
\subsubsection{OS Coding}

Coding is a foundational activity in the operating system ecosystem, encompassing tasks such as kernel development, bug fixing, and vulnerability repair. These activities are traditionally labor-intensive and require deep expertise, given the complexity and criticality of OS code. Recent advances in AI tools, particularly LLMs, have shown strong potential to augment and automate these processes. Since source code can be treated as a specialized form of natural language, LLMs are well-suited for tasks such as specification synthesis, code generation, debugging, vulnerability repair, and fault localization. By leveraging the knowledge encoded from large-scale codebases and documentation, AI-driven approaches can reduce manual effort, improve developer productivity, and enhance the security and reliability of operating system software. 

The ability to modify and extend an operating system is essential for improving security, reliability, and performance. The extended Berkeley Packet Filter (eBPF) ecosystem has become the de facto standard for extending the Linux kernel. eBPF programs enable developers to inject new logic into the kernel that executes before or after existing functionality. Despite its flexibility, writing eBPF programs requires deep knowledge of operating system internals and compliance with strict control flow and data access constraints enforced by the eBPF verifier. To address this challenge, Zheng et al.~\cite{zheng2023kenkernelextensionsusing} proposed KEN, a framework that allows kernel extensions to be written directly in natural language. KEN leverages LLMs to synthesize eBPF programs from English prompts and ensures semantic alignment between the generated code and the user's intent through a combination of program comprehension, symbolic execution, and iterative feedback loops. The novelty of KEN lies in its integration of symbolic execution with LLM-based synthesis and comprehension, enabling more reliable program generation. Evaluation results show that KEN achieves 80\% correctness on a newly curated corpus of natural language prompts for eBPF programs, representing a 2.67$\times$ improvement over an LLM-only synthesis baseline.

Another critical challenge in OS coding is fault localization (FL), which aims to identify buggy code elements. While recent LLM agents have shown promising results on benchmarks such as SWE-bench, their effectiveness on the Linux kernel remains unclear due to the kernel’s massive codebase and complex dependencies. To address this, LinuxFLBench~\cite{zhou2025benchmarking} introduces a benchmark constructed from real-world Linux kernel bugs, enabling systematic evaluation of FL methods. An empirical study reveals that state-of-the-art LLM agents achieve only 41.6\% top-1 accuracy at the file level. To improve this, the authors propose \textit{LinuxFL\^+}, an enhancement framework that boosts FL accuracy by 7.2\%–11.2\% with minimal overhead.

Beyond bug discovery, AI tools are also being applied to vulnerability repair. SecRepair~\cite{islam2024llmpoweredcodevulnerabilityrepair} is a system designed to identify and automatically fix code vulnerabilities. Powered by LLMs, it integrates reinforcement learning and a semantic reward mechanism to enhance repair effectiveness. The authors also release a comprehensive instruction-based vulnerability dataset to support training and evaluation. Furthermore, SecRepair generates concise and contextually appropriate commit messages to explain the fixes, offering developers both automated vulnerability resolution and human-readable documentation of the patching process. This not only accelerates the repair workflow but also improves code security and maintainability.

\paragraph{Insights and Challenge} Recent advances demonstrate that AI tools can substantially improve the efficiency, security, and reliability of OS coding. The success of tools shows that AI can support multiple stages of the coding workflow: from vulnerability discovery in both general-purpose and embedded OSs, to automated patching with human-readable commit explanations, to systematic benchmarking and enhancement of fault localization. An important insight is that AI tools are not limited to a single coding task but can serve as versatile assistants across the software lifecycle, offering developers both automation and interpretability.  

Despite these achievements, significant challenges remain. First, the complexity and scale of production kernels like Linux pose difficulties for accuracy and scalability, as models trained on smaller benchmarks often underperform when applied to real-world code. Second, ensuring correctness and security of AI-generated outputs is critical: mis-specified fuzzing inputs, incorrect patches, or erroneous fault localization may introduce new risks instead of resolving existing ones. Third, the diversity of operating system environments—including general-purpose, embedded, and domain-specific OSs—complicates the generalization of AI-based coding tools, often requiring retraining, adaptation, or domain-specific enhancements. Finally, integrating these tools into established development pipelines (e.g., kernel patch submission, continuous integration, and security auditing) demands careful consideration of trust, transparency, and maintainability. Addressing these challenges will be essential to realize the full potential of AI-enhanced OS coding in practice.

\subsubsection{OS Ops\&Maintenance}

OS Ops are crucial for ensuring that systems function reliably and securely, thereby supporting the smooth operation of businesses and services that depend on them. Given the increasing complexity and scale of modern IT infrastructures, the potential for leveraging AI to improve OS Ops and maintenance is significant. 

A common starting point for such monitoring efforts is the analysis of system logs, which are the most accessible resources for engineers seeking clues about an OS's status. Building on this idea, Dusane et al.~\cite{9453065} introduce LogEA, a specialized tool for forensic investigations of Linux-based systems. LogEA uses natural language processing (NLP) and machine learning to analyze system logs, identifying security breaches, unauthorized access, and other anomalies. By detecting log messages with negative sentiment, LogEA flags and alerts users to potential events of interest, offering a targeted approach to anomaly detection. This focus on log analysis highlights the importance of leveraging existing OS outputs to uncover irregularities.

Also based on logs, while LogEA focuses on forensic analysis and post-event investigation, Akram et al.~\cite{10.1145/3208040.3208051} take a more proactive approach with Desh, a framework designed to diagnose and predict failures with short lead times. Desh applies a novel three-phase deep learning pipeline to logs, not only identifying chains of events that lead to failures but also predicting when and where failures are likely to occur. In the first phase, sequences of phrases leading to node failures are extracted, and LSTM Phase 1 is trained to identify and recognize such chains based on the training data. In the second phase, the chains formulated in Phase 1 are fed into LSTM Phase 2, which learns to account for the cumulative time differences ($\delta$ times) of the phrases within a failure chain, relative to the terminal phrase in that chain. During the inference phase, the chains learned in Phase 2 are used to estimate the lead times of future failures, along with predictions of the likely failure locations, based on test data that is completely independent of the training data. Desh utilizes an RNN with input/output layers and multiple hidden layers that form a stacked LSTM architecture, enabling effective training and prediction on system logs. Unlike LogEA, which alerts users to negative events retrospectively, Desh enables preemptive actions, such as migrating computation to healthy nodes, to minimize downtime. This shift from reactive forensics to proactive failure prediction showcases the versatility of ML in OS monitoring, as it can both analyze past events and anticipate future issues.
\begin{figure}[h]
    \centering
    \includegraphics[width=0.90\linewidth]{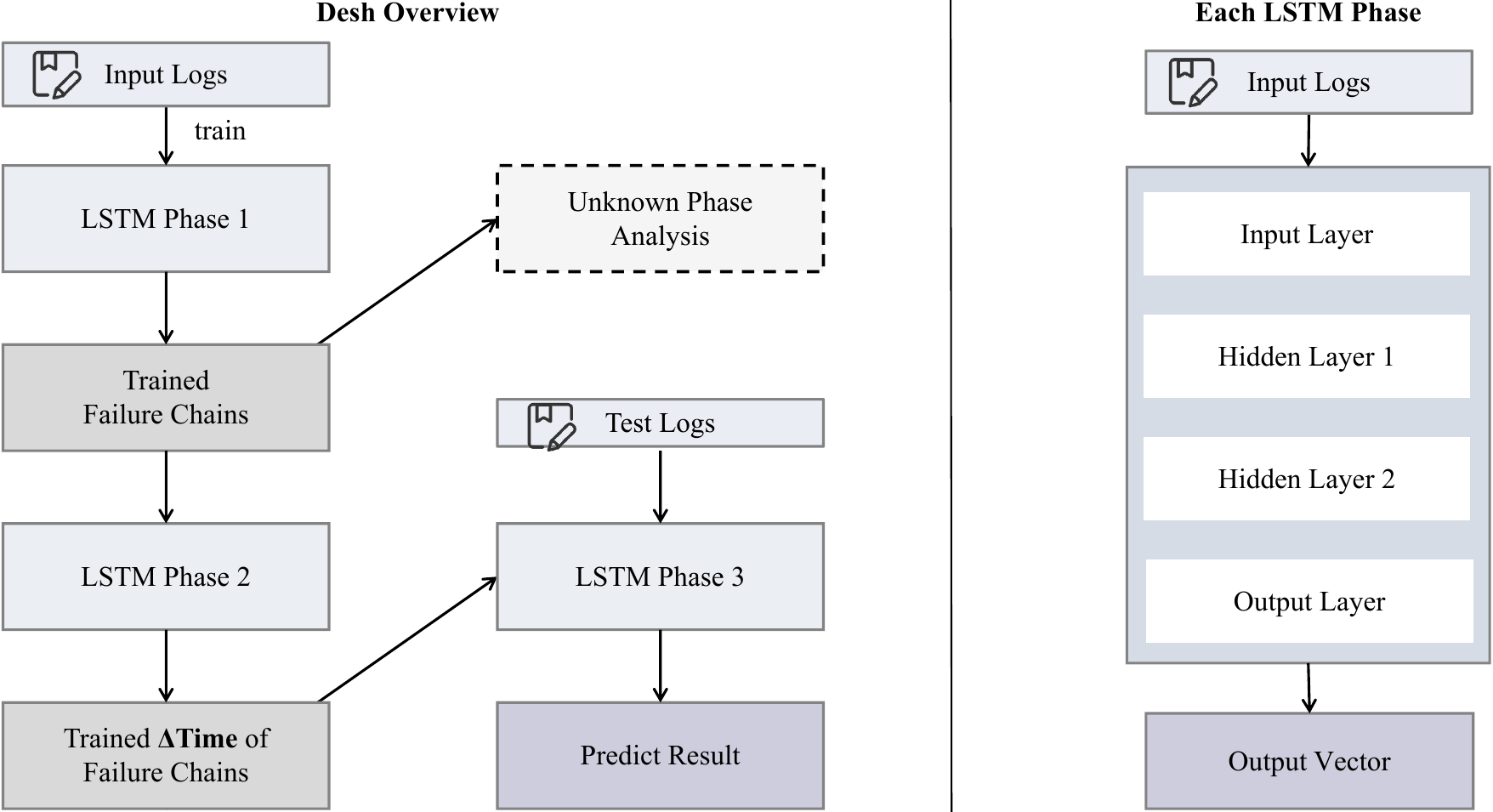}
    \caption{Workflow for Desh.}
    \label{fig:desh}
\end{figure}

In contrast to the log-based approaches above, Zhang et al.~\cite{8758626} propose a hardware-based solution for real-time workload forensics. Their method analyzes process execution flows by dividing them into consecutive frames and extracting Translation Lookaside Buffer (TLB) utilization features. These features are then fed into machine learning models to identify workloads at the granularity of individual processes. Unlike software-based workload forensics, which depend on OS services and can be compromised by software attacks, this hardware-based approach operates independently of the OS, offering a secure and efficient solution for monitoring workload behavior in real-time. This distinction highlights a critical difference: while tools like LogEA, Desh, and Maasmi’s pipeline leverage OS-level data for analysis, Zhang et al.'s method bypasses the OS entirely, focusing on hardware-level monitoring to enhance security and reliability.

Kim et al.~\cite{kim2025logs} introduced SAN2PATCH, which generates patches using only sanitizer logs and source code, thereby eliminating costly manual analysis. SAN2PATCH leverages multi-stage reasoning with LLMs to decompose the repair process into four tasks: vulnerability comprehension, fault localization, fix strategy formulation, and patch generation. By employing tree-structured prompting and rigorous validation, the system can produce diverse and functionally correct patches. Empirical results demonstrate its practicality: SAN2PATCH achieves a 79.5\% success rate on the VulnLoc dataset, significantly outperforming state-of-the-art tools such as ExtractFix and VulnFix, and further attains a 63\% success rate on the newly curated SAN2VULN dataset. Particularly noteworthy is its effectiveness in handling complex memory-related vulnerabilities, such as buffer overflows, with an 81.8\% repair rate. 

Recent work has also explored energy monitoring as a critical aspect of system operations and maintenance. Guan et al. present WattScope~\cite{guan2024wattscope}, a system for non-intrusively estimating the power consumption of individual applications in multi-tenant datacenters. Unlike existing application-level monitoring techniques that require privileged access and coordinated hardware–software support, WattScope operates using only aggregate server- and rack-level power measurements, which are already available in production environments. The key insight is that datacenter workloads often exhibit low variability, low magnitude, and high periodicity in their power characteristics, making them well suited to ML-based disaggregation techniques originally developed for building power monitoring. By adapting and extending these methods, WattScope is able to attribute server-level energy consumption to resident applications with high accuracy, achieving normalized mean absolute error around 10\% in evaluation with production traces. This capability provides datacenters with a practical tool for externally monitoring and managing application-level energy usage without intrusive instrumentation.


\paragraph{Insights and Challenges} Using AI tools as an external monitor for OS operations offers new opportunities to enhance reliability, security, and efficiency. This external monitoring uses logs, behaviors, and workload patterns to provide insights and predictions, enabling both forensic analysis and proactive failure prevention. However, challenges persist. Log-based methods depend on high-quality logs, and ensuring low overhead while maintaining real-time responsiveness can be difficult. Generalizability of these tools across different workloads and hardware configurations remains limited, requiring extensive retraining. Additionally, security and interpretability of ML solutions must be improved to prevent adversarial exploitation and ensure trust in critical systems. Monitoring is just one part of OS Ops. To truly reduce the burden on engineers, future AI-driven tools need to be more intelligent, general-purpose, and capable of self-healing, enabling systems to automatically detect, diagnose, and resolve issues without human intervention.

\subsubsection{OS Tuning}
Tuning an operating system involves configuring parameters and kernel options to balance performance, resource efficiency, and reliability across diverse workloads and hardware platforms. This task is notoriously complex: modern kernels expose thousands of parameters and configuration options, many of which interact in nonlinear ways that are difficult for human experts to optimize manually. Suboptimal tuning can lead to degraded performance, wasted resources, or even instability. 

Cui et al.~\cite{cui2022linux} propose an ML-based parameter recommendation model for Linux I/O scheduling. By combining the XGBoost prediction algorithm with Bayesian optimization, the system automatically tunes critical parameters of the Deadline scheduler (e.g., \texttt{read\_expire}, \texttt{write\_expire}, \texttt{fifo\_batch}, \texttt{writes\_starved}, and \texttt{nr\_requests}), which are otherwise difficult to optimize manually. Experimental results on Filebench workloads show that this approach improves throughput by 21.6\% and IOPS by 24.1\% under high concurrency, demonstrating the effectiveness of ML-guided parameter tuning in the I/O path.

Chen et al.'s AutoOS~\cite{chen2024autoos} utilize LLMs to address challenges in operating system configuration optimization, but they focus on different domains and employ distinct methodologies. AutoOS tackles the challenge of Linux kernel configuration optimization, a critical task for AIoT devices with constrained hardware resources. AutoOS models the kernel configuration space as a dynamic tree structure and employs a state-machine-based framework to traverse, prune, propose, and correct configuration options. The LLM frontend identifies performance-critical options, while the backend applies these refinements to boost system performance. Experiments showed that AutoOS could enhance UnixBench scores by up to 25\% compared to vendor-provided defaults, achieving superior results with minimal human input. By automating the debugging and refinement of kernel configurations, AutoOS reduces the time and expertise required for manual optimization, addressing the complexity of configuring over 15,000 kernel options. AutoOS leverage LLMs to simplify traditionally complex OS tasks, minimizing the need for manual tuning and expert knowledge. It also demonstrate the ability of LLMs to adapt dynamically to changing requirements. Moreover, this framework utilizes LLMs' reasoning and zero-shot learning capabilities to autonomously make decisions, showcasing their effectiveness in reducing development and administrative overhead.

Lin et al.~\cite{lin2025byos} introduce BYOS, a knowledge-driven framework that leverages LLMs to automatically generate optimized Linux kernel configurations tailored to user requirements. It introduces an OS-oriented dual-layer knowledge graph that encodes both configuration dependencies and conceptual system knowledge, enabling accurate alignment of user needs with kernel options. Through reasoning and iterative exploration guided by the knowledge graph , BYOS avoids hallucinations and reduces tuning overhead. Experiments across benchmarks, Linux distributions, and real-world applications show that BYOS significantly outperforms default and baseline methods, achieving performance improvements from 7.1\% up to 155.4\%.

\paragraph{Insights and Challenges}  
Recent progress demonstrates that AI tools can effectively reduce the complexity of OS tuning by automating the exploration of large and interdependent configuration spaces. Learning-based schedulers, such as Cui et al.’s parameter recommendation model, show that targeted performance gains can be achieved by automatically adjusting critical I/O parameters. More general frameworks like AutoOS and BYOS highlight the broader potential of LLMs and knowledge-driven methods to reason about kernel configuration dependencies, prune infeasible options, and generate tailored configurations that outperform vendor defaults. Together, these works provide the insight that AI-enhanced tuning can shift OS optimization from a manual, expert-driven process to an automated, adaptive pipeline, thereby improving both efficiency and accessibility.  

Nonetheless, several challenges remain. First, the sheer size and heterogeneity of kernel configuration spaces make scalability a persistent issue, particularly when configurations must adapt across diverse workloads, hardware platforms, and deployment environments. Second, ensuring correctness and stability of AI-generated configurations is critical, as misconfigurations can lead to crashes, performance regressions, or security risks. Third, many methods rely on benchmark-driven evaluations, raising questions about their generalization to production workloads with more complex and dynamic resource demands. Finally, integration into real-world development and deployment pipelines requires attention to usability, transparency, and trust, so that system administrators and developers can confidently adopt AI-generated tuning recommendations. Addressing these challenges will be key to realizing the full benefits of AI-assisted OS tuning at scale.

\subsubsection{Verification}
Beyond code and configuration, LLMs are also being harnessed for formal verification, which is essential for building trustworthy OS kernels. Selene~\cite{zhang2024selene} introduces a benchmark for automated proof generation in software verification, based on the industrial-grade seL4 operating system microkernel. By leveraging large language models and a novel lemma isolation technique, Selene enables efficient end-to-end proof generation and rapid correctness checking. Experimental results show that LLMs like GPT-4 can automatically generate valid proofs for simpler OS verification tasks, though complex proofs remain challenging. Selene demonstrates the potential of LLMs to assist in formal verification for OS development, paving the way for more automated and reliable system engineering.

For example, Yang et al.~\cite{yang2023kernelgptenhancedkernelfuzzing} present \textit{KernelGPT}, a novel approach that leverages LLMs to automatically synthesize syscall specifications for operating system kernel fuzzing. Unlike traditional methods that require extensive manual effort to create syscall specifications, KernelGPT iteratively generates, validates, and refines specifications with minimal human intervention. Experimental results show that KernelGPT produces more new and valid syscall specifications and achieves higher code coverage compared to state-of-the-art techniques. Notably, KernelGPT has already helped discover 24 unique bugs in the Linux kernel, including 12 that have been fixed and 11 assigned CVE numbers, and its generated specifications have been adopted by the Syzkaller kernel fuzzer. This demonstrates the significant potential of LLM-powered automation to improve kernel security and reliability at scale.

Extending this line of research to the embedded domain, ECG~\cite{zhang2024ecg} introduces an LLM-powered fuzzer for Embedded OSs, which are widely deployed in critical infrastructure but are generally less tested than general-purpose OSs. ECG addresses the challenges of limited documentation and low-quality payload generation by automatically synthesizing input specifications from source code and documentation, guiding fuzzing with execution feedback, and refining inputs interactively. In evaluations, ECG uncovered 32 new vulnerabilities across three open-source Embedded OSs (RT-Linux, RaspiOS, OpenWrt) and detected 10 bugs in a commercial Embedded OS. Compared to existing fuzzers such as Syzkaller, Moonshine, KernelGPT, Rtkaller, and DRLF, ECG achieved an average of 16.02\% higher kernel code coverage, underscoring its effectiveness in enhancing Embedded OS security.

\subsubsection{Education}
OS education constitutes the talent-support layer of the OS ecosystem, cultivating the competencies needed to design, implement, and maintain modern systems. However, OS courses are among the most demanding in computer science curricula due to complex internal mechanisms, heterogeneous hardware and runtime environments, and the strong emphasis on hands-on practice. Artificial intelligence (AI) can mitigate these challenges by delivering personalized guidance, adaptive learning resources, and automated evaluation, thereby enhancing instructional scalability and learning outcomes.

Zhang et al.~\cite{zhang2025sortinghat} present SortingHat, a digital teaching assistant tailored to OS education. Combining a multimodal 3D digital-human interface with LLM-based reasoning agents, SortingHat provides empathetic, context-aware, and domain-specific guidance. It automatically generates exercises aligned with each student’s learning history and academic performance, reinforcing weaknesses while introducing advanced concepts. The system further incorporates a multi-agent reinforcement learning (MARL)–based evaluation pipeline to ensure consistent, fair, and unbiased grading, accompanied by detailed, personalized feedback. By uniting personalized guidance, adaptive content generation, and automated assessment, SortingHat makes OS learning more engaging, scalable, and effective for both students and instructors.

Complementing this assistant-centric perspective, Lin et al.\cite{lin2025empowering} conduct a study at Xiamen University of Technology investigating the use of large models in experimental OS instruction via Cursor, an IDE augmented with multimodal AI. Using Linux courses as the case context, the authors employ Cursor to support open-ended, project-driven laboratories covering environment setup, shell-command practice, web-server deployment, and Linux-based programming. With interactive features such as intelligent completion, error prompts, and configuration guidance, Cursor helps students master both foundational and advanced tasks. The study reports that the Cursor-based approach significantly improves student efficiency, lab completion quality, and performance on high-difficulty tasks compared with traditional methods, while also fostering self-directed learning and innovation.

\section{What Are the Categories of AI Tools in OS Enhancement?}
\label{sec:tool}
In the development of intelligent operating systems, understanding and categorizing existing AI tools is essential for driving optimization and innovation. Different AI tools vary in methodology, application scope, and the way they integrate with system architectures. In this section, we present three major categories of AI tools: Traditional Machine Learning, Large Language Models, and Agent-based Intelligence. 
Traditional ML encompasses statistical and predictive techniques such as regression, classification, clustering, and reinforcement learning, which have long been used for workload prediction, anomaly detection, and optimization within OS components.  
LLMs extend this paradigm by introducing large-scale pretraining and generative reasoning capabilities.  
Agent-based Intelligence builds on LLMs and other reasoning engines to achieve higher-level autonomy. Agents are not merely single models but goal-driven entities capable of perceiving environments, invoking multiple tools (including LLMs), and planning or coordinating actions toward system-level objectives.  
These categories represent an evolutionary trajectory from statistical learning to generative reasoning and finally to autonomous decision-making, providing multi-layered support for building adaptive and intelligent operating systems.

\subsection{Traditional Machine Learning}
Traditional ML methods—including decision trees, support vector machines, random forests, and neural networks—have been widely applied in operating system research. These approaches typically rely on structured data and employ supervised, unsupervised, or reinforcement learning paradigms to model system behavior.
\subsubsection{Applications}

Traditional ML methods have been applied across a wide range of OS modules to enhance performance, security, and adaptability. Unlike heuristic approaches, ML-driven solutions exploit runtime data to predict system behaviors, optimize resource allocation, and detect anomalies. Below we summarize representative applications of ML in different OS subsystems.

\paragraph{Process Scheduling} 
Scheduling is one of the most critical OS subsystems, directly affecting fairness, latency, and throughput. Conventional schedulers, such as Linux’s CFS, rely on static heuristics that often fail in heterogeneous and multicore environments. Chen et al.~\cite{chen2020machine} proposed a resource-aware load balancer that integrates a multi-layer perceptron (MLP) into the Linux kernel, imitating CFS decisions with low inference overhead and improved efficiency. Similarly, Goodarzy et al.~\cite{goodarzy2021smartos} introduced \textit{SmartOS}, which leverages reinforcement learning to allocate CPU, memory, I/O, and network bandwidth based on user preferences. For HPC clusters, Springborg et al.~\cite{aaen2023automatic} developed \textit{Chronus}, which uses ML to predict energy-efficient configurations under SLURM, achieving significant energy savings. These studies demonstrate how ML can transform scheduling from rule-based heuristics to adaptive, workload-aware strategies.

\paragraph{I/O Scheduling} 
The Linux I/O subsystem is highly sensitive to workload variability, and heuristic-based policies often struggle to balance throughput and latency. Hao et al.~\cite{hao2020linnos} proposed \textit{LinnOS}, embedding a lightweight neural network in the kernel I/O path to predict SSD behavior in real-time, reducing latency by up to 40\%. More recently, Kurniawan et al.~\cite{kurniawan2025heimdall} introduced \textit{Heimdall}, an ML-powered I/O admission controller with a full pipeline of labeling, feature engineering, and quantized inference, achieving 93\% decision accuracy with sub-µs latency. These works highlight that ML enables predictable, adaptive I/O management, crucial for modern storage-intensive workloads.

\paragraph{Storage Management}  
At the stoage management level, ML has been used to optimize flash translation and mitigate hardware-specific failures. Wang et al.~\cite{wang2024learnedftl} designed \textit{LearnedFTL}, which employs learned indexes for page-level flash translation, reducing double reads by 55.5\% and improving P99 latency by 12.2×. Wu et al.~\cite{wu2024mitigating} proposed \textit{LearnWD}, which applies clustering to mitigate write-disturbance errors in NVMs, reducing errors by 20.1\% and extending device endurance by 21.9\%. These innovations show how ML can push intelligence closer to hardware, improving both performance and reliability.

\paragraph{Memory Management} 
Advanced memory technologies such as 3D-stacked DRAM and far memory introduce new opportunities for ML-enhanced management. Zhang et al.~\cite{zhang2022software} proposed \textit{Software-Defined Address Mapping} (SDAM), which uses LSTM models to identify access patterns and optimize memory address mapping dynamically. Lagar-Cavilla et al.~\cite{lagar2019software} deployed a software-defined far memory system at Google, employing ML-based autotuning with Gaussian Process Bandits to balance cost savings and performance. Miura et al.~\cite{9355915} further applied LSTM models to cache management in virtualized systems, improving hit ratios and throughput. These approaches illustrate how ML can adapt memory systems to workload diversity and dynamic environments.

\paragraph{Network Management} 
ML has also been applied to network traffic classification, rate adaptation, and resilience. Kwon et al.~\cite{kwon2023efficient} deployed DNNs in the ONOS platform for packet classification, while Li et al.~\cite{9123561} proposed ML-based rate adaptation (MLRA) for IEEE 802.11ac, achieving up to 658\% improvement over heuristic methods. For resilience, Veliyath et al.~\cite{10677020} combined SDN and ML to predict disaster-affected areas and reconfigure networks in real-time. These works highlight how ML enhances both performance and robustness in increasingly dynamic networking environments.

\paragraph{Security} 
Operating system security is one of the most active domains for ML applications. Benabderrahmane et al.~\cite{benabderrahmane2024hack} introduced AE-APT, an autoencoder-based framework for detecting advanced persistent threats in imbalanced datasets. Qin et al.~\cite{qin2019msndroid} and Alzaylaee et al.~\cite{Alzaylaee_2020} applied deep learning to Android malware detection, achieving accuracy above 97\%. Wihar et al.~\cite{wihar2024novel} proposed an LSTM-based ransomware detector integrated into the Linux kernel, enabling real-time anomaly detection. These works demonstrate that ML can move OS security from reactive defense to proactive, adaptive protection.

\subsubsection{Analysis}
\begin{figure}[t]
    \centering
    \includegraphics[width=1\linewidth]{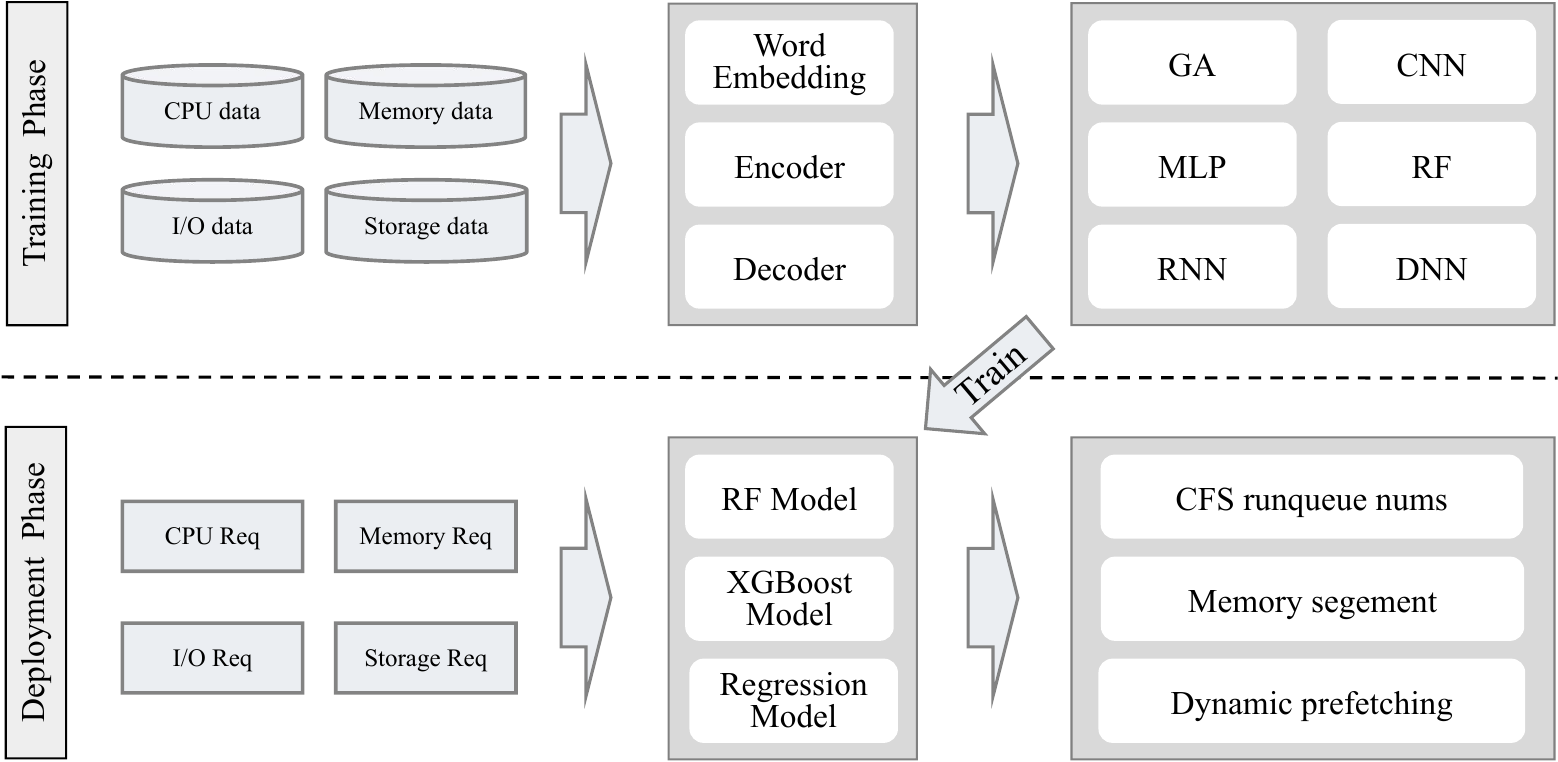}
    \caption{Workflow for enhancing operating systems with machine learning. In the training phase, data collected is preprocessed and used to train various models using advanced machine learning methods. These trained models are then deployed, where they process new data and provide optimization results for key system parameters}
    \label{fig:auto-tune}
\end{figure}
Overall, machine learning applications in operating systems span multiple layers, from low-level schedulers and memory managers to higher-level security and networking subsystems. The common insight is that ML enables operating systems to evolve from static, heuristic-driven mechanisms toward adaptive, data-driven strategies. Despite challenges in scalability, overhead, and generalization, ML-enhanced OS modules have already achieved measurable improvements in performance, reliability, and security. Figure \ref{fig:auto-tune} showed the general workflow for ML tools to assist OS. The workflow for enhancing operating systems through machine learning is structured into two distinct phases: training and deployment. In the training phase, diverse datasets are collected from key system components, including the CPU, memory, storage, and I/O operations, serving as the foundation for model development. This raw data undergoes extensive preprocessing and transformation to ensure it is properly structured and optimized for training purposes. Advanced machine learning techniques are then employed to develop specialized models designed to address specific tasks, leveraging methods such as XGBoost, and various regression techniques.

In the deployment phase, the trained models are integrated into the operating system and applied to real-time operational data. By analyzing this incoming data, the system uses the predictive capabilities of the models to optimize performance dynamically. This phase involves fine-tuning configurations and generating actionable insights to improve key areas, including memory segment optimization, task prioritization, and dynamic resource reconfiguration. These enhancements illustrate the transformative potential of machine learning when systematically applied to operating systems, enabling more efficient, adaptive, and intelligent resource management.
\subsubsection{How to Choose Proper Algorithms to Improve OS?}
 Among the previous studies surveyed, we figure out some popular algorithms, include Random Forests (RF/DT), Recurrent Neural Networks (RNN/LSTM), K-Nearest Neighbors (KNN), Reinforcement Learning (RL) and Multi-Layer Perceptrons (MLP). Those work that clearly state their method is list in Table.~\ref{tab:mltool}.
\renewcommand{\tabularxcolumn}[1]{m{#1}}

\begin{table*}[t]
\scriptsize
\caption{Machine Learning Tools in Papers}
\label{tab:mltool}
\begin{center}
\begin{tabularx}{\textwidth}{>{\centering\arraybackslash}m{0.18\textwidth} X}
\toprule
\textbf{Tools} & \multicolumn{1}{c}{\textbf{Representative Works}} \\ 
\midrule
RL & Cardoso et al.~\cite{cardoso2023evaluation}; White et al.~\cite{white2019improving}; Lin et al.~\cite{lin2025r1} \\
\hline
RF(DT) & Sun et al.~\cite{sun2021linux}; Benabderrahmane et al.~\cite{benabderrahmane2024hack}; Panman et al.~\cite{panman2022dynamic}; Dusane et al.~\cite{9453065}; Ahmed et al.~\cite{ahmed2022heterogeneous}; Ongun et al.~\cite{ongun2021living}; Cruz et al.~\cite{cruz2023patching}; Metzger et al.~\cite{metzger2021device}; Chowdhury et al.~\cite{chowdhury2021novel}\\
\hline
RNN(LSTM) & Fingler et al.~\cite{fingler2023towards}, Zhang et al.~\cite{zhang2022software}, Miura et al.~\cite{9355915}; Benabderrahmane et al.~\cite{benabderrahmane2024hack}; Wihar et al.~\cite{wihar2024novel}; Akram et al.~\cite{10.1145/3208040.3208051}; Chowdhury et al.~\cite{chowdhury2021novel}; Qi et al.~\cite{qi2021efficient}  \\
\hline
KNN & Fingler et al.~\cite{fingler2023towards};Yang et al.~\cite{yang2022improvement}; Panman et al.~\cite{panman2022dynamic}; Dusane et al.~\cite{9453065}; Ahmed et al.~\cite{ahmed2022heterogeneous} \\
\hline
MLP & Chen et al.~\cite{chen2020machine}; Qiu et al.~\cite{qiu2021toward} \\
\bottomrule
\end{tabularx}
\end{center}
\end{table*}

\paragraph{RF(DT)}A decision tree is a type of ML model used when the relationship between a set of predictor variables and a response variable is nonlinear,while random forest is essentially a collection of decision trees. It is quick to fit to a dataset and easy to interpret. Benabderrahmane et al.~\cite{benabderrahmane2024hack} used RF as one if the models to  accurately detect the attack from the network traffic. To construct the random forest classifier, the authors employed the Random Forest Regressor, which serves as both a regressor and a meta estimator. It accomplishes this by fitting multiple decision trees to different subsets of the dataset. For this specific model, the authors opted for a forest containing 1000 trees. Ahmed et al.~\cite{ahmed2022heterogeneous} used RF to build a device fitness model,based on the Dataset collected during runtime and statically.  De Wit et al.~\cite{panman2022dynamic} trained a statistical classifier able to recognize malware signatures in any log data collected on a smartphone. The classifier was trained, cross-validated, and tested using the dataset described above and RF classifier had a better performance. Metzge el at.~\cite{metzger2021device} used RF to get a optimal kernel runtime switching slice size. The model is a random forest regressor with 50 decision trees with a depth of two for the GPU model. Ongun et al.~\cite{ongun2021living} used RF to  get the probability of a command being malicious,based on labels dataset. Sun et al.~\cite{sun2021linux} used RF to filter storage I/O parameters demonstrates that tuning only the important parameters can nearly achieve the high performance obtained by tuning all parameters. Known for their versatility in handling nonlinear relationships and minimal tuning requirements, RF/DT are used for tasks such as attack detection and configuration settings. They are chosen for their ability to quickly fit to datasets and provide interpretable models.

\paragraph{RNN(LSTM)}RNNs are a class of neural networks designed to handle sequential data. They have feedback connections, allowing them to maintain an internal state or memory. Each step in an RNN processes an input and updates its hidden state based on the current input and the previous hidden state. While LSTM, a type of RNN architecture with a more complex cell structure,were introduced to address the vanishing gradient problem. Motivated by the problem that exploiting 3D-stacking memory's performance is challenging because bandwidth utilization heavily depends on address mapping in the memory controller, Zhang et al.~\cite{zhang2022software} used a software-defined address mapping, allowing user programs to directly control low-level memory hardware in an intelligent and fine-grained manner. LSTM is used to in a method to get access pattern information to select an address mapping. It identify the major variables that significantly contribute to external memory access and have a substantial impact on memory traffic and data movement. Miura et al.~\cite{9355915} proposed to apply LSTM for time series data, for cache replacement in the second cache. Designed to handle sequential data, RNN/LSTM are particularly suited for tasks that require capturing temporal dependencies, such as managing memory access patterns in 3D-stacking memory systems.

\paragraph{KNN} KNN is a supervised ML algorithm used for both classification and regression tasks.
The fundamental idea behind KNN is simple: neighbors influence each other. If you’re surrounded by similar things, you’re likely similar too. It is widely applicable in pattern recognition, data mining, and intrusion detection. Yang et al.~\cite{yang2022improvement} introduce a  KNN-based ML algorithms can accurately predict the Turnaround-time (TaT) of a process. It can effectively reduce the TaT of the process and reduce the number of process context switches.

\paragraph{MLP} An MLP is a type of feedforward neural network used for supervised learning tasks, such as classification and regression. Chen et al.~\cite{chen2020machine} argues that traditional Linux CFS scheduler maximizes the utilization of processing time but overlooks the contention for lower-level hardware resources and try to solve the above problem using an ML-based resource-aware load balancer.They employed supervised imitation learning to replace a portion of its internal logic with an MLP model. This trained MLP model emulates the kernel’s load balancing decisions. MLP is chosen because this current work doesn’t require a very complex model and MLP has a relatively simple implementation compared to the other models. Based on this work, Qiu et al.~\cite{qiu2021toward} propose the concept of reconfigurable kernel datapaths that enables kernels to self-optimize dynamically to reduce the cost of kernel. The authors also used MLP ML model that can mimic Linux CFS decisions.

\subsubsection{Pros and Cons for ML Enhanced OS}

\paragraph{Pros} 
Traditional machine learning offers several clear advantages when applied to operating systems.  
First, ML algorithms are relatively lightweight compared to modern LLMs, making them easier to integrate into kernel modules or low-level system components where computational budgets are tight.  
Second, these models can be trained on structured runtime data (e.g., CPU traces, I/O logs, memory access patterns) and achieve good predictive accuracy with limited resources.  
Third, ML methods are interpretable in many cases (e.g., decision trees, random forests), which helps developers and system administrators understand why a particular scheduling or configuration decision is made.  
Finally, ML-enhanced OS components have already demonstrated significant improvements in performance, security, and resource management, with examples including adaptive I/O tuning, anomaly detection, and memory access optimization.  

\paragraph{Cons} 
Despite these benefits, ML-based approaches also face notable limitations.  
First, most ML models rely heavily on feature engineering and carefully curated training datasets, which can limit generalization to unseen workloads or rapidly evolving hardware environments.  
Second, models such as KNN or RF scale poorly with very large datasets or high-dimensional features, making them less suitable for complex or dynamic OS contexts.  
Third, sequential models like RNN/LSTM suffer from training difficulties (e.g., vanishing gradients) and require substantial tuning, which increases deployment complexity.  
Fourth, unlike rule-based kernel logic, ML models may produce non-deterministic or hard-to-explain decisions, raising concerns for correctness, predictability, and trust in safety-critical systems.  
Finally, ML approaches are often task-specific: a model optimized for scheduling may not transfer well to security or memory management, leading to fragmented solutions rather than a unified framework.

\subsection{Large Language Models}
The rapid development of LLMs has introduced transformative capabilities in natural language understanding and generation, making them a promising tool for advancing OS design and interaction. LLMs, such as GPT-4~\cite{achiam2023gpt}, LLaMA~\cite{touvron2023llama}, Gemini~\cite{team2023gemini} and DeepSeek~\cite{liu2024deepseek}, have demonstrated remarkable proficiency in processing and generating human-like text, understanding complex instructions, and adapting to a wide variety of tasks with minimal or no task-specific training. These models are built on advances in deep learning architectures, particularly Transformer-based designs, and trained on vast datasets that enable them to capture context, semantics, and nuanced relationships between concepts.

One of the defining features of LLMs is their ability to act as general-purpose reasoning and interaction tools. Unlike traditional machine learning models, which are often task-specific, LLMs are highly adaptable, capable of handling a broad range of tasks through natural language interfaces. Their ability to understand intent, provide detailed explanations, and generate contextualized responses makes them an ideal complement to OS functionalities. As operating systems serve as the backbone for user-computer interaction, integrating LLMs can enable more intuitive, conversational interfaces, where users can communicate their needs and preferences in natural language rather than relying on rigid command structures or graphical user interfaces. The integration of LLMs into OSs presents a significant opportunity to enhance the user experience and overall system functionality ~\cite{chan2023consistency,wu2024copilot}. LLMs can transform the way users interact with and manage their computing environments~\cite{hè2024perospersonalizedselfadaptingoperating}.

\subsubsection{Applications}  
LLMs have been applied across many layers of the operating system stack as well as its broader ecosystem, covering tasks from low-level kernel engineering to education and user interaction.

\paragraph{Coding}  
Large language models have begun to reshape the way operating systems are developed and maintained, particularly in the areas of kernel extensions, bug fixing, and vulnerability repair. Zheng et al.~\cite{zheng2023kenkernelextensionsusing} proposed KEN, a framework that synthesizes eBPF programs directly from natural language prompts. By combining LLM-based code generation with symbolic execution and semantic feedback, KEN ensures that the generated kernel extensions align with user intent and comply with eBPF’s strict safety rules. Building on this direction, Islam et al.~\cite{islam2024llmpoweredcodevulnerabilityrepair} developed SecRepair, which leverages reinforcement learning and semantic reward mechanisms to automatically repair vulnerable code. In addition to generating patches, SecRepair also produces concise commit messages, enabling both automated remediation and clear human-readable documentation.

These examples illustrate how LLMs can lower the expertise barrier for OS coding by translating natural language into enforceable kernel extensions and by automating vulnerability repair pipelines.

\paragraph{Ops and Maintenance}  
Large language models are increasingly applied to operations and maintenance tasks, where their ability to interpret complex logs and reason over system behavior provides clear advantages. Kim et al.~\cite{kim2025logs} introduced SAN2PATCH, a framework that leverages Tree-of-Thought prompting and multi-stage reasoning to automatically generate vulnerability patches directly from sanitizer logs and source code, reducing the need for costly manual diagnosis. Zhong et al.~\cite{zhong2024logparser} proposed Logparser-LLM, which applies LLMs to log parsing tasks, significantly improving the extraction of structured information from diverse and irregular system log formats. To further enhance accuracy and consistency, Huang et al.~\cite{huang2025logrules} developed LogRules, a rule-augmented approach that guides LLMs in log analysis, mitigating inconsistencies and improving reliability in real-world operational settings. These studies demonstrate that LLMs can go beyond traditional anomaly detection or log filtering, offering end-to-end solutions that encompass log understanding, failure diagnosis, and even automated patch generation. This highlights their growing role in building adaptive, intelligent, and highly automated operational workflows within modern operating systems.

\paragraph{Tuning} 
LLMs also aid in configuration and performance optimization. AutoOS~\cite{chen2024autoos} employs a state-machine framework to optimize Linux kernel configurations for AIoT devices, achieving up to 25\% higher UnixBench scores. BYOS~\cite{lin2025byos} integrates a dual-layer knowledge graph with LLM reasoning, generating optimized kernel configurations with improvements up to 155.4\%. These works reduce the complexity of tuning thousands of kernel options by automating parameter selection.

\paragraph{Verification} 
Verification is a critical component of building trustworthy operating systems, and recent work shows that large language models can play an increasingly meaningful role in this domain. KernelGPT~\cite{yang2023kernelgptenhancedkernelfuzzing} automatically synthesizes syscall specifications for Linux fuzzing, significantly improving coverage and bug discovery. ECG~\cite{zhang2024ecg} extends this idea to Embedded OSs, uncovering vulnerabilities in real-world systems. In formal verification, Selene~\cite{zhang2024selene} introduces a benchmark for LLM-assisted proof generation based on the seL4 microkernel. By applying lemma isolation and leveraging GPT-4, Selene shows that LLMs can automatically generate valid proofs for simpler tasks, though more complex obligations remain challenging. This demonstrates the potential of LLMs to reduce manual proof engineering. Collectively, these efforts indicate that LLMs are not yet a full replacement for rigorous formal methods, but they already show promise for reducing manual proof engineering and lowering the entry barrier to applying verification in OS research and practice.

\paragraph{Education} 
LLMs are reshaping OS pedagogy. Cursor~\cite{lin2025empowering} enhances Linux experimental teaching by providing intelligent completion, error prompts, and interactive guidance. Both systems show that LLMs can tailor education to diverse students while improving efficiency and engagement.

\paragraph{User Interaction} 
Beyond internal mechanisms, LLMs are redefining OS–user interfaces. Ge et al.~\cite{ge2023llm} proposed AIOS, embedding LLMs into the OS kernel to act as an “OS with a soul.” Shi et al.~\cite{shi2024commandspromptsllmbasedsemantic} designed LSFS, enabling natural language file management, while Xu et al.~\cite{xu2024aios} developed CoRE to interpret natural language pseudo-code for execution. These projects illustrate the potential of LLMs to democratize OS interaction through conversational programming and semantic services.

\subsubsection{Analysis}  
Compared to traditional ML models that are narrowly optimized for specific tasks such as scheduling or anomaly detection, LLMs provide a more general reasoning capability that spans code, documentation, configuration, and user interaction. Their ability to interpret system requirements, synthesize code, and interact through natural language allows them to unify multiple OS-related tasks under a single framework. This versatility reduces the need for bespoke models and lowers the barrier for users to configure or extend systems. However, the same generality brings challenges: reasoning over massive codebases like the Linux kernel often exceeds the context-handling capacity of current LLMs, while correctness and safety of their outputs remain difficult to guarantee. As a result, LLMs complement but do not replace ML approaches, positioning themselves as flexible assistants where interpretability, trust, and scalability are still open challenges.

\subsubsection{Pros and Cons for LLM Enhanced OS}  
\paragraph{Pros}The major strength of LLMs lies in their versatility and adaptability. They can automate complex OS engineering tasks, from fuzzing and debugging to configuration tuning and verification, significantly reducing manual effort and accelerating development cycles. Their natural language interfaces democratize OS usage by allowing non-experts to interact with system components without deep technical knowledge. Moreover, their contextual reasoning capabilities enable them to adapt to evolving workloads and environments, while also offering substantial pedagogical value by generating tailored exercises and feedback for students.

\paragraph{Cons}Despite these benefits, significant limitations persist. LLM outputs are often opaque and may suffer from hallucinations, raising concerns about reliability and interpretability in safety-critical contexts such as kernel code. Their computational demands limit deployment in embedded or resource-constrained environments, and applying them to massive codebases remains expensive and error-prone. Furthermore, building high-quality, domain-specific datasets for tasks like vulnerability repair or formal verification is costly, making fine-tuning difficult. These challenges highlight that while LLMs are reshaping the role of operating systems toward more intelligent and user-centric paradigms, ensuring correctness, scalability, and trust remains a core obstacle for their practical adoption.

\subsubsection{Differnce Compared to ML-enhanced OS:}
LLMs and traditional ML techniques enhance OS in distinct ways, reflecting their unique capabilities and methodologies.

Traditional ML techniques are often more task-specific and rely on structured data for training. These methods are typically employed for optimization tasks, such as system performance tuning, resource allocation, and configuration management. Additionally, ML models are frequently viewed as black boxes, making it challenging for users to understand the underlying principles and the relationship between the input training data and the resulting outputs. This lack of interpretability can hinder trust and usability, especially in critical applications where comprehending the rationale behind decisions is essential. In contrast, LLMs primarily focus on natural language understanding and generation, enabling more intuitive user interactions through conversational interfaces. They excel in interpreting user intent, managing complex tasks through natural language commands, and providing personalized experiences based on user input. This integration allows LLMs to transform how users engage with the OS, making technology more accessible and user-friendly. 

While there has been extensive research exploring the integration of traditional ML techniques into operating systems, where they effectively delve into the complexities of system performance, resource allocation, and configuration management, the role of LLMs remains more superficial. LLMs primarily function as an additional layer that enhances user interaction with the OS through natural language processing, enabling more intuitive and conversational interfaces.

\subsection{Agent-based Intelligence}
An agent, in this context, denotes an autonomous component that can perceive its environment, reason about objectives, and interact with system resources or other agents through defined actions. A single agent may handle end-to-end workflows such as debugging or user support, whereas multi-agent systems introduce specialization and collaboration across distinct roles.

\subsubsection{Applications}  
Agent-based intelligence extends the role of LLMs by treating them, or systems built around them, as autonomous agents capable of perceiving environments, planning actions, and adapting over time. Compared with standalone LLMs, agent-based approaches emphasize autonomy, specialization, and coordination, enabling adaptive and sustained operation in evolving system contexts.

\paragraph{Coding}  
In OS development, agent-based methods improve fault localization and vulnerability repair by distributing tasks among specialized agents. Zhou et al.~\cite{zhou2025benchmarking} introduced LinuxFLBench and LinuxFL\^+, which coordinate multiple reasoning modules to enhance debugging accuracy in Linux. SecRepair~\cite{islam2024llmpoweredcodevulnerabilityrepair} also benefits from agent-based extensions, where reinforcement learning–enhanced LLM agents autonomously repair vulnerabilities and generate commit messages, reducing the need for manual intervention. These works show that agents can orchestrate code analysis, patch generation, and validation within a collaborative framework.

\paragraph{Education}  
SortingHat~\cite{zhang2025sortinghat} applies multi-agent reinforcement learning (MARL) to OS pedagogy. Its agents work together to grade student submissions fairly, reduce bias, and generate personalized exercises and feedback. Through collaboration, SortingHat ensures both fairness and adaptability, marking a shift from static digital teaching assistants to dynamic, agent-driven learning environments.

\paragraph{User Interaction}  
LLM-powered agents are also reshaping how users interact with operating systems, turning them into more intuitive and intelligent copilots. The AIOS series~\cite{mei2024aios,ge2023llm} pioneers this vision by embedding LLMs into the operating system itself, enabling natural language interaction and seamless integration of system-level tasks with AI reasoning capabilities. Building on this direction, Liu et al.~\cite{xu2024osagent} introduced OSAgent, a memory-enhanced LLM agent designed to assist users in operating their devices. By leveraging an external memory of past requests—enriched with semantic and intent vectors—OSAgent retrieves relevant examples for prompting, performs task planning, and safely invokes system tools. Extensive experiments across mobile operating systems demonstrate its strong ability to handle AI capabilities, third-party applications, system settings, and data resources. Pushing further into graphical interfaces, AgentCPM-GUI~\cite{zhang2025agentcpm} proposes an 8B-parameter agent tailored for robust and efficient GUI-based interaction in both Chinese and English mobile ecosystems. Its training pipeline combines grounding-aware pretraining, supervised fine-tuning on high-quality multilingual trajectories, and reinforcement optimization to improve reasoning. Together, these works illustrate how agent-based LLMs are moving OS interaction beyond command lines and scripts, toward natural, multimodal, and culturally adaptive experiences for end users.

\subsubsection{Analysis}  
Compared with ML models and standalone LLMs, agent-based intelligence broadens the design space by supporting both single-agent and multi-agent paradigms. A single agent, equipped with memory, planning, and tool-use capabilities, can already outperform plain LLMs by sustaining long-horizon workflows and autonomously completing tasks such as debugging or device operation. Multi-agent systems further extend this idea by distributing roles among specialized agents that collaborate on complex workflows, bringing benefits in scalability, adaptability, and workload diversity. However, the agent paradigm also introduces new challenges: single agents face limits in coverage and robustness, while multi-agent systems add orchestration overhead and consistency risks. Overall, agent-based intelligence offers a promising yet complex direction for applying LLMs to OS research, balancing autonomy, specialization, and coordination across different deployment settings.

\subsubsection{Pros and Cons for Agent Enhanced OS}  
\paragraph{Pros}The main strength of agent-based intelligence lies in its ability to decompose tasks into specialized roles and coordinate them effectively. This enables OSs to autonomously tackle workflows that require sustained reasoning, such as kernel debugging, configuration tuning, or personalized education. Multi-agent collaboration also improves robustness by providing redundancy and mitigating hallucinations, while natural language–driven interfaces like AIOS democratize OS usage by making system management conversational and task-oriented.

\paragraph{Cons}Nonetheless, these advantages come with trade-offs. Multi-agent orchestration incurs coordination overhead, which can hinder performance in latency-sensitive OS tasks. Ensuring correctness, security, and trust across autonomous agents is difficult, as misaligned behaviors or compounding errors can degrade reliability rather than improve it. Furthermore, the computational cost of maintaining agent ecosystems is significantly higher than that of standalone models, limiting applicability in embedded or resource-constrained environments. Thus, while agent-based intelligence pushes operating systems toward greater autonomy and adaptability, its practical deployment still faces challenges of efficiency, verification, and scalability.

\subsection{Comparison of AI tools for OS Enhancement}
\renewcommand{\arraystretch}{1.35}

\begin{table}
\scriptsize
\centering
\caption{Representative works at the intersection of AI and OS.}
\label{tab:ai-os-representative}
\resizebox{\textwidth}{!}{
\begin{tabular}{@{}%
    m{2cm} m{2cm} m{2cm} m{2cm} m{1.8cm} m{2cm} m{2cm} m{4cm}@{}}
\toprule
\textbf{Representative Work} & 
\textbf{Module} & 
\textbf{Tools} & 
\textbf{Deployment Site} & 
\textbf{Mode} & 
\textbf{Evaluation Metrics} & 
\textbf{Code} & 
\textbf{Limitations} \\
\midrule
~\cite{chen2020machine} & Scheduler & MLP, Adam & VM & Offline train, online infer & Accuracy, latency  & Open source & Scenario-dependent\\
~\cite{goodarzy2021smartos} &User-space scheduler & RL & VM & Online RL & Overhead, throughput & Not provided & Limited real-world validation; lacks guardrails and cross-platform support. \\
~\cite{tetzlaff2010intelligent} & Scheduler & Classifier & Compiler toolchain & Offline train, online infer & MAE & Not provided & Limited evaluation，lack generalization \\
~\cite{hao2020linnos} & I\/O & NN & Kernel block layer & Offline train, online infer & Predictability, stability & Open source & Focuses on read-path predictability \\

~\cite{kurniawan2025heimdall} & I\/O & NN & User-space + Ceph & Offline train, online infer & Latency, accuracy, throughput & Open source at GitHub & Long-term drift; limited retraining; trade-off between accuracy and throughput \\

~\cite{wu2024mitigating} & Storage & K-means, MinHash & DRAM + OS layer & Offline train, online infer & Performance, generality & Not provided & Retraining overhead; sensitive to data/workloads; depends on GC and stale blocks \\

~\cite{wang2024learnedftl} & Storage & Linear Regression & SSD firmware (FTL)	 & Offline train, online infer & Throughput, latency, energy & Open source & Emulator-based validation; model size overhead; performance tied to GC/rewrite \\

~\cite{zhang2022software} & Memory & K-means, LSTM-AE & Memory controller &Offline train, online infer & Performance, bandwidth & Not provided & Limited generality; hardware-bound and non-transparent \\

~\cite{lagar2019software} & Memory & Gaussian Process & Warehouse-scale infra & Offline train, online infer & Capacity, performance, robustness & Not provided & Conservative SLOs; limited parameter coverage; compression-only far memory \\

~\cite{de2022using} & Memory & ANN & NUMA servers & Offline train, online infer & Accuracy, energy & Not provided & Limited generalization; dataset imbalance; algorithm-awareness gaps \\

~\cite{9355915} & Cache & LSTM & VM & Offline train, online infer & Hit ratio, throughput & Not provided & Constrained model/features; limited deployment scope and generality \\

~\cite{9123561} & Network &ANN & NIC driver & Offline train, online infer & Goodput, overhead & Not provided & High training cost; simple model; limited visibility and generalization \\

~\cite{qin2019msndroid} & Malware detection & DBN &  Android malware env.	 & Offline only & Accuracy & Not provided &  Limited feature robustness and semantics; static-only malware detection \\

~\cite{shi2024commandspromptsllmbasedsemantic} & UI & LLM & FS semantic layer & Offline train, online infer & Accuracy, scalability & Open source & Limited modality and file-type scope; privacy and retrieval issues \\

~\cite{zhang2024enhanced} & UI & LLM, IML, DL & UI layer & Offline train, online infer  & Coverage, latency & Not provided & Lacks quantitative evaluation and explainability; weak OS integration \\

~\cite{rivard2025neuralos} & UI & RNN, Diffusion, LLM & VM & Offline train, online infer & Accuracy, latency & Open source &  Limited resolution and speed; partial input coverage; controllability gaps \\

~\cite{zhou2025benchmarking} & Coding & Agent & LLM benchmark env. & Benchmark & Recall, accuracy & Open source & Limited LLM diversity and knowledge coverage; scaling and robustness issues \\

\bottomrule
\end{tabular}}
\end{table}

This section examines the evolving role of intelligence in OS design by comparing three paradigms: traditional machine learning (ML), large language models (LLMs), and agent-based intelligence. Table~\ref{tab:ai-os-representative} summarizes their representative implementations across core OS modules. Each entry highlights the applied AI technique, its deployment site, operational mode, evaluation metrics, and observed limitations. Collectively, these studies reveal a clear technological transition—from compact, domain-specific ML models toward more semantic, adaptive, and autonomous forms of intelligence that interact with the OS at increasingly higher abstraction levels.

\subsubsection{Efficiency and Overhead}
ML offers the highest efficiency with minimal overhead by exploiting compact models and structured features. These characteristics make it particularly effective for system-level optimization tasks such as scheduling, caching, and performance prediction. LLMs achieve moderate efficiency with lower overhead than agents, supported by recent developments in model quantization and inference acceleration. Agent-based systems, while highly adaptable, are the least efficient and incur substantial runtime overhead due to orchestration, multi-stage reasoning, and iterative decision-making.

\subsubsection{Versatility}
LLMs and agent-based approaches demonstrate significantly greater versatility than ML. Through natural-language or semantic interfaces, they can communicate with diverse OS components and system tools. This capability enables a broad spectrum of applications—ranging from vulnerability detection and auto-repair to kernel tuning, service orchestration, and user education. In contrast, traditional ML remains confined to narrowly defined, numerically intensive domains such as anomaly detection, workload forensics, and resource forecasting, where feature engineering and labeled data pipelines are well established.

\subsubsection{Interpretability}
Interpretability also differs substantially among the three paradigms. Traditional ML provides limited transparency, as many statistical or neural models act as black boxes despite their predictive accuracy. LLMs enhance interpretability by producing human-readable explanations and rationales for their outputs. Agent-based systems further advance this dimension through explicit reasoning traces, tool-call histories, and auditable decision processes, thereby improving trust and transparency in OS-level interactions.

\subsubsection{Numerical Capability}
In terms of numerical reasoning, ML retains a clear advantage. Well-calibrated ML models provide precise and deterministic quantitative predictions, which are crucial for tasks involving latency, throughput, or energy estimation. Agents can achieve comparable or even superior performance by invoking external solvers or simulation tools to ensure numerical consistency beyond the capacity of statistical models. In contrast, LLMs remain relatively weak in this dimension—although capable of approximate arithmetic reasoning, they often struggle to maintain stable quantitative accuracy across long reasoning sequences.

\subsubsection{Discussion}
Beyond these specific dimensions, hybrid intelligence paradigms are reshaping OS research and development. LLM- and agent-based approaches particularly excel in tasks that require cross-module reasoning, contextual understanding, and iterative adaptation. Their pretrained knowledge and natural-language interfaces allow them to interpret documentation, logs, and configurations, facilitating autonomous maintenance, self-debugging, and dynamic optimization. Agent frameworks further extend these abilities by coordinating multi-tool pipelines and refining their strategies over time, making them highly suitable for complex, evolving operating environments.

Nevertheless, traditional ML remains the most practical and reliable choice in latency-critical, resource-constrained, or safety-sensitive scenarios. Its concise model structure, deterministic execution, and predictable timing make it well suited for kernel-level components such as scheduling, caching, and device management, where bounded overhead and real-time constraints are essential. In these contexts, LLMs and agent systems may still introduce unacceptable delays or uncertainty. Accordingly, a promising research direction lies in hybrid designs that combine both paradigms—using ML for low-level, realtime control and relying on LLMs or agents for high-level reasoning, monitoring, and adaptive orchestration across the OS stack.

\section{How Does OS Provide Enhancement for AI?}
\label{sec:structure}

While much of the existing research has focused on how AI can enhance operating systems, an equally important but sometimes overlooked perspective is the reverse relationship: how advances in operating systems can, in turn, empower and accelerate AI systems themselves. As AI workloads become increasingly complex, data-intensive, and heterogeneous, they place unprecedented demands on the underlying computing infrastructure. Traditional OS designs, originally optimized for general-purpose workloads, often become bottlenecks when faced with the unique requirements of large-scale machine learning, deep learning, and large language models.

Modern AI applications require not only high throughput and low latency, but also efficient resource sharing, scalability across heterogeneous hardware, and robust isolation for multi-tenant environments. To meet these demands, operating systems are being enhanced along two complementary directions. At the \textbf{component level}, optimization works within the existing OS architecture. It refines internal subsystems—such as scheduling, memory management, and I/O handling—to improve performance, energy efficiency, or responsiveness, without altering the kernel’s structural model. At the \textbf{architecture level}, optimization reconsiders the OS structure itself. Here the goal is to reshape how the kernel is organized, how components are isolated or composed, and how the OS interfaces with heterogeneous hardware and services.
Table~\ref{tab:ai_os_focus} summarizes representative AI workloads and their corresponding OS‑level optimizations.
\begin{table}
\scriptsize
\centering
\caption{Representative AI workloads, OS levels, and optimization focus of representative systems.}
\label{tab:ai_os_focus}
\resizebox{\textwidth}{!}{%
\begin{tabular}{@{} l c c l @{}} 
\toprule
\textbf{Workload Type} & \textbf{Work} & \textbf{OS Level} & \textbf{Optimization Focus} \\ 
\midrule

\multirow{4}{*}{Short-context inference} 
& \cite{oh2024exegpt} & Component & Low latency and execution efficiency. \\
& \cite{shen2025xsched} & Component & Scheduling efficiency and GPU utilization. \\
& \cite{xu2025camel} & Component & Energy efficiency and performance balance. \\
& \cite{patel2024splitwise} & Component & Inference throughput and efficiency. \\

\midrule
\multirow{4}{*}{Long-context inference} 
& \cite{kwon2023efficient} & Component & Memory bandwidth and cache efficiency. \\
& \cite{prabhu2025vattention} & Component & Attention efficiency and computation throughput. \\
& \cite{gao2024attentionstore} & Component & Context scalability and memory utilization. \\
& \cite{lin2024infinite} & Component & Context scalability and memory efficiency. \\

\midrule
\multirow{3}{*}{Distributed training} 
& \cite{mei2024helix} & Component & Parallelism and synchronization efficiency. \\
& \cite{zhang2024speedloader} & Component & Data-loading efficiency and GPU utilization. \\
& \cite{zhao2024hetegen} & Architecture & Heterogeneous scalability and coordination efficiency. \\

\midrule
\multirow{2}{*}{Edge inference} 
& \cite{xu2025fast} & Component & Latency and energy efficiency. \\
& \cite{chen2025characterizing} & Component & Heterogeneous utilization and energy balance. \\

\midrule
\multirow{2}{*}{ML workloads} 
& \cite{fingler2023towards} & Architecture & Compute efficiency and scheduling optimization. \\
& \cite{qureshi2023gpu} & Architecture & Device utilization and system throughput. \\

\midrule
\multirow{2}{*}{Datacenter-scale AI Services} 
& \cite{zhang2021demikernel} & Architecture & I/O latency and heterogeneous scalability. \\
& \cite{lefeuvre2022flexos} & Architecture & Isolation and energy–performance balance. \\

\bottomrule
\end{tabular}%
}
\end{table}

\subsection{Component-level Perspective}
Component-level enhancements focus on improving individual subsystems of the operating system that directly affect the execution of AI workloads. Instead of redesigning the entire OS architecture, these efforts target specific mechanisms such as scheduling, memory management, I/O processing, and security. This perspective is particularly important because many performance bottlenecks in AI applications stem not from high-level system design, but from the fine-grained policies and data paths within these subsystems. The following subsections examine representative workloads—short‑context inference, long‑context or retrieval‑augmented generation, distributed training within a node, and edge inference—to illustrate how existing OS components can be extended or tuned to serve AI workloads without altering the overall system architecture.

\paragraph{Short-Context Inference} Short‑context inference encompasses latency‑critical AI services such as conversational response, completion, and small‑batch model serving.
Each request involves only a brief sequence of tokens, so performance is dominated by the per‑request overhead of scheduling, memory access, and GPU kernel dispatch. ExeGPT~\cite{oh2024exegpt} accelerates short text inference by optimizing execution pipelines and reducing kernel‑launch latency through improved CPU–GPU coordination.
XSched~\cite{shen2025xsched} redesigns scheduling policies to maximize GPU utilization while preserving low tail latency under highly concurrent request streams.
Camel~\cite{xu2025camel} balances performance and energy efficiency through adaptive batching and dynamic resource scaling, whereas SplitWise~\cite{patel2024splitwise} increases throughput via fine-grained model partitioning and runtime overlap. These systems optimize process scheduling, adaptive batching, and request admission while preserving the existing kernel architecture. The primary bottlenecks are scheduling latency from frequent kernel invocations, poor memory locality for hot model weights and the key–value (KV) cache, and tail-latency control under bursty arrivals. Common evaluation metrics include P99 latency, queries per second (QPS), GPU occupancy, and energy per inference.

\paragraph{Long‑Context Inference}
Long‑context Inference including retrieval‑augmented generation (RAG) inference deal with inputs that extend over tens of thousands or even millions of tokens, often coupled with continuous retrieval of external documents or embeddings.
The execution time and memory footprint scale non‑linearly with context length, making the workload heavily bound by memory bandwidth and I/O throughput rather than compute capacity.
PagedAttention~\cite{kwon2023efficient}, VAttention~\cite{prabhu2025vattention}, AttentionStore~\cite{gao2024attentionstore}, and InfiniteContext~\cite{lin2024infinite} improve attention computation and cache management through optimized memory allocation, asynchronous I/O streaming, and overlap of computation with retrieval.
These works focus on optimizing memory and I/O behavior while preserving the conventional OS architecture, tuning page‑cache policies, DMA paths, and data staging to avoid bandwidth stalls and excessive page faults.
The main pressure points lie in memory hierarchy management, cross‑device data movement, and efficient I/O scheduling for retrieval traffic.
Evaluation metrics typically include context‑length scalability, memory‑bandwidth utilization, and end‑to‑end latency per token or per request.

\paragraph{Distributed Training}
Distributed training scales model optimization across multi‑GPU servers or multi‑node clusters, where massive gradient synchronization and parameter exchange dominate system cost.
Each training step incurs tight communication, synchronization, and storage workloads, stressing the kernel communication stack and I/O pipelines.
Helix~\cite{mei2024helix} improves throughput via communication–computation overlap, SpeedLoader~\cite{zhang2024speedloader} reduces the data‑loading bottleneck and improves GPU utilization.
These systems primarily refine distributed data pipelines and communication scheduling, remaining mostly within existing kernel abstractions except where heterogeneity demands deeper architectural support.
The major bottlenecks lie in synchronization latency, network contention, and storage throughput for checkpointing and recovery.
Typical evaluation metrics include scaling efficiency with GPU count, per‑step execution latency, and network or I/O utilization rate.

\paragraph{Edge Inference}
Edge inference executes compact or quantized AI models on mobile and embedded hardware that combine CPUs, GPUs, and NPUs under strict power and thermal constraints.
Workloads feature diverse real‑time requirements and frequent task switching across heterogeneous accelerators.
llm.npu~\cite{xu2025fast} reduces inference latency through coordinated CPU–NPU scheduling under energy constraints, while HeteroInfer~\cite{chen2025characterizing} enables GPU–NPU cooperative LLM inference via unified‑memory synchronization, substantially improving resource utilization without modifying the underlying OS architecture.
These works focus on efficient hardware utilization and energy‑aware scheduling inside the existing OS framework, aiming to minimize idle power and manage dynamic voltage and frequency scaling.
Key pressure points include lightweight data transfer between processors, predictable scheduling under thermal throttling, and fairness across concurrent tasks.
Common evaluation metrics are inference latency under power caps, throughput per watt, and QoS violation rate.

\subsubsection{Analysis}
To elucidate how operating systems should evolve for AI workloads, we further shift perspective from workload-specific optimization to the underlying system modules that sustain them. The evolution can be abstracted into three interdependent dimensions—\emph{scheduling}, \emph{memory management}, and \emph{heterogeneous adaptation}—each defining a foundational axis of future AI-oriented OS design.

\paragraph{Scheduling}
Among all components, the scheduler is the primary control point determining both latency and throughput. Short-context inference stresses micro‑scale dispatch latency—the interval between request arrival and accelerator launch—whereas distributed training emphasizes macro‑scale synchronization across nodes. These conflicting demands reveal the limits of batch-oriented schedulers that optimize average throughput but ignore tail latency and cross‑device balance. Recent studies converge toward semantic- and hardware-aware coordination, viewing each inference request not as a process but as a dependency graph spanning prefill, decoding, and post-processing stages. 
Accordingly, the OS should expose a standardized interface for performance hints (e.g., batch size, stage dependency, criticality) and uniform preemption control across heterogeneous accelerators. 
By decoupling \emph{policy} from \emph{mechanism}, this design enables runtime-level innovation without kernel refactoring. 

\paragraph{Memory Management}
Scalability for long-context inference pivots on efficient memory behavior. Unlike traditional NUMA-style pressure, large models suffer from temporal locality degradation as context length grows—the key challenge becomes determining which states remain resident and on which device. 
Existing virtual memory primitives such as paging and prefetching are still valuable when reinterpreted as elastic attention caches that migrate dynamically across host and accelerator memory. 
Future OS memory managers are trending toward a \emph{distributed working-set abstraction}, coordinating GPU high-bandwidth memory, host DRAM, and even remote tiers through page-granular placement and reuse semantics. 

\paragraph{Heterogeneous Adaptation}
As CPUs, GPUs, and NPUs coexist within diverse platforms, resource scheduling and memory management must converge around hardware heterogeneity. 
The OS’s success will hinge on treating hardware diversity as a schedulable capability rather than a fixed constraint. 
Static placements cannot adapt to fluctuating bandwidth or energy budgets; thus, the kernel’s resource model should support adaptive heterogeneity, allowing tasks to declare requirements (e.g., latency, precision, power) that guide runtime decisions across compute units. 
Operationally, heterogeneity awareness should surface at the OS interface, while device-specific policies remain in user space. Such separation enables hybrid GPU–NPU inference and mixed-precision training to flourish atop commodity kernels, balancing flexibility with maintainability.

\paragraph{Summary}
In essence, the OS of the AI era must progress from rigid, heuristic orchestration toward unified, hint-driven coordination across time (scheduling), space (memory), and device (heterogeneity). 
These three dimensions collectively define the core substrate on which scalable, efficient, and intelligent AI workloads can coexist with general-purpose computing.

\subsection{Architecture-level Perspective}
\paragraph{Distributed Training}
Distributed training exercises all aspects of system scalability, from network communication to heterogeneous coordination.
While most implementations rely on user‑space collectives atop existing kernels, HeteGen~\cite{zhao2024hetegen} advances to the architectural layer by coupling multiple specialized kernels into a unified framework.
It integrates CPUs and GPUs through cooperative scheduling and cross‑domain data movement, enabling hybrid parallelism that overlaps computation with I/O synchronization.
At this level, scaling efficiency depends not on one kernel’s policy but on multi‑kernel coordination, where each device runs a minimal kernel orchestrated by a global control plane.
Such architecture redefines distributed training as an OS‑level composition problem, aligning thread and memory managers across heterogeneous nodes to sustain throughput and reduce synchronization latency.

\paragraph{ML Workloads}
General ML workloads such as large‑scale analytics and model inference often stress the OS data path rather than its scheduling logic.
Two lines of research address this through kernel‑bypass and device‑driven architectures.
Fingler et al.~\cite{fingler2023towards} explore integrating accelerator control directly into the kernel to give ML jobs nanosecond‑level access to compute engines, collapsing the boundary between device and OS threads.
Complementarily, BaM~\cite{qureshi2023gpu} enables GPUs to initiate and manage I/O requests directly to storage, removing the CPU from the critical path.
These designs eliminate context switches and mediation layers that historically separated computation, I/O, and scheduling.
Architecturally, they lay the foundation for a device‑centric OS structure, where accelerators act as autonomous schedulers and the kernel evolves into a lightweight coordination substrate.

\paragraph{Datacenter‑scale AI Services}
AI services deployed at datacenter scale operate under stringent latency and throughput constraints,
while simultaneously requiring strong isolation and resource accounting to support multi‑tenant deployment.
Systems such as Demikernel~\cite{zhang2021demikernel} and FlexOS~\cite{lefeuvre2022flexos} exemplify this class of architecture‑level innovation.
Demikernel represents a Library OS (LibOS) approach that relocates I/O, networking, and synchronization into user space, thus bypassing the traditional kernel data path and achieving microsecond‑level latency for AI serving workloads.
FlexOS, on the other hand, introduces a modular and configurable kernel framework that generates per‑application OS instances with tunable protection domains and isolation profiles.
At the architectural level, these systems converge on the same principle of \emph{vertical specialization}: 
collapsing generic OS abstractions into application‑aware building blocks that can be customized for different tenants or workload profiles.
Together, they demonstrate how datacenter‑scale and multi‑tenant AI services motivate operating systems that are both high‑performance and reconfigurable, providing low‑latency I/O paths, composable security boundaries, and flexible per‑service kernels.

\subsubsection{Analysis}
Across these architecture‑level directions, two fundamental design philosophies can be observed.
\paragraph{Kernel-bypass OS Structure}
Kernel-bypass architectures fundamentally aim to eliminate mediation overhead and expose hardware acceleration directly to applications. Fingler et al.~\cite{fingler2023towards} demonstrate this by enabling kernel‑space access to accelerator APIs through an API‑remoting mechanism, allowing kernel modules to invoke ML accelerators without complex re‑engineering. Raza et al.~\cite{raza2023unikernel} extend the concept of unikernels into Linux (UKL), linking optimized processes directly with the kernel to bypass system‑call overhead and reduce tail latency. Similarly, Cadden et al.~\cite{cadden2020seuss} propose SEUSS, which employs unikernel snapshots for rapid and memory‑efficient deployment of serverless functions. Moving beyond CPU mediation, Qureshi et al.~\cite{qureshi2023gpu} introduce BaM, enabling GPUs to initiate and manage storage I/O directly, achieving up to 5.3×peedups for data‑intensive workloads. More recently, Fried et al.~\cite{fried2024making} present Junction, a high‑density kernel‑bypass system that maintains full Linux compatibility while improving throughput and reducing CPU consumption. Collectively, these systems illustrate how kernel‑bypass designs—ranging from accelerated API access to device‑driven I/O and lightweight unikernels—deliver substantial latency and efficiency gains, forming a practical foundation for AI‑oriented operating‑system architectures.

\paragraph{Modular Kernel Organization}
Modern modular kernel designs seek to balance performance, isolation, and scalability through decomposition and specialization. Demikernel~\cite{zhang2021demikernel} exemplifies this approach with a datacenter‑oriented Library OS architecture that bypasses traditional kernel I/O paths via zero‑copy mechanisms and hardware‑accelerated operations, achieving microsecond‑scale latency. FlexOS~\cite{lefeuvre2022flexos} extends this idea by introducing a configurable LibOS framework that generates per‑application kernel variants with customizable protection domains and isolation strategies, enabling secure yet efficient multi‑tenant deployment. LegoOS~\cite{shan2018legoos} adopts a split‑kernel model to disaggregate OS functionalities across compute and memory nodes, supporting independent scaling and management of heterogeneous hardware components. Beyond these, DBOS~\cite{skiadopoulos2021dbos} demonstrates deeper vertical specialization by embedding OS services directly into the database engine, replacing conventional abstractions with domain‑specific logic. Collectively, these systems illustrate how modular and specialization‑oriented architectures redefine the OS as a composable substrate, enabling tailored performance and protection for diverse AI‑driven workloads.

An overarching observation is that, despite being motivated by specific workloads, e.g., distributed training, general ML pipelines, and datacenter‑scale AI services, these architectural restructurings are not tightly coupled to individual workload characteristics.
Instead, they redefine the infrastructure substrate upon which diverse AI computations can operate efficiently.
In essence, architecture‑level innovation in AI‑native operating systems focuses less on tailoring mechanisms for each workload and more on providing a universal, high‑performance, and composable foundation that reduces mediation overhead and exposes greater flexibility across computing domains.

\section{A Maturity Roadmap for AI-Assisted OS}
\label{sec:stages}

\begin{table}[t]
\scriptsize
\begin{tabularx}{\linewidth}{lXcc}
\toprule
\textbf{Type} & \textbf{Description} & \textbf{Maturity} & \textbf{Representative Work} \\
\midrule
AI-Powered & Intelligent module to enhance OS & High & Fingler et al.~\cite{fingler2023towards}, Chen et al.~\cite{chen2024autoos} \\
\hline
AI-Refactored & Intelligent mechanisms for smart OS upgrade & Medium & Mei et al.~\cite{mei2024aios}, Shi et al.~\cite{shi2024commandspromptsllmbasedsemantic}, Ge et al.~\cite{ge2023llm} \\
\hline
AI-Driven & Embedded AI models adaptively manage OS & Low & Kamath et al.~\cite{kamath2024herding}, Rivard et al.~\cite{rivard2025neuralos}\\
\bottomrule
\end{tabularx}
\caption{Summary of the integration stages between AI and OS, including the level of maturity and representative works for each stage.}
\label{tab:aios}
\end{table}

The evolution of AI-OS integration can be delineated into three progressive stages as shown in Table.\ref{tab:aios}, each marking significant advancements in the integration and utilization of AI within the foundational software that manages computer hardware resources and provides common services for computer programs.

\subsection{AI-Powered OS}
AI-powered Operating Systems build upon traditional architectures by integrating AI tools to enhance user experiences and optimize resource management efficiency. These enhancements include providing applications with common, system-level intelligent services such as understanding user needs~\cite{chan2023consistency}, learning habits and preferences, and recommending applications. Additionally, these OSs more efficient and intelligent resource management services that are independent of hardware type and distribution, facilitating cross-device application execution, cross-architecture application migration, and dynamic hardware allocation based on operational data analysis. Furthermore, they support real-time fault detection and repair, dynamic monitoring of system operational states, and real-time adjustment of internal policies to ensure optimal operational performance. Lastly, they leverage AI in software engineering to automate or correct operating system code, thereby improving the efficiency and quality of OS development via automation and intelligence.
Currently, most previous jobs focus on this stages.

\subsection{AI-Refactored OS}
The transition to an AI-Refactored OS involves a profound architectural transformation where intelligent mechanisms are deeply integrated into the OS framework~\cite{ge2023llm,mei2024aios,packer2023memgpt}. This advancement leads to a sophisticated level of software intelligence, fostering the development of intelligent, modularized OS structures.

This stage paves the way for leveraging AI to conduct a sophisticated analysis of assets derived from open-source operating systems. AI algorithms formulate methodologies for constructing micro-libraries and facilitate the decoupling of existing systems, thereby maximizing the utilization of the current operating system ecosystem. In alignment with the generation requirements of the OS, AI identifies optimal combinations or configurations of these micro-libraries. Additionally, AI enables precise sensing of the operational state of the OS, which supports self-adaptive optimization during runtime, fostering the autonomous growth of the OS.

Moreover, AI-Refactored OS facilitates the strategic planning of coordinated evolutionary pathways among multiple operating systems, promoting a collective process of self-evolution. Ultimately, this approach establishes an ecological paradigm grounded in shared capabilities, positioning itself competitively in the ongoing technological race.

\subsection{AI-Driven OS}

The AI-Driven OS signifies a radical departure from conventional frameworks by placing AI at the core of OS. It replaces the traditional operational logic with AI-driven logic and employs AI to manage and schedule various computing resources uniformly~\cite{kamath2024herding}. This phase may see AI wholly or partially supplanting the traditional kernel, replacing heuristic-based decision-making systems with AI-driven operations, and supporting auto-generation and evolution of kernel modules under AI guidance. Additionally, it utilizes AI for comprehensive scheduling of diverse computing resources, achieving unified management and orchestration.

\subsection{Analysis}
The three stages outlined above represent a logical progression in the depth of AI integration within operating systems. The AI-Powered stage provides incremental enhancements by attaching AI functions to existing OS structures, improving usability and efficiency without altering core designs. The AI-Refactored stage emerges when these incremental additions expose architectural limitations, prompting deeper restructuring of kernel components and services to achieve modularity, adaptability, and long-term maintainability. The AI-Driven stage reflects the most ambitious vision, where AI transitions from a supportive module to the primary mechanism for decision-making and orchestration. 

This evolutionary path is shaped by both technical capacity and practical constraints. In the short term, AI is best suited for auxiliary services, as replacing mission-critical kernel logic remains risky. As AI techniques mature and concerns over robustness, interpretability, and safety are addressed, deeper refactoring becomes viable. The long-term aspiration of AI-Driven OSs rests on the expectation that AI can manage dynamic scheduling and enable self-evolving kernels more effectively than static heuristics. Each stage, however, entails trade-offs: opportunistic integration risks fragmentation in AI-Powered systems; AI-Refactored systems demand substantial engineering effort and raise compatibility issues; and AI-Driven systems face challenges of security, verification, and trust. Overall, the progression from augmentation to refactoring and ultimately to AI-centered operating systems reflects a gradual but directional evolution.

\section{Pitfalls}
\label{sec:pitfalls}

The rapid integration of artificial intelligence into operating systems introduces transformative capabilities in automation, optimization, and adaptability.  
Yet our survey reveals persistent pitfalls that hinder the maturity and large-scale adoption of AI-enhanced operating systems.  
These challenges manifest across three interdependent layers—\textit{methodological}, \textit{engineering}, and \textit{governance}—each exposing distinct sources of technical fragility and systemic risk.

\subsection{Methodological Pitfalls}
At the methodological layer, AI-augmented operating systems suffer from the inherent instability and opacity of data-driven models.  
Most existing studies train machine learning or language models on limited or static workload traces, leading to overfitting and poor generalization across hardware and usage contexts.  
Misclassification, hallucination, and prediction errors can propagate through tightly coupled kernel paths, amplifying small inaccuracies into large-scale performance or stability failures.  
A primary cause of these failures is the scarcity of high-quality and representative training data; models often rely on narrow or biased workload traces that inadequately capture the diversity and dynamics of real-world operating environments.
Consequently, models embedded in the OS often lose relevance over time, exhibit unpredictable behaviors under unseen workloads, and provide insufficient grounds for reproducible evaluation.

\subsection{Engineering Pitfalls}
From an engineering perspective, integrating AI components into operating systems introduces substantial structural and lifecycle challenges.  
Legacy kernels were never designed for modular AI integration, resulting in tangled dependencies, poor observability, and limited debugging transparency.  
Embedding machine learning or language models into performance-critical paths further increases complexity—expanding attack surfaces, disrupting timing determinism, and complicating verification.  
In addition, AI components contend with user processes for scarce CPU, memory, and I/O resources, often degrading performance in multi-tenant or latency-sensitive environments.  
Deployment also remains a critical weakness: most research prototypes neglect production-stage requirements such as gating policies, rollback mechanisms, and workload-aware testing.  
Consequently, AI-augmented kernels largely persist as fragile prototypes rather than maintainable or evolvable system infrastructures.

\subsection{Governance Pitfalls}
At the governance level, AI-driven operating systems disrupt traditional expectations of transparency, traceability, and regulatory compliance.  
Unlike conventional OS modules that produce deterministic results, model-based decisions—particularly those derived from large language models and multi-agent systems—are often opaque and difficult to reproduce.  
Developers and operators frequently lack visibility into the reasoning behind specific scheduling or fault-localization outcomes, eroding trust and complicating post-incident analysis.  
Equally concerning are the privacy risks associated with large-scale data collection: logs, system traces, and telemetry used for model training can inadvertently expose sensitive or confidential information.  
Together with the absence of standardized auditing frameworks, these factors create accountability gaps in which neither the AI component nor the operating system kernel can be clearly held responsible for erroneous or unsafe actions.  
This combination of opacity and legal uncertainty presents a significant barrier to the dependable deployment of AI-augmented operating systems in safety- and security-critical environments.

\section{Future Works}
\label{sec:future}

Building upon the methodological, engineering, and governance pitfalls identified in Section~\ref{sec:pitfalls}, future research on AI-enhanced operating systems should develop systematic responses across these three layers. Each layer corresponds to a distinct category of risks—model behavior, system integration, and trust governance—and thus requires its own set of technical and institutional advances.

\subsection{Toward Methodological Robustness and Model Feedback Stability}
AI-enhanced operating systems must evolve from isolated, heuristic-driven modules into robust, data-driven systems capable of generalizing across diverse workloads and hardware architectures.  
Future research can advance this goal through three complementary directions:

\subsubsection{Unified and Continually Updated AI Toolchains}
To mitigate model drift, fragmentation, and overfitting, operating-system research should prioritize unified AI toolchains that integrate machine learning, large language models (LLMs), and agent-based methods within a unified training–evaluation framework.  
Standardized datasets and benchmarks for OS-centric scenarios will enable reproducible assessment and foster cross-model synergy.  
Continuous or federated retraining can further maintain model freshness and adaptability by incorporating data from evolving workload patterns while preserving diversity and representativeness.

\subsubsection{Hybrid Rules + AI Decision Frameworks}
To avoid false alarms, hallucinations, and unsafe automation, hybrid inference frameworks should merge deterministic rule-based mechanisms with learned models.  
The rule layer establishes invariant system boundaries, whereas the AI layer delivers adaptive prediction and optimization.  
This dual-guardrail approach allows the OS to balance determinism and flexibility, ensuring that AI-driven decisions remain explainable, stable, and verifiable.

\subsubsection{High-Quality Training Data and Feedback Reliability}
The robustness of AI components fundamentally depends on the quality and coverage of their training data.  
Future operating systems should therefore emphasize systematic data curation—collecting diverse, representative, and noise-filtered workload traces that accurately capture runtime dynamics.  
Transparent data provenance, labeling consistency, and unified logging schemas will further strengthen feedback fidelity, enabling models to learn from real-world conditions without inheriting bias or distortion.  
Such efforts are essential for building trustworthy, continuously improving AI models within operating systems.

\subsection{Toward Engineering Reliability and Lifecycle Management}
From an engineering standpoint, the next generation of AI-ready operating systems must be modular, resource-aware, and reliable throughout the entire lifecycle—from design and development to deployment and maintenance.

\subsubsection{Modular and AI-Ready Kernel Architecture}
Legacy monolithic kernels restrict the safe integration of AI models.  
Future systems should adopt modular or microkernel-inspired architectures that encapsulate AI functionalities as microservices or user-space daemons communicating with the kernel through secure system calls.  
This architectural separation keeps kernel functions lightweight, auditable, and replaceable, thereby reducing maintenance complexity and supporting incremental evolution toward AI-native operating systems.

\subsubsection{AI-Aware Resource Orchestration and Real-Time Assurance}
When embedded AI modules contend with user applications for limited CPU, GPU, and memory resources, the operating system must dynamically manage such competition.  
Emerging schedulers and memory managers should incorporate AI-aware orchestration mechanisms that adapt to concurrent workloads while maintaining bounded latency and deterministic performance.  
Further research into kernel-integrated inference engines, real-time model compression, and hardware–software co-design can minimize computational overhead and satisfy timing requirements in safety-critical environments.

\subsubsection{Safe Deployment and Continuous Adaptation}
Engineering reliability also depends on robust deployment and maintenance pipelines.  
Staged rollout mechanisms—such as canary or shadow deployments—should be used to evaluate AI components under realistic workloads prior to full activation.  
Continuous drift detection across system metrics can automatically trigger model retraining or fallback to deterministic rule-based policies when performance regressions occur.  
Such mechanisms enable sustained lifecycle management, ensuring resilient and fault-tolerant integration of AI subsystems within production operating environments.

\subsection{Toward Governance, Trust, and Explainable Intelligence}
As AI becomes a central decision-making component within operating systems, ensuring transparency, compliance, and trustworthiness grows increasingly critical.  
This governance layer links technical accountability with ethical and regulatory obligations, establishing the foundation for reliable and responsible AI-driven system behavior.

\subsubsection{Explainable and Auditable OS Intelligence}
Future operating systems should integrate explainable-AI (XAI) mechanisms that generate traceable rationales for decisions such as scheduling, anomaly detection, and fault localization.  
Lightweight provenance logs and structured audit trails can support post-mortem analysis without imposing significant runtime overhead.  
Such designs enable developers and auditors to clearly distinguish conventional kernel faults from AI-induced misjudgments, thereby improving interpretability and accountability.

\subsubsection{Privacy-Preserving and Compliant Data Management}
Because operating systems continuously collect sensitive telemetry data, research should advance privacy-preserving learning techniques such as differential logging, selective data minimization, and on-device federated training.  
These approaches comply with contemporary data-protection regulations while sustaining sufficient observability for performance monitoring and continuous learning, balancing system transparency with user privacy.

\subsubsection{Data–Intelligence Integration for System-Level Governance}
Beyond local explainability, long-term trust depends on robust cross-domain data governance.  
Future operating systems can function as intelligent orchestrators that unify CPUs, GPUs, NPUs, and storage resources into secure, auditable data-processing platforms.  
By exposing standardized APIs incorporating compliance logic, monitoring, and policy enforcement, the OS can ensure that AI-driven decisions remain transparent, verifiable, and legally traceable.  
Incorporating blockchain-based provenance or policy-as-code frameworks could further reinforce this foundation of accountable and trustworthy system governance.

\section{Conclusion}
\label{sec:conclusion}
This survey presents a comprehensive overview of the evolving intersection between AI and OSs. We examined how a wide range of AI techniques—from traditional machine learning to large language models and agent-based intelligence—have been integrated into core OS modules, including scheduling, memory, I/O, storage, networking, and security, as well as into the broader ecosystem of coding, verification, tuning, and education. In parallel, we discussed how advances in OS design enhance the efficiency, scalability, and adaptability of modern AI workloads. Building on this dual perspective, we proposed a developmental roadmap spanning AI-powered, AI-refactored, and AI-driven operating systems. Despite notable progress, significant challenges persist in scalability, reliability, real-time performance, and interpretability. Future research should prioritize modular AI-ready architectures, unified toolchains, hybrid rule–AI decision frameworks, and trustworthy resource management. By consolidating existing work and identifying open directions, this survey aims to guide researchers and practitioners in developing next-generation intelligent operating systems capable of adapting to increasingly complex and heterogeneous computing environments.

\newpage

\bibliographystyle{elsarticle-num}
\bibliography{ref}

@article{li2023batch,
  title={Batch jobs load balancing scheduling in cloud computing using distributional reinforcement learning},
  author={Li, Tiangang and Ying, Shi and Zhao, Yishi and Shang, Jianga},
  journal={IEEE Transactions on Parallel and Distributed Systems},
  volume={35},
  number={1},
  pages={169--185},
  year={2023},
  publisher={IEEE}
}

@inproceedings{chen2020machine,
  title={Machine learning for load balancing in the linux kernel},
  author={Chen, Jingde and Banerjee, Subho S and Kalbarczyk, Zbigniew T and Iyer, Ravishankar K},
  booktitle={Proceedings of the 11th ACM SIGOPS Asia-Pacific Workshop on Systems},
  pages={67--74},
  year={2020}
}

@book{peterson1985operating,
  title={Operating system concepts},
  author={Peterson, James L and Silberschatz, Abraham},
  year={1985},
  publisher={Addison-Wesley Longman Publishing Co., Inc.}
}

@book{hansen1973operating,
  title={Operating system principles},
  author={Hansen, Per Brinch},
  year={1973},
  publisher={Prentice-Hall, Inc.}
}

@book{tanenbaum1997operating,
  title={Operating systems: design and implementation},
  author={Tanenbaum, Andrew S and Woodhull, Albert S and others},
  volume={68},
  year={1997},
  publisher={Prentice Hall Englewood Cliffs}
}

@inproceedings{shen2025xsched,
  title={XSched: Preemptive Scheduling for Diverse XPUs},
  author={Shen, Weihang and Han, Mingcong and Liu, Jialong and Chen, Rong and Chen, Haibo},
  booktitle={19th USENIX Symposium on Operating Systems Design and Implementation (OSDI 25)},
  pages={671--692},
  year={2025}
}

@article{xu2025camel,
  title={Camel: Energy-Aware LLM Inference on Resource-Constrained Devices},
  author={Xu, Hao and Peng, Long and Song, Shezheng and Liu, Xiaodong and Jun, Ma and Li, Shasha and Yu, Jie and Mao, Xiaoguang},
  journal={arXiv preprint arXiv:2508.09173},
  year={2025}
}

@article{zhao2024hetegen,
  title={Hetegen: Efficient heterogeneous parallel inference for large language models on resource-constrained devices},
  author={Zhao, Xuanlei and Jia, Bin and Zhou, Haotian and Liu, Ziming and Cheng, Shenggan and You, Yang},
  journal={Proceedings of Machine Learning and Systems},
  volume={6},
  pages={162--172},
  year={2024}
}

@inproceedings{xu2025fast,
  title={Fast on-device LLM inference with npus},
  author={Xu, Daliang and Zhang, Hao and Yang, Liming and Liu, Ruiqi and Huang, Gang and Xu, Mengwei and Liu, Xuanzhe},
  booktitle={Proceedings of the 30th ACM International Conference on Architectural Support for Programming Languages and Operating Systems, Volume 1},
  pages={445--462},
  year={2025}
}

@article{zhang2024speedloader,
  title={SpeedLoader: An I/O efficient scheme for heterogeneous and distributed LLM operation},
  author={Zhang, Yiqi and You, Yang},
  journal={Advances in Neural Information Processing Systems},
  volume={37},
  pages={34637--34655},
  year={2024}
}

@article{mei2024helix,
  title={Helix: Distributed serving of large language models via max-flow on heterogeneous gpus},
  author={Mei, Yixuan and Zhuang, Yonghao and Miao, Xupeng and Yang, Juncheng and Jia, Zhihao and Vinayak, Rashmi},
  journal={Parameters},
  volume={4},
  number={A100s},
  pages={H100s},
  year={2024}
}

@inproceedings{patel2024splitwise,
  title={Splitwise: Efficient generative llm inference using phase splitting},
  author={Patel, Pratyush and Choukse, Esha and Zhang, Chaojie and Shah, Aashaka and Goiri, {\'I}{\~n}igo and Maleki, Saeed and Bianchini, Ricardo},
  booktitle={2024 ACM/IEEE 51st Annual International Symposium on Computer Architecture (ISCA)},
  pages={118--132},
  year={2024},
  organization={IEEE}
}

@inproceedings{kwon2023efficient,
  title={Efficient memory management for large language model serving with pagedattention},
  author={Kwon, Woosuk and Li, Zhuohan and Zhuang, Siyuan and Sheng, Ying and Zheng, Lianmin and Yu, Cody Hao and Gonzalez, Joseph and Zhang, Hao and Stoica, Ion},
  booktitle={Proceedings of the 29th Symposium on Operating Systems Principles},
  pages={611--626},
  year={2023}
}

@article{lin2024infinite,
  title={Infinite-llm: Efficient llm service for long context with distattention and distributed kvcache},
  author={Lin, Bin and Zhang, Chen and Peng, Tao and Zhao, Hanyu and Xiao, Wencong and Sun, Minmin and Liu, Anmin and Zhang, Zhipeng and Li, Lanbo and Qiu, Xiafei and others},
  journal={arXiv preprint arXiv:2401.02669},
  year={2024}
}

@article{gao2024attentionstore,
  title={Attentionstore: Cost-effective attention reuse across multi-turn conversations in large language model serving},
  author={Gao, Bin and He, Zhuomin and Sharma, Puru and Kang, Qingxuan and Jevdjic, Djordje and Deng, Junbo and Yang, Xingkun and Yu, Zhou and Zuo, Pengfei},
  journal={arXiv preprint arXiv:2403.19708},
  volume={52},
  pages={20--38},
  year={2024}
}

@inproceedings{prabhu2025vattention,
  title={vattention: Dynamic memory management for serving llms without pagedattention},
  author={Prabhu, Ramya and Nayak, Ajay and Mohan, Jayashree and Ramjee, Ramachandran and Panwar, Ashish},
  booktitle={Proceedings of the 30th ACM International Conference on Architectural Support for Programming Languages and Operating Systems, Volume 1},
  pages={1133--1150},
  year={2025}
}

@inproceedings{fingler2023towards,
  title={Towards a Machine Learning-Assisted Kernel with LAKE},
  author={Fingler, Henrique and Tarte, Isha and Yu, Hangchen and Szekely, Ariel and Hu, Bodun and Akella, Aditya and Rossbach, Christopher J},
  booktitle={Proceedings of the 28th ACM International Conference on Architectural Support for Programming Languages and Operating Systems, Volume 2},
  pages={846--861},
  year={2023}
}

@inproceedings{oh2024exegpt,
  title={Exegpt: Constraint-aware resource scheduling for llm inference},
  author={Oh, Hyungjun and Kim, Kihong and Kim, Jaemin and Kim, Sungkyun and Lee, Junyeol and Chang, Du-seong and Seo, Jiwon},
  booktitle={Proceedings of the 29th ACM International Conference on Architectural Support for Programming Languages and Operating Systems, Volume 2},
  pages={369--384},
  year={2024}
}

@article{li2023towards,
  title={Towards general and efficient online tuning for spark},
  author={Li, Yang and Jiang, Huaijun and Shen, Yu and Fang, Yide and Yang, Xiaofeng and Huang, Danqing and Zhang, Xinyi and Zhang, Wentao and Zhang, Ce and Chen, Peng and others},
  journal={arXiv preprint arXiv:2309.01901},
  year={2023}
}

@inproceedings{hao2020linnos,
  title={LinnOS: Predictability on unpredictable flash storage with a light neural network},
  author={Hao, Mingzhe and Toksoz, Levent and Li, Nanqinqin and Halim, Edward Edberg and Hoffmann, Henry and Gunawi, Haryadi S},
  booktitle={14th USENIX Symposium on Operating Systems Design and Implementation (OSDI 20)},
  pages={173--190},
  year={2020}
}

@inproceedings{goodarzy2021smartos,
  title={Smartos: Towards automated learning and user-adaptive resource allocation in operating systems},
  author={Goodarzy, Sepideh and Nazari, Maziyar and Han, Richard and Keller, Eric and Rozner, Eric},
  booktitle={Proceedings of the 12th ACM SIGOPS Asia-Pacific Workshop on Systems},
  pages={48--55},
  year={2021}
}

@inproceedings{de2022using,
  title={Using machine learning to optimize graph execution on numa machines},
  author={de A. Rocha, Hiago Mayk G and Schwarzrock, Janaina and Lorenzon, Arthur F and Beck, Antonio Carlos S},
  booktitle={Proceedings of the 59th ACM/IEEE Design Automation Conference},
  pages={1027--1032},
  year={2022}
}

@article{liu2022multi,
  title={Multi-job intelligent scheduling with cross-device federated learning},
  author={Liu, Ji and Jia, Juncheng and Ma, Beichen and Zhou, Chendi and Zhou, Jingbo and Zhou, Yang and Dai, Huaiyu and Dou, Dejing},
  journal={IEEE Transactions on Parallel and Distributed Systems},
  volume={34},
  number={2},
  pages={535--551},
  year={2022},
  publisher={IEEE}
}

@inproceedings{zhang2022software,
  title={Software-defined address mapping: a case on 3d memory},
  author={Zhang, Jialiang and Swift, Michael and Li, Jing},
  booktitle={Proceedings of the 27th ACM International Conference on Architectural Support for Programming Languages and Operating Systems},
  pages={70--83},
  year={2022}
}

@article{ahmed2022heterogeneous,
  title={Heterogeneous energy-aware load balancing for industry 4.0 and IoT environments},
  author={Ahmed, Usman and Lin, Jerry Chun-Wei and Srivastava, Gautam},
  journal={ACM Transactions on Management Information Systems (TMIS)},
  volume={13},
  number={4},
  pages={1--23},
  year={2022},
  publisher={ACM New York, NY}
}

@article{zhang2024selene,
  title={Selene: Pioneering Automated Proof in Software Verification},
  author={Zhang, Lichen and Lu, Shuai and Duan, Nan},
  journal={arXiv preprint arXiv:2401.07663},
  year={2024}
}

@article{benabderrahmane2024hack,
  title={Hack me if you can: Aggregating autoencoders for countering persistent access threats within highly imbalanced data},
  author={Benabderrahmane, Sidahmed and Hoang, Ngoc and Valtchev, Petko and Cheney, James and Rahwan, Talal},
  journal={Future Generation Computer Systems},
  volume={160},
  pages={926--941},
  year={2024},
  publisher={Elsevier}
}

@inproceedings{qin2019msndroid,
  title={MSNdroid: the Android malware detector based on multi-class features and deep belief network},
  author={Qin, Xiaoxia and Zeng, Fangping and Zhang, Yu},
  booktitle={Proceedings of the ACM Turing Celebration Conference-China},
  pages={1--5},
  year={2019}
}

@inproceedings{ongun2021living,
  title={Living-off-the-land command detection using active learning},
  author={Ongun, Talha and Stokes, Jack W and Or, Jonathan Bar and Tian, Ke and Tajaddodianfar, Farid and Neil, Joshua and Seifert, Christian and Oprea, Alina and Platt, John C},
  booktitle={Proceedings of the 24th International Symposium on Research in Attacks, Intrusions and Defenses},
  pages={442--455},
  year={2021}
}

@article{jurevckova2024online,
  title={Online Clustering of Known and Emerging Malware Families},
  author={Jure{\v{c}}kov{\'a}, Olha and Jure{\v{c}}ek, Martin and Stamp, Mark},
  journal={arXiv preprint arXiv:2405.03298},
  year={2024}
}

@article{panman2022dynamic,
  title={Dynamic detection of mobile malware using smartphone data and machine learning},
  author={Panman de Wit, JS and Bucur, Doina and van der Ham, Jeroen},
  journal={Digital Threats: Research and Practice (DTRAP)},
  volume={3},
  number={2},
  pages={1--24},
  year={2022},
  publisher={ACM New York, NY}
}

@article{cruz2023patching,
  title={Patching locking bugs statically with crayons},
  author={Cruz-Carlon, Juan and Varshosaz, Mahsa and Le Goues, Claire and Wasowski, Andrzej},
  journal={ACM Transactions on Software Engineering and Methodology},
  volume={32},
  number={3},
  pages={1--28},
  year={2023},
  publisher={ACM New York, NY}
}

@inproceedings{yang2022improvement,
  title={Improvement of lottery scheduling algorithm based on machine learning algorithm},
  author={Yang, Xiaofeng and Bai, Zhijiang},
  booktitle={Proceedings of the 2022 2nd International Conference on Control and Intelligent Robotics},
  pages={894--897},
  year={2022}
}

@inproceedings{zhang2024enhanced,
  title={Enhanced user interaction in operating systems through machine learning language models},
  author={Zhang, Chenwei and Lu, Wenran and Ni, Chunhe and Wang, Hongbo and Wu, Jiang},
  booktitle={International Conference on Image, Signal Processing, and Pattern Recognition (ISPP 2024)},
  volume={13180},
  pages={1623--1630},
  year={2024},
  organization={SPIE}
}

@inproceedings{balkanski2024energyefficientschedulingpredictions,
  title={Energy-efficient scheduling with predictions},
  author={Balkanski, Eric and Perivier, Noemie and Stein, Clifford and Wei, Hao-Ting},
  booktitle={Advances in Neural Information Processing Systems},
  volume={36},
  pages={79012--79023},
  year={2023}
}

@article{metzger2021device,
  title={Device Hopping: Transparent Mid-Kernel Runtime Switching for Heterogeneous Systems},
  author={Metzger, Paul and Seeker, Volker and Fensch, Christian and Cole, Murray},
  journal={ACM Transactions on Architecture and Code Optimization (TACO)},
  volume={18},
  number={4},
  pages={1--25},
  year={2021},
  publisher={ACM New York, NY, USA}
}

@article{Alzaylaee_2020,
   title={DL-Droid: Deep learning based android malware detection using real devices},
   volume={89},
   ISSN={0167-4048},
   DOI={10.1016/j.cose.2019.101663},
   journal={Computers Security},
   publisher={Elsevier BV},
   author={Alzaylaee, Mohammed K. and Yerima, Suleiman Y. and Sezer, Sakir},
   year={2020},
   month=feb, pages={101663} 
}

@inproceedings{zhang2021demikernel,
  title={The demikernel datapath os architecture for microsecond-scale datacenter systems},
  author={Zhang, Irene and Raybuck, Amanda and Patel, Pratyush and Olynyk, Kirk and Nelson, Jacob and Leija, Omar S Navarro and Martinez, Ashlie and Liu, Jing and Simpson, Anna Kornfeld and Jayakar, Sujay and others},
  booktitle={Proceedings of the ACM SIGOPS 28th Symposium on Operating Systems Principles},
  pages={195--211},
  year={2021}
}

@inproceedings{raza2023unikernel,
  title={Unikernel linux (ukl)},
  author={Raza, Ali and Unger, Thomas and Boyd, Matthew and Munson, Eric B and Sohal, Parul and Drepper, Ulrich and Jones, Richard and De Oliveira, Daniel Bristot and Woodman, Larry and Mancuso, Renato and others},
  booktitle={Proceedings of the Eighteenth European Conference on Computer Systems},
  pages={590--605},
  year={2023}
}

@article{kamath2024herding,
  title={Herding LLaMaS: Using LLMs as an OS Module},
  author={Kamath, Aditya K and Yadalam, Sujay},
  journal={arXiv preprint arXiv:2401.08908},
  year={2024}
}

@article{shankar2025machine,
  title={Machine Learning for Linux Kernel Optimization: Current Trends and Future Directions},
  author={Shankar, Vasuki},
  journal={International Journal of Computer Sciences and Engineering},
  volume={13},
  number={3},
  pages={56--64},
  year={2025}
}

@article{ge2023llm,
  title={LLM as OS, agents as apps: Envisioning AIOS, agents and the AIOS-agent ecosystem},
  author={Ge, Yingqiang and Ren, Yujie and Hua, Wenyue and Xu, Shuyuan and Tan, Juntao and Zhang, Yongfeng},
  journal={arXiv e-prints},
  pages={arXiv--2312},
  year={2023}
}

@article{mei2024aios,
  title={AIOS: LLM agent operating system},
  author={Mei, Kai and Li, Zelong and Xu, Shuyuan and Ye, Ruosong and Ge, Yingqiang and Zhang, Yongfeng},
  journal={arXiv e-prints, pp. arXiv--2403},
  year={2024}
}

@article{packer2023memgpt,
  title={Memgpt: Towards llms as operating systems},
  author={Packer, Charles and Wooders, Sarah and Lin, Kevin and Fang, Vivian and Patil, Shishir G and Stoica, Ion and Gonzalez, Joseph E},
  journal={arXiv preprint arXiv:2310.08560},
  year={2023}
}

@article{islam2024llmpoweredcodevulnerabilityrepair,
  title={Llm-powered code vulnerability repair with reinforcement learning and semantic reward},
  author={Islam, Nafis Tanveer and Khoury, Joseph and Seong, Andrew and Karkevandi, Mohammad Bahrami and Parra, Gonzalo De La Torre and Bou-Harb, Elias and Najafirad, Peyman},
  journal={arXiv preprint arXiv:2401.03374},
  year={2024}
}

@inproceedings{cadden2020seuss,
  title={SEUSS: skip redundant paths to make serverless fast},
  author={Cadden, James and Unger, Thomas and Awad, Yara and Dong, Han and Krieger, Orran and Appavoo, Jonathan},
  booktitle={Proceedings of the Fifteenth European Conference on Computer Systems},
  pages={1--15},
  year={2020}
}

@article{skiadopoulos2021dbos,
  title={DBOS: A dbms-oriented operating system},
  author={Skiadopoulos, Athinagoras and Li, Qian and Kraft, Peter and Kaffes, Kostis and Hong, Daniel and Mathew, Shana and Bestor, David and Cafarella, Michael and Gadepally, Vijay and Graefe, Goetz and others},
  year={2021},
  publisher={VLDB Endowment}
}

@inproceedings{shan2018legoos,
  title={LegoOS: A disseminated, distributed OS for hardware resource disaggregation},
  author={Shan, Yizhou and Huang, Yutong and Chen, Yilun and Zhang, Yiying},
  booktitle={13th USENIX Symposium on Operating Systems Design and Implementation (OSDI 18)},
  pages={69--87},
  year={2018}
}

@article{achiam2023gpt,
  title={Gpt-4 technical report},
  author={Achiam, Josh and Adler, Steven and Agarwal, Sandhini and Ahmad, Lama and Akkaya, Ilge and Aleman, Florencia Leoni and Almeida, Diogo and Altenschmidt, Janko and Altman, Sam and Anadkat, Shyamal and others},
  journal={arXiv preprint arXiv:2303.08774},
  year={2023}
}

@article{touvron2023llama,
  title={Llama: Open and efficient foundation language models},
  author={Touvron, Hugo and Lavril, Thibaut and Izacard, Gautier and Martinet, Xavier and Lachaux, Marie-Anne and Lacroix, Timoth{\'e}e and Rozi{\`e}re, Baptiste and Goyal, Naman and Hambro, Eric and Azhar, Faisal and others},
  journal={arXiv preprint arXiv:2302.13971},
  year={2023}
}

@article{team2023gemini,
  title={Gemini: a family of highly capable multimodal models},
  author={Team, Gemini and Anil, Rohan and Borgeaud, Sebastian and Wu, Yonghui and Alayrac, Jean-Baptiste and Yu, Jiahui and Soricut, Radu and Schalkwyk, Johan and Dai, Andrew M and Hauth, Anja and others},
  journal={arXiv preprint arXiv:2312.11805},
  year={2023}
}

@article{chan2023consistency,
  title={The Consistency between Popular Generative Artificial Intelligence (AI) Robots in Evaluating the User Experience of Mobile Device Operating Systems},
  author={Chan, Victor KY},
  journal={Artificial Intelligence, Social Computing and Wearable Technologies},
  volume={113},
  number={113},
  year={2023},
  publisher={AHFE Open Acces}
}

@article{wu2024copilot,
  title={Os-copilot: Towards generalist computer agents with self-improvement},
  author={Wu, Zhiyong and Han, Chengcheng and Ding, Zichen and Weng, Zhenmin and Liu, Zhoumianze and Yao, Shunyu and Yu, Tao and Kong, Lingpeng},
  journal={arXiv preprint arXiv:2402.07456},
  year={2024}
}

@inproceedings{yang2023kernelgptenhancedkernelfuzzing,
  title={Kernelgpt: Enhanced kernel fuzzing via large language models},
  author={Yang, Chenyuan and Zhao, Zijie and Zhang, Lingming},
  booktitle={Proceedings of the 30th ACM International Conference on Architectural Support for Programming Languages and Operating Systems, Volume 2},
  pages={560--573},
  year={2025}
}

@article{zheng2023kenkernelextensionsusing,
  title={KEN: Kernel extensions using natural language},
  author={Zheng, Yusheng and Yang, Yiwei and Chen, Maolin and Quinn, Andrew},
  journal={arXiv preprint arXiv:2312.05531},
  year={2023}
}

@article{hè2024perospersonalizedselfadaptingoperating,
    title={PerOS: Personalized Self-Adapting Operating Systems in the Cloud},
    author={H{\`e}, Hongyu},
    journal={arXiv preprint arXiv:2404.00057},
    year={2024}
}

@inproceedings{lefeuvre2022flexos,
  title={FlexOS: towards flexible OS isolation},
  author={Lefeuvre, Hugo and B{\u{a}}doiu, Vlad-Andrei and Jung, Alexander and Teodorescu, Stefan Lucian and Rauch, Sebastian and Huici, Felipe and Raiciu, Costin and Olivier, Pierre},
  booktitle={Proceedings of the 27th ACM International Conference on Architectural Support for Programming Languages and Operating Systems},
  pages={467--482},
  year={2022}
}

@inproceedings{wu2024mitigating,
  title={Mitigating Write Disturbance in Non-Volatile Memory via Coupling Machine Learning with Out-of-Place Updates},
  author={Wu, Ronglong and Shen, Zhirong and Yang, Zhiwei and Shu, Jiwu},
  booktitle={2024 IEEE International Symposium on High-Performance Computer Architecture (HPCA)},
  pages={1184--1198},
  year={2024},
  organization={IEEE}
}

@inproceedings{aaen2023automatic,
  title={Automatic Energy-Efficient Job Scheduling in HPC: A Novel SLURM Plugin Approach},
  author={Aaen Springborg, Anders and Albano, Michele and Xavier-de-Souza, Samuel},
  booktitle={Proceedings of the SC'23 Workshops of The International Conference on High Performance Computing, Network, Storage, and Analysis},
  pages={1831--1838},
  year={2023}
}

@inproceedings{wang2024learnedftl,
  title={LearnedFTL: A Learning-Based Page-Level FTL for Reducing Double Reads in Flash-Based SSDs},
  author={Wang, Shengzhe and Lin, Zihang and Wu, Suzhen and Jiang, Hong and Zhang, Jie and Mao, Bo},
  booktitle={2024 IEEE International Symposium on High-Performance Computer Architecture (HPCA)},
  pages={616--629},
  year={2024},
  organization={IEEE}
}

@inproceedings{lagar2019software,
  title={Software-defined far memory in warehouse-scale computers},
  author={Lagar-Cavilla, Andres and Ahn, Junwhan and Souhlal, Suleiman and Agarwal, Neha and Burny, Radoslaw and Butt, Shakeel and Chang, Jichuan and Chaugule, Ashwin and Deng, Nan and Shahid, Junaid and others},
  booktitle={Proceedings of the Twenty-Fourth International Conference on Architectural Support for Programming Languages and Operating Systems},
  pages={317--330},
  year={2019}
}

@article{guan2024wattscope,
  title={WattScope: Non-intrusive Application-level Power Disaggregation in Datacenters},
  author={Guan, Xiaoding and Bashir, Noman and Irwin, David and Shenoy, Prashant},
  journal={ACM SIGMETRICS Performance Evaluation Review},
  volume={51},
  number={4},
  pages={24--25},
  year={2024},
  publisher={ACM New York, NY, USA}
}

@inproceedings{qiu2021toward,
  title={Toward reconfigurable kernel datapaths with learned optimizations},
  author={Qiu, Yiming and Liu, Hongyi and Anderson, Thomas and Lin, Yingyan and Chen, Ang},
  booktitle={Proceedings of the Workshop on Hot Topics in Operating Systems},
  pages={175--182},
  year={2021}
}

@article{cheng2024lightweight,
  title={A lightweight authentication-driven trusted management framework for IoT collaboration},
  author={Cheng, Guanjie and Wang, Yewei and Deng, Shuiguang and Xiang, Zhengzhe and Yan, Xueqiang and Zhao, Peng and Dustdar, Schahram},
  journal={IEEE Transactions on Services Computing},
  volume={17},
  number={3},
  pages={747--760},
  year={2024},
  publisher={IEEE}
}

@article{cheng2023conditional,
  title={Conditional privacy-preserving multi-domain authentication and pseudonym management for 6G-enabled IoV},
  author={Cheng, Guanjie and Huang, Junqin and Wang, Yewei and Zhao, Jun and Kong, Linghe and Deng, Shuiguang and Yan, Xueqiang},
  journal={IEEE Transactions on Information Forensics and Security},
  year={2023},
  publisher={IEEE}
}

@article{su2025secure,
  title={Secure and Efficient Personalized Multi-Receiver Data Sharing with Cross-Domain Authentication for Internet of Vehicles},
  author={Su, Taolong and Cheng, Guanjie and Huang, Junqin and Zhao, Xinkui and Deng, Shuiguang},
  journal={IEEE Transactions on Dependable and Secure Computing},
  year={2025},
  publisher={IEEE}
}

@inproceedings{biswas2023machine,
  title={A Machine Learning Approach for Predicting Efficient CPU Scheduling Algorithm},
  author={Biswas, Swapnil and Ahmed, Md Shakil and Rahman, Md Jobayer and Khaer, Anika and Islam, Md Motaharul},
  booktitle={2023 5th International Conference on Sustainable Technologies for Industry 5.0 (STI)},
  pages={1--6},
  year={2023},
  organization={IEEE}
}

@inproceedings{sun2021linux,
  title={Linux Storage IO Important Parameters Filtering Model Based on Random Forest},
  author={Sun, Zhangpin and Chen, Lijun},
  booktitle={2021 16th International Conference on Intelligent Systems and Knowledge Engineering (ISKE)},
  pages={340--346},
  year={2021},
  organization={IEEE}
}

@article{agarwal2023artificial,
  title={Artificial Intelligence and Qubit-Based Operating Systems: Current Progress and Future Perspectives},
  author={Agarwal, Tejashwa and Tyagi, Amit Kumar},
  journal={Quantum Computing in Cybersecurity},
  pages={121--136},
  year={2023},
  publisher={Wiley Online Library}
}

@misc{wohlrabapplication,
  title={Application of Object-Oriented and Artificial Intelligence Methods for Structuring and Optimising Operating Systems},
  author={Wohlrab, Lutz},
  publisher={Citeseer}
}

@inproceedings{korshun2023automation,
  title={Automation and Management in Operating Systems: The Role of Artificial Intelligence and Machine Learning.},
  author={Korshun, Nataliia and Myshko, Ivan and Tkachenko, Olha},
  booktitle={DSMSI},
  pages={59--68},
  year={2023}
}

@misc{ranasinghe2009artificial,
  title={Artificial Intelligence in Distributed Operating Systems},
  author={Ranasinghe, Nadeesha O},
  journal={2009},
  publisher={Citeseer}
}

@inproceedings{Vishwakarma2021AIBO,
  title={AI based OS Future of Operating System},
  author={Shivam Vishwakarma},
  year={2021}}

@article{safarzadeh2021artificial,
  title={Artificial Intelligence in the Low-Level Realm--A Survey},
  author={Safarzadeh, Vahid Mohammadi and Loghmani, Hamed Ghasr},
  journal={arXiv preprint arXiv:2111.00881},
  year={2021}
}

@article{zhang2019learned,
  title={" Learned" Operating Systems},
  author={Zhang, Yiying and Huang, Yutong},
  journal={ACM SIGOPS Operating Systems Review},
  volume={53},
  number={1},
  pages={40--45},
  year={2019},
  publisher={ACM New York, NY, USA}
}

@inproceedings{lozi2016linux,
  title={The Linux scheduler: a decade of wasted cores},
  author={Lozi, Jean-Pierre and Lepers, Baptiste and Funston, Justin and Gaud, Fabien and Qu{\'e}ma, Vivien and Fedorova, Alexandra},
  booktitle={Proceedings of the Eleventh European Conference on Computer Systems},
  pages={1--16},
  year={2016}
}

@article{kim2020memory,
  title={Memory-aware fair-share scheduling for improved performance isolation in the Linux kernel},
  author={Kim, Jungho and Shin, Philkyue and Kim, Myungsun and Hong, Seongsoo},
  journal={IEEE Access},
  volume={8},
  pages={98874--98886},
  year={2020},
  publisher={IEEE}
}

@inproceedings{cardoso2023evaluation,
  title={Evaluation of Automatic Test Case Generation for the Android Operating System using Deep Reinforcement Learning},
  author={Cardoso, Ana Paula and Santos, Cleicy Priscilla and Collins, Eliane and Lima, Kelen and Quiroga, Pablo and Griego, Marlon},
  booktitle={Proceedings of the XXII Brazilian Symposium on Software Quality},
  pages={228--235},
  year={2023}
}

@inproceedings{white2019improving,
  title={Improving random GUI testing with image-based widget detection},
  author={White, Thomas D and Fraser, Gordon and Brown, Guy J},
  booktitle={Proceedings of the 28th ACM SIGSOFT International Symposium on Software Testing and Analysis},
  pages={307--317},
  year={2019}
}

@article{chowdhury2021novel,
  title={A novel insider attack and machine learning based detection for the internet of things},
  author={Chowdhury, Morshed and Ray, Biplob and Chowdhury, Sujan and Rajasegarar, Sutharshan},
  journal={ACM Transactions on Internet of Things},
  volume={2},
  number={4},
  pages={1--23},
  year={2021},
  publisher={ACM New York, NY, USA}
}

@inproceedings{qi2021efficient,
  title={An Efficient Method for Analyzing Widget Intent of Android System},
  author={Qi, Chunhao and Shao, Shuai and Guo, Yanhui and Peng, Junhao and Xu, Guosheng},
  booktitle={Proceedings of the 2021 9th International Conference on Communications and Broadband Networking},
  pages={78--85},
  year={2021}
}

@article{wihar2024novel,
  title={A Novel Approach of LSTM-Based Ransomware Detection in the Linux Operating System Kernel},
  author={Wihar, Javier and Mathur, Rajiv and Northington, Khalid and Ortega, Alejandro},
  journal={Authorea Preprints},
  year={2024},
  publisher={Authorea}
}

@inproceedings{sibai2023,
author = {Sibai, Fadi and Asaduzzaman, Abu and Sibai, Ahmad},
year = {2023},
month = {05},
pages = {184-188},
title = {A Comparative Study of Machine Learning Methods for Intrusion Detection},
doi = {10.1109/ICEEE59925.2023.00041}
}

@article{qiu2020survey,
  title={A survey of android malware detection with deep neural models},
  author={Qiu, Junyang and Zhang, Jun and Luo, Wei and Pan, Lei and Nepal, Surya and Xiang, Yang},
  journal={ACM Computing Surveys (CSUR)},
  volume={53},
  number={6},
  pages={1--36},
  year={2020},
  publisher={ACM New York, NY, USA}
}

@article{pereira2021learning,
  title={Learning software configuration spaces: A systematic literature review},
  author={Pereira, Juliana Alves and Acher, Mathieu and Martin, Hugo and J{\'e}z{\'e}quel, Jean-Marc and Botterweck, Goetz and Ventresque, Anthony},
  journal={Journal of Systems and Software},
  volume={182},
  pages={111044},
  year={2021},
  publisher={Elsevier}
}

@inproceedings{sk2022literature,
  title={A literature review on android mobile malware detection using machine learning techniques},
  author={Sk, Heena Kauser and others},
  booktitle={2022 6th International Conference on Computing Methodologies and Communication (ICCMC)},
  pages={986--991},
  year={2022},
  organization={IEEE}
}

@article{zhao2023survey,
  title={A survey of large language models},
  author={Zhao, Wayne Xin and Zhou, Kun and Li, Junyi and Tang, Tianyi and Wang, Xiaolei and Hou, Yupeng and Min, Yingqian and Zhang, Beichen and Zhang, Junjie and Dong, Zican and others},
  journal={arXiv preprint arXiv:2303.18223},
  year={2023}
}

@phdthesis{schupbach2012tackling,
  title={Tackling OS Complexity with Declarative Techniques},
  author={Sch{\"u}pbach, Adrian L},
  year={2012},
  school={ETH Zurich}
}

@INPROCEEDINGS{9289257,
  author={Kwon, Jungmin and Lee, Jungjin and Yu, Miseon and Park, Hyunggon},
  booktitle={2020 International Conference on Information and Communication Technology Convergence (ICTC)}, 
  title={Automatic Classification of Network Traffic Data based on Deep Learning in ONOS Platform}, 
  year={2020},
  volume={},
  number={},
  pages={1028-1030},
  doi={10.1109/ICTC49870.2020.9289257}}

@INPROCEEDINGS{9355915,
  author={Miura, Taisei and Kourai, Kenichi and Yamaguchi, Saneyasu},
  booktitle={2020 Eighth International Symposium on Computing and Networking Workshops (CANDARW)}, 
  title={Cache Replacement Based on LSTM in the Second Cache in Virtualized Environment}, 
  year={2020},
  volume={},
  number={},
  pages={421-424},
  doi={10.1109/CANDARW51189.2020.00086}
}

@inproceedings{10.1145/3447555.3466566,
author = {Herzog, Benedict and Reif, Stefan and H\"{u}gel, Fabian and H\"{o}nig, Timo and Schr\"{o}der-Preikschat, Wolfgang},
title = {Towards Automated System-Level Energy-Efficiency Optimisation using Machine Learning: Poster},
year = {2021},
isbn = {9781450383332},
publisher = {Association for Computing Machinery},
address = {New York, NY, USA},
doi = {10.1145/3447555.3466566},
booktitle = {Proceedings of the Twelfth ACM International Conference on Future Energy Systems},
pages = {274–275},
numpages = {2},
location = {Virtual Event, Italy},
series = {e-Energy '21}
}

@article{10.1145/3568429,
author = {Akgun, Ibrahim Umit and Aydin, Ali Selman and Burford, Andrew and McNeill, Michael and Arkhangelskiy, Michael and Zadok, Erez},
title = {Improving Storage Systems Using Machine Learning},
year = {2023},
issue_date = {February 2023},
publisher = {Association for Computing Machinery},
address = {New York, NY, USA},
volume = {19},
number = {1},
issn = {1553-3077},
doi = {10.1145/3568429},
journal = {ACM Trans. Storage},
month = jan,
articleno = {9},
numpages = {30}
}

@article{9123561,
  author={Li, Chi-Yu and Chen, Syuan-Cheng and Kuo, Chien-Ting and Chiu, Chui-Hao},
  journal={IEEE Transactions on Vehicular Technology}, 
  title={Practical Machine Learning-Based Rate Adaptation Solution for Wi-Fi NICs: IEEE 802.11ac as a Case Study}, 
  year={2020},
  volume={69},
  number={9},
  pages={10264-10277},
  doi={10.1109/TVT.2020.3004471}}

@INPROCEEDINGS{8869545,
  author={Horstmann, Leonardo Passig and Hoffmann, José Luis Conradi and Fröhlich, Antônio Augusto},
  booktitle={2019 24th IEEE International Conference on Emerging Technologies and Factory Automation (ETFA)}, 
  title={A Framework to Design and Implement Real-time Multicore Schedulers using Machine Learning}, 
  year={2019},
  volume={},
  number={},
  pages={251-258},
  doi={10.1109/ETFA.2019.8869545}}

@INPROCEEDINGS{9665597,
  author={Maasmi, Fatema and Morcos, Martina and al Hamadi, Hussam and Damiani, Ernesto},
  booktitle={2021 28th IEEE International Conference on Electronics, Circuits, and Systems (ICECS)}, 
  title={Identifying Applications' State via System Calls Activity: A Pipeline Approach}, 
  year={2021},
  volume={},
  number={},
  pages={1-6},
  doi={10.1109/ICECS53924.2021.9665597}}

@INPROCEEDINGS{9110260,
  author={Coronado, Estefanía and Thomas, Abin and Bayhan, Suzan and Riggio, Roberto},
  booktitle={NOMS 2020 - 2020 IEEE/IFIP Network Operations and Management Symposium}, 
  title={aiOS: An Intelligence Layer for SD-WLANs}, 
  year={2020},
  volume={},
  number={},
  pages={1-9},
  doi={10.1109/NOMS47738.2020.9110260}}

@article{9078880,
  author={Wu, Ji-Yan and Wu, Kaishun and Wang, Ming},
  journal={IEEE Transactions on Mobile Computing}, 
  title={Power-Constrained Quality Optimization for Mobile Video Chatting With Coding-Transmission Adaptation}, 
  year={2021},
  volume={20},
  number={9},
  pages={2862-2876},
  doi={10.1109/TMC.2020.2990374}}

@INPROCEEDINGS{8758626,
  author={Zhang, Yunjie and Zhou, Liwei and Makris, Yiorgos},
  booktitle={2019 IEEE 37th VLSI Test Symposium (VTS)}, 
  title={Hardware-based Real-time Workload Forensics via Frame-level TLB Profiling}, 
  year={2019},
  volume={},
  number={},
  pages={1-6},
  doi={10.1109/VTS.2019.8758626}}

@inproceedings{zhong2024logparser,
  title={Logparser-llm: Advancing efficient log parsing with large language models},
  author={Zhong, Aoxiao and Mo, Dengyao and Liu, Guiyang and Liu, Jinbu and Lu, Qingda and Zhou, Qi and Wu, Jiesheng and Li, Quanzheng and Wen, Qingsong},
  booktitle={Proceedings of the 30th ACM SIGKDD Conference on Knowledge Discovery and Data Mining},
  pages={4559--4570},
  year={2024}
}

@inproceedings{cortez2017resource,
  title={Resource central: Understanding and predicting workloads for improved resource management in large cloud platforms},
  author={Cortez, Eli and Bonde, Anand and Muzio, Alexandre and Russinovich, Mark and Fontoura, Marcus and Bianchini, Ricardo},
  booktitle={Proceedings of the 26th Symposium on Operating Systems Principles},
  pages={153--167},
  year={2017}
}

@inproceedings{huang2025logrules,
  title={LogRules: Enhancing Log Analysis Capability of Large Language Models through Rules},
  author={Huang, Xin and Zhang, Ting and Zhao, Wen},
  booktitle={Findings of the Association for Computational Linguistics: NAACL 2025},
  pages={452--470},
  year={2025}
}

@inproceedings{kim2025logs,
  title={Logs In, Patches Out: Automated Vulnerability Repair via Tree-of-ThoughtLLM Analysis},
  author={Kim, Youngjoon and Shin, Sunguk and Kim, Hyoungshick and Yoon, Jiwon},
  booktitle={34th USENIX Security Symposium (USENIX Security 25)},
  pages={4401--4419},
  year={2025}
}

@INPROCEEDINGS{9453065,
  author={Dusane, Palash and Sujatha, G.},
  booktitle={2021 5th International Conference on Trends in Electronics and Informatics (ICOEI)}, 
  title={LogEA: Log Extraction and Analysis Tool to Support Forensic Investigation of Linux-based System}, 
  year={2021},
  volume={},
  number={},
  pages={909-916},
  doi={10.1109/ICOEI51242.2021.9453065}}

@INPROCEEDINGS{10184243,
  author={Al-Saleh, Mohammed I. and Alkouz, Akram and Alarabeyyat, Abdulsalam and Bouchahma, Majed},
  booktitle={2023 9th International Conference on Information Technology Trends (ITT)}, 
  title={Towards Classifying File Segments in Memory Using Machine-Learning}, 
  year={2023},
  volume={},
  number={},
  pages={44-49},
  doi={10.1109/ITT59889.2023.10184243}}

@article{10649557,
  author={Zhong, Yuan and Chen, Pengfei and Zhang, HuXing},
  journal={IEEE Access}, 
  title={ESX: A Self-generated Control Policy for Remote Access with SSH based on eBPF}, 
  year={2024},
  volume={},
  number={},
  pages={1-1},
  doi={10.1109/ACCESS.2024.3450496}}

@inproceedings{bateni2020neuos,
  title={NeuOS: A Latency-Predictable Multi-Dimensional Optimization Framework for DNN-driven Autonomous Systems},
  author={Bateni, Soroush and Liu, Cong},
  booktitle={2020 USENIX Annual Technical Conference (USENIX ATC 20)},
  pages={371--385},
  year={2020}
}

@inproceedings{hedayati2019multi,
  title={Multi-Queue fair queuing},
  author={Hedayati, Mohammad and Shen, Kai and Scott, Michael L and Marty, Mike},
  booktitle={2019 USENIX Annual Technical Conference (USENIX ATC 19)},
  pages={301--314},
  year={2019}
}

@inproceedings{hu2022primo,
  title={Primo: Practical Learning-Augmented Systems with Interpretable Models},
  author={Hu, Qinghao and Nori, Harsha and Sun, Peng and Wen, Yonggang and Zhang, Tianwei},
  booktitle={2022 USENIX Annual Technical Conference (USENIX ATC 22)},
  pages={519--538},
  year={2022}
}

@article{zikria2019internet,
  title={Internet of Things (IoT) operating systems management: Opportunities, challenges, and solution},
  author={Zikria, Yousaf Bin and Kim, Sung Won and Hahm, Oliver and Afzal, Muhammad Khalil and Aalsalem, Mohammed Y},
  journal={Sensors},
  volume={19},
  number={8},
  pages={1793},
  year={2019},
  publisher={MDPI}
}

@article{javed2018internet,
  title={Internet of Things (IoT) operating systems support, networking technologies, applications, and challenges: A comparative review},
  author={Javed, Farhana and Afzal, Muhamamd Khalil and Sharif, Muhammad and Kim, Byung-Seo},
  journal={IEEE Communications Surveys \& Tutorials},
  volume={20},
  number={3},
  pages={2062--2100},
  year={2018},
  publisher={IEEE}
}

@article{okediran2014mobile,
  title={Mobile operating systems and application development platforms: A survey},
  author={Okediran, OO and Arulogun, OT and Ganiyu, RA and Oyeleye, CA},
  journal={International Journal of Advanced Networking and Application},
  volume={6},
  number={1},
  pages={2195},
  year={2014},
  publisher={Eswar Publications}
}

@inproceedings{li2010operating,
  title={Operating system support for overlapping-ISA heterogeneous multi-core architectures},
  author={Li, Tong and Brett, Paul and Knauerhase, Rob and Koufaty, David and Reddy, Dheeraj and Hahn, Scott},
  booktitle={HPCA-16 2010 The Sixteenth International Symposium on High-Performance Computer Architecture},
  pages={1--12},
  year={2010},
  organization={IEEE}
}

@inproceedings{musse2016cloud,
  title={Cloud computing: Architecture and operating system},
  author={Musse, Hodan M and Alamro, Lama A},
  booktitle={2016 Global Summit on Computer \& Information Technology (GSCIT)},
  pages={3--8},
  year={2016},
  organization={IEEE}
}

@inproceedings{pianese2010toward,
  title={Toward a cloud operating system},
  author={Pianese, Fabio and Bosch, Peter and Duminuco, Alessandro and Janssens, Nico and Stathopoulos, Thanos and Steiner, Moritz},
  booktitle={2010 IEEE/IFIP Network Operations and Management Symposium Workshops},
  pages={335--342},
  year={2010},
  organization={IEEE}
}

@inproceedings{li2007efficient,
  title={Efficient operating system scheduling for performance-asymmetric multi-core architectures},
  author={Li, Tong and Baumberger, Dan and Koufaty, David A and Hahn, Scott},
  booktitle={Proceedings of the 2007 ACM/IEEE Conference on Supercomputing},
  pages={1--11},
  year={2007}
}

@article{hall2009operating,
  title={Operating systems for mobile computing},
  author={Hall, Sharon P and Anderson, Eric},
  journal={Journal of Computing Sciences in Colleges},
  volume={25},
  number={2},
  pages={64--71},
  year={2009},
  publisher={Consortium for Computing Sciences in Colleges}
}

@inproceedings{liang2019cognitive,
  title={Cognitive SSD: A deep learning engine for In-Storage data retrieval},
  author={Liang, Shengwen and Wang, Ying and Lu, Youyou and Yang, Zhe and Li, Huawei and Li, Xiaowei},
  booktitle={2019 USENIX Annual Technical Conference (USENIX ATC 19)},
  pages={395--410},
  year={2019}
}

@inproceedings{choi2022serving,
  title={Serving heterogeneous machine learning models on Multi-GPU servers with Spatio-Temporal sharing},
  author={Choi, Seungbeom and Lee, Sunho and Kim, Yeonjae and Park, Jongse and Kwon, Youngjin and Huh, Jaehyuk},
  booktitle={2022 USENIX Annual Technical Conference (USENIX ATC 22)},
  pages={199--216},
  year={2022}
}

@inproceedings{kim2023dream,
  title={DREAM: A Dynamic Scheduler for Dynamic Real-time Multi-model ML Workloads},
  author={Kim, Seah and Kwon, Hyoukjun and Song, Jinook and Jo, Jihyuck and Chen, Yu-Hsin and Lai, Liangzhen and Chandra, Vikas},
  booktitle={Proceedings of the 28th ACM International Conference on Architectural Support for Programming Languages and Operating Systems, Volume 4},
  pages={73--86},
  year={2023}
}

@inproceedings{hanel2020vortex,
  title={Vortex: Extreme-performance memory abstractions for data-intensive streaming applications},
  author={Hanel, Carson and Arman, Arif and Xiao, Di and Keech, John and Loguinov, Dmitri},
  booktitle={Proceedings of the Twenty-Fifth International Conference on Architectural Support for Programming Languages and Operating Systems},
  pages={623--638},
  year={2020}
}

@inproceedings{blocher2021switches,
  title={Switches for HIRE: Resource scheduling for data center in-network computing},
  author={Bl{\"o}cher, Marcel and Wang, Lin and Eugster, Patrick and Schmidt, Max},
  booktitle={Proceedings of the 26th ACM International Conference on Architectural Support for Programming Languages and Operating Systems},
  pages={268--285},
  year={2021}
}

@inproceedings{hildebrand2020autotm,
  title={Autotm: Automatic tensor movement in heterogeneous memory systems using integer linear programming},
  author={Hildebrand, Mark and Khan, Jawad and Trika, Sanjeev and Lowe-Power, Jason and Akella, Venkatesh},
  booktitle={Proceedings of the Twenty-Fifth International Conference on Architectural Support for Programming Languages and Operating Systems},
  pages={875--890},
  year={2020}
}

@inproceedings{qiu2020firm,
  title={FIRM: An intelligent fine-grained resource management framework for SLO-Oriented microservices},
  author={Qiu, Haoran and Banerjee, Subho S and Jha, Saurabh and Kalbarczyk, Zbigniew T and Iyer, Ravishankar K},
  booktitle={14th USENIX Symposium on Operating Systems Design and Implementation (OSDI 20)},
  pages={805--825},
  year={2020}
}

@inproceedings{gupta2024relief,
  title={RELIEF: Relieving Memory Pressure In SoCs Via Data Movement-Aware Accelerator Scheduling},
  author={Gupta, Sudhanshu and Dwarkadas, Sandhya},
  booktitle={2024 IEEE International Symposium on High-Performance Computer Architecture (HPCA)},
  pages={1063--1079},
  year={2024},
  organization={IEEE}
}

@inproceedings{yuan2022adaptive,
  title={Adaptive security support for heterogeneous memory on gpus},
  author={Yuan, Shougang and Awad, Amro and Yudha, Ardhi Wiratama Baskara and Solihin, Yan and Zhou, Huiyang},
  booktitle={2022 IEEE International Symposium on High-Performance Computer Architecture (HPCA)},
  pages={213--228},
  year={2022},
  organization={IEEE}
}

@article{youngmin2023deeplearningbasedmodeling,
      title={Deep Learning based Modeling of Wireless Communication Channel with Fading}, 
      author={Lee Youngmin and Ma Xiaomin and Lang S. I. D. Andrew and Valderrama-Araya F. Enrique and Chapuis L. Andrew},
      year={2023},
      journal={IEEE},
      eprint={2312.06849},
      archivePrefix={arXiv},
      primaryClass={cs.IT}
}

@article{arnold2019enablingfddmassivemimo,
      title={Enabling FDD Massive MIMO through Deep Learning-based Channel Prediction}, 
      author={Maximilian Arnold and Sebastian Dörner and Sebastian Cammerer and Sarah Yan and Jakob Hoydis and Stephan ten Brink},
      year={2019},
      eprint={1901.03664},
      archivePrefix={arXiv},
      primaryClass={cs.IT},
      journal={SPAWC2019}
}

@article{shi2024commandspromptsllmbasedsemantic,
  title={From Commands to Prompts: LLM-based Semantic File System for AIOS},
  author={Shi, Zeru and Mei, Kai and Jin, Mingyu and Su, Yongye and Zuo, Chaoji and Hua, Wenyue and Xu, Wujiang and Ren, Yujie and Liu, Zirui and Du, Mengnan and others},
  journal={arXiv preprint arXiv:2410.11843},
  year={2024}
}

@article{delvecchio2024dynamiccodeorchestrationharnessing,
  title={Dynamic Code Orchestration: Harnessing the Power of Large Language Models for Adaptive Script Execution},
  author={Del Vecchio, Justin and Perreault, Andrew and Furmanek, Eliana},
  journal={arXiv preprint arXiv:2408.11060},
  year={2024}
}

@book{kavitha2019operating,
  title={Operating System Design Sensor Networks Using Artificial Intelligence},
  author={Kavitha, V and BHUVANESH, A and JOSHUA, J and MANI, M JOSHUA and SUBBIAH, K AJAY},
  year={2019},
  publisher={SSRN}
}

@article{zhang2025agentcpm,
  title={AgentCPM-GUI: Building Mobile-Use Agents with Reinforcement Fine-Tuning},
  author={Zhang, Zhong and Lu, Yaxi and Fu, Yikun and Huo, Yupeng and Yang, Shenzhi and Wu, Yesai and Si, Han and Cong, Xin and Chen, Haotian and Lin, Yankai and others},
  journal={arXiv preprint arXiv:2506.01391},
  year={2025}
}

@article{ye2025mobile,
  title={Mobile-Agent-v3: Foundamental Agents for GUI Automation},
  author={Ye, Jiabo and Zhang, Xi and Xu, Haiyang and Liu, Haowei and Wang, Junyang and Zhu, Zhaoqing and Zheng, Ziwei and Gao, Feiyu and Cao, Junjie and Lu, Zhengxi and others},
  journal={arXiv preprint arXiv:2508.15144},
  year={2025}
}

@inproceedings{zheng2024kgent,
  title={Kgent: Kernel extensions large language model agent},
  author={Zheng, Yusheng and Yang, Yiwei and Chen, Maolin and Quinn, Andrew},
  booktitle={Proceedings of the ACM SIGCOMM 2024 Workshop on eBPF and Kernel Extensions},
  pages={30--36},
  year={2024}
}

@inproceedings{xu2024osagent,
  title={OSAgent: Copiloting Operating System with LLM-based Agent},
  author={Xu, Jiaming and Guo, Kaibin and Gong, Wuxuan and Shi, Runyu},
  booktitle={2024 International Joint Conference on Neural Networks (IJCNN)},
  pages={1--9},
  year={2024},
  organization={IEEE}
}

@article{zhang2024ecg,
  title={ECG: Augmenting embedded operating system fuzzing via LLM-based corpus generation},
  author={Zhang, Qiang and Shen, Yuheng and Liu, Jianzhong and Xu, Yiru and Shi, Heyuan and Jiang, Yu and Chang, Wanli},
  journal={IEEE Transactions on Computer-Aided Design of Integrated Circuits and Systems},
  volume={43},
  number={11},
  pages={4238--4249},
  year={2024},
  publisher={IEEE}
}

@inproceedings{zhang2025sortinghat,
author = {Zhang, Yifan and Zhao, Xinkui and Wang, Zuxin and Zhou, Zhengyi and Cheng, Guanjie and Deng, Shuiguang and Yin, Jianwei},
title = {SortingHat: Redefining Operating Systems Education with a Tailored Digital Teaching Assistant},
year = {2025},
isbn = {9798400713316},
publisher = {Association for Computing Machinery},
address = {New York, NY, USA},
doi = {10.1145/3701716.3715199},
booktitle = {Companion Proceedings of the ACM on Web Conference 2025},
pages = {2951–2954},
numpages = {4},
location = {Sydney NSW, Australia},
series = {WWW '25}
}

@inproceedings{lin2025empowering,
  title={Empowering Linux Experimental Teaching with Cursor: Ecological Connotations and Practical Approaches},
  author={Lin, Pengcheng},
  booktitle={Proceedings of the 2024 4th International Conference on Education, Language and Art (ICELA 2024)},
  volume={907},
  pages={252},
  year={2025},
  organization={Springer Nature}
}

@article{zhou2025benchmarking,
  title={Benchmarking and Enhancing LLM Agents in Localizing Linux Kernel Bugs},
  author={Zhou, Zhenhao and Huang, Zhuochen and He, Yike and Wang, Chong and Wang, Jiajun and Wu, Yijian and Peng, Xin and Lou, Yiling},
  journal={arXiv preprint arXiv:2505.19489},
  year={2025}
}

@article{zhao2025stackpilot,
  title={StackPilot: Autonomous Function Agents for Scalable and Environment-Free Code Execution},
  author={Zhao, Xinkui and Zhang, Yifan and Zhou, Zhengyi and Xu, Yueshen},
  journal={arXiv preprint arXiv:2508.11665},
  year={2025}
}

@article{10.1145/3697835,
author = {Goel, Shikha and Kedia, Rajesh and Sen, Rijurekha and Balakrishnan, M},
title = {EXPRESS: A Framework for Execution Time Prediction of Concurrent CNNs on Xilinx DPU Accelerator},
year = {2024},
publisher = {Association for Computing Machinery},
address = {New York, NY, USA},
issn = {1539-9087},
doi = {10.1145/3697835},
note = {Just Accepted},
journal = {ACM Trans. Embed. Comput. Syst.},
month = oct
}

@inproceedings{10.1145/3625549.3658659,
author = {Chang, Juneseo and Doh, Wanju and Moon, Yaebin and Lee, Eojin and Ahn, Jung Ho},
title = {IDT: Intelligent Data Placement for Multi-tiered Main Memory with Reinforcement Learning},
year = {2024},
isbn = {9798400704130},
publisher = {Association for Computing Machinery},
address = {New York, NY, USA},
doi = {10.1145/3625549.3658659},
booktitle = {Proceedings of the 33rd International Symposium on High-Performance Parallel and Distributed Computing},
pages = {69–82},
numpages = {14},
location = {Pisa, Italy},
series = {HPDC '24}
}

@INPROCEEDINGS{10575154,
  author={Sedláček, Ondřej and Bartoš, Václav},
  booktitle={NOMS 2024-2024 IEEE Network Operations and Management Symposium}, 
  title={Fusing Heterogeneous Data for Network Asset Classification – A Two-layer Approach}, 
  year={2024},
  volume={},
  number={},
  pages={1-6},
  doi={10.1109/NOMS59830.2024.10575154}}

@INPROCEEDINGS{10677020,
  author={Veliyath, Alan Joji and Abraham, Ashish Binoy and Abraham, Adarsh Liju and Poddar, Aditya and Giri, Animesh},
  booktitle={2024 IEEE International Conference on Electronics, Computing and Communication Technologies (CONECCT)}, 
  title={Enhancing Network Resilience for Flood Response and Rehabilitation Using SDN}, 
  year={2024},
  volume={},
  number={},
  pages={1-6},
  doi={10.1109/CONECCT62155.2024.10677020}}

@inproceedings{10.1145/3208040.3208051,
author = {Das, Anwesha and Mueller, Frank and Siegel, Charles and Vishnu, Abhinav},
title = {Desh: deep learning for system health prediction of lead times to failure in HPC},
year = {2018},
isbn = {9781450357852},
publisher = {Association for Computing Machinery},
address = {New York, NY, USA},
doi = {10.1145/3208040.3208051},
booktitle = {Proceedings of the 27th International Symposium on High-Performance Parallel and Distributed Computing},
pages = {40–51},
numpages = {12},
location = {Tempe, Arizona},
series = {HPDC '18}
}

@article{7185326,
  author={Wang, Xiaofei and Li, Xiuhua and Leung, Victor C. M.},
  journal={IEEE Access}, 
  title={Artificial Intelligence-Based Techniques for Emerging Heterogeneous Network: State of the Arts, Opportunities, and Challenges}, 
  year={2015},
  volume={3},
  number={},
  pages={1379-1391},
  doi={10.1109/ACCESS.2015.2467174}}

@article{dean2013tail,
  title={The tail at scale},
  author={Dean, Jeffrey and Barroso, Luiz Andr{\'e}},
  journal={Communications of the ACM},
  volume={56},
  number={2},
  pages={74--80},
  year={2013},
  publisher={ACM New York, NY, USA}
}

@inproceedings{hauser2018predictability,
  title={Predictability of resource intensive big data and hpc jobs in cloud data centres},
  author={Hauser, Christopher B and Domaschka, J{\"o}rg and Wesner, Stefan},
  booktitle={2018 IEEE International Conference on Software Quality, Reliability and Security Companion (QRS-C)},
  pages={358--365},
  year={2018},
  organization={IEEE}
}

@inproceedings{kurniawan2025heimdall,
  title={Heimdall: Optimizing Storage I/O Admission with Extensive Machine Learning Pipeline},
  author={Kurniawan, Daniar H and Putri, Rani Ayu and Qin, Peiran and Zulkifli, Kahfi S and Sinurat, Ray AO and Bhimani, Janki and Madireddy, Sandeep and Kistijantoro, Achmad Imam and Gunawi, Haryadi S},
  booktitle={Proceedings of the Twentieth European Conference on Computer Systems},
  pages={1109--1125},
  year={2025}
}

@inproceedings{cui2022linux,
  title={Linux Storage I/O Performance Optimization Based on Machine Learning},
  author={Cui, Pengcheng and Liu, Zhaoyuan and Bai, Jiaqing},
  booktitle={2022 4th International Conference on Natural Language Processing (ICNLP)},
  pages={552--557},
  year={2022},
  organization={IEEE}
}

@article{christianos2024lightweight,
  title={Lightweight neural app control},
  author={Christianos, Filippos and Papoudakis, Georgios and Coste, Thomas and Hao, Jianye and Wang, Jun and Shao, Kun},
  journal={arXiv preprint arXiv:2410.17883},
  year={2024}
}

@article{wang2025mobilea3gent,
  title={MobileA3gent: Training Mobile GUI Agents Using Decentralized Self-Sourced Data from Diverse Users},
  author={Wang, Wenhao and Yuan, Mengying and Yu, Zijie and Liu, Guangyi and Ye, Rui and Jin, Tian and Chen, Siheng and Wang, Yanfeng},
  journal={arXiv preprint arXiv:2502.02982},
  year={2025}
}

@article{lin2025r1,
  title={Os-r1: Agentic operating system kernel tuning with reinforcement learning},
  author={Lin, Hongyu and Li, Yuchen and Luo, Haoran and Yao, Kaichun and Zhang, Libo and Xing, Mingjie and Wu, Yanjun},
  journal={arXiv preprint arXiv:2508.12551},
  year={2025}
}

@article{he2024pc,
  title={PC Agent: While You Sleep, AI Works--A Cognitive Journey into Digital World},
  author={He, Yanheng and Jin, Jiahe and Xia, Shijie and Su, Jiadi and Fan, Runze and Zou, Haoyang and Hu, Xiangkun and Liu, Pengfei},
  journal={arXiv preprint arXiv:2412.17589},
  year={2024}
}

@inproceedings{10.1145/3317550.3321433,
author = {Aguilera, Marcos K. and Keeton, Kimberly and Novakovic, Stanko and Singhal, Sharad},
title = {Designing Far Memory Data Structures: Think Outside the Box},
year = {2019},
isbn = {9781450367271},
publisher = {Association for Computing Machinery},
address = {New York, NY, USA},
doi = {10.1145/3317550.3321433},
booktitle = {Proceedings of the Workshop on Hot Topics in Operating Systems},
pages = {120–126},
numpages = {7},
location = {Bertinoro, Italy},
series = {HotOS '19}
}

@inproceedings{
chen2024autoos,
title={Auto{OS}: Make Your {OS} More Powerful by Exploiting Large Language Models},
author={Huilai Chen and Yuanbo Wen and Limin Cheng and Shouxu Kuang and Yumeng Liu and Weijia Li and Ling Li and Rui Zhang and Xinkai Song and Wei Li and Qi Guo and Yunji Chen},
booktitle={Forty-first International Conference on Machine Learning},
year={2024}
}

@inproceedings{chen2025characterizing,
  title={Characterizing Mobile SoC for Accelerating Heterogeneous LLM Inference},
  author={Chen, Le and Feng, Dahu and Feng, Erhu and Wang, Yingrui and Zhao, Rong and Xia, Yubin and Xu, Pinjie and Chen, Haibo},
  booktitle={Proceedings of the ACM SIGOPS 31st Symposium on Operating Systems Principles},
  pages={359--374},
  year={2025}
}

@article{rivard2025neuralos,
  title={NeuralOS: Towards Simulating Operating Systems via Neural Generative Models},
  author={Rivard, Luke and Sun, Sun and Guo, Hongyu and Chen, Wenhu and Deng, Yuntian},
  journal={arXiv preprint arXiv:2507.08800},
  year={2025}
}

@article{xu2024aios,
  title={AIOS Compiler: LLM as Interpreter for Natural Language Programming and Flow Programming of AI Agents},
  author={Xu, Shuyuan and Li, Zelong and Mei, Kai and Zhang, Yongfeng},
  journal={CoRR},
  year={2024}
}

@article{perez2022development,
  title={Development of an Operating System Fingerprinting Tool Based on Artificial Intelligence},
  author={P{\'e}rez-Jove, Rub{\'e}n},
  year={2022}
}

@inproceedings{qureshi2023gpu,
  title={GPU-initiated on-demand high-throughput storage access in the BaM system architecture},
  author={Qureshi, Zaid and Mailthody, Vikram Sharma and Gelado, Isaac and Min, Seungwon and Masood, Amna and Park, Jeongmin and Xiong, Jinjun and Newburn, Chris J and Vainbrand, Dmitri and Chung, I-Hsin and others},
  booktitle={Proceedings of the 28th ACM International Conference on Architectural Support for Programming Languages and Operating Systems, Volume 2},
  pages={325--339},
  year={2023}
}

@inproceedings{fried2024making,
  title={Making kernel bypass practical for the cloud with junction},
  author={Fried, Joshua and Chaudhry, Gohar Irfan and Saurez, Enrique and Choukse, Esha and Goiri, {\'I}{\~n}igo and Elnikety, Sameh and Fonseca, Rodrigo and Belay, Adam},
  booktitle={21st USENIX Symposium on Networked Systems Design and Implementation (NSDI 24)},
  pages={55--73},
  year={2024}
}

@book{tanenbaum2015modern,
  title={Modern operating systems},
  author={Tanenbaum, Andrew S and Bos, Herbert},
  year={2015},
  publisher={Pearson Education, Inc.}
}

@book{stallings2011operating,
  title={Operating systems: internals and design principles},
  author={Stallings, William},
  year={2011},
  publisher={Prentice Hall Press}
}

@book{mckusick1996design,
  title={The design and implementation of the 4.4 BSD operating system},
  author={McKusick, Marshall Kirk and Bostic, Keith and Karels, Michael J and Quarterman, John S},
  volume={2},
  year={1996},
  publisher={Addison-Wesley Reading, MA}
}

@book{ceruzzi2003history,
  title={A history of modern computing},
  author={Ceruzzi, Paul E},
  year={2003},
  publisher={MIT Press}
}

@article{ritchie1974unix,
  title={The UNIX time-sharing system},
  author={Ritchie, Dennis M and Thompson, Ken},
  journal={Communications of the ACM},
  volume={17},
  number={7},
  pages={365--375},
  year={1974},
  publisher={ACM New York, NY, USA}
}

@article{satyanarayanan2017emergence,
  title={The emergence of edge computing},
  author={Satyanarayanan, Mahadev},
  journal={Computer},
  volume={50},
  number={1},
  pages={30--39},
  year={2017},
  publisher={IEEE}
}

@article{heiser2016l4,
  title={L4 microkernels: The lessons from 20 years of research and deployment},
  author={Heiser, Gernot and Elphinstone, Kevin},
  journal={ACM Transactions on Computer Systems (TOCS)},
  volume={34},
  number={1},
  pages={1--29},
  year={2016},
  publisher={ACM New York, NY, USA}
}

@article{ekmecic1996survey,
  title={A survey of heterogeneous computing: concepts and systems},
  author={Ekmecic, Ilija and Tartalja, Igor and Milutinovic, Veljko},
  journal={Proceedings of the IEEE},
  volume={84},
  number={8},
  pages={1127--1144},
  year={1996},
  publisher={IEEE}
}

@article{shelepov2009hass,
  title={HASS: A scheduler for heterogeneous multicore systems},
  author={Shelepov, Daniel and Saez Alcaide, Juan Carlos and Jeffery, Stacey and Fedorova, Alexandra and Perez, Nestor and Huang, Zhi Feng and Blagodurov, Sergey and Kumar, Viren},
  journal={ACM SIGOPS Operating Systems Review},
  volume={43},
  number={2},
  pages={66--75},
  year={2009},
  publisher={ACM New York, NY, USA}
}

@article{wentzlaff2009factored,
  title={Factored operating systems (fos) the case for a scalable operating system for multicores},
  author={Wentzlaff, David and Agarwal, Anant},
  journal={ACM SIGOPS Operating Systems Review},
  volume={43},
  number={2},
  pages={76--85},
  year={2009},
  publisher={ACM New York, NY, USA}
}

@inproceedings{beckman2006influence,
  title={The influence of operating systems on the performance of collective operations at extreme scale},
  author={Beckman, Pete and Iskra, Kamil and Yoshii, Kazutomo and Coghlan, Susan},
  booktitle={2006 IEEE International Conference on Cluster Computing},
  pages={1--12},
  year={2006},
  organization={IEEE}
}

@inproceedings{barak2005design,
  title={Design principles of operating systems for large scale multicomputers},
  author={Barak, Amnon and Kornatzky, Yoram},
  booktitle={Experiences with Distributed Systems: International Workshop Kaiserslautern, FRG, September 28--30, 1987 Proceedings},
  pages={104--123},
  year={2005},
  organization={Springer}
}

@inproceedings{matias2013operating,
  title={Operating system reliability from the quality of experience viewpoint: an exploratory study},
  author={Matias Jr, Rivalino and Oliveira, Geycy Dyany and de Araujo, Lucio Borges},
  booktitle={Proceedings of the 28th Annual ACM Symposium on Applied Computing},
  pages={1644--1649},
  year={2013}
}

@inproceedings{wong2000operating,
  title={Operating System Support for Multi-User, Remote, Graphical Interaction},
  author={Wong, Alexander Ya-li and Seltzer, Margo},
  booktitle={2000 USENIX Annual Technical Conference (USENIX ATC 00)},
  year={2000}
}

@inproceedings{mao2016resource,
  title={Resource management with deep reinforcement learning},
  author={Mao, Hongzi and Alizadeh, Mohammad and Menache, Ishai and Kandula, Srikanth},
  booktitle={Proceedings of the 15th ACM Workshop on Hot Topics in Networks},
  pages={50--56},
  year={2016}
}

@article{casas2021drsir,
  title={DRSIR: A deep reinforcement learning approach for routing in software-defined networking},
  author={Casas-Velasco, Daniela M and Rendon, Oscar Mauricio Caicedo and da Fonseca, Nelson LS},
  journal={IEEE Transactions on Network and Service Management},
  volume={19},
  number={4},
  pages={4807--4820},
  year={2021},
  publisher={IEEE}
}

@article{rajpurkar2022ai,
  title={AI in health and medicine},
  author={Rajpurkar, Pranav and Chen, Emma and Banerjee, Oishi and Topol, Eric J},
  journal={Nature Medicine},
  volume={28},
  number={1},
  pages={31--38},
  year={2022},
  publisher={Nature Publishing Group US New York}
}

@article{hamet2017artificial,
  title={Artificial intelligence in medicine},
  author={Hamet, Pavel and Tremblay, Johanne},
  journal={Metabolism},
  volume={69},
  pages={S36--S40},
  year={2017},
  publisher={Elsevier}
}

@article{cao2022ai,
  title={Ai in finance: challenges, techniques, and opportunities},
  author={Cao, Longbing},
  journal={ACM Computing Surveys (CSUR)},
  volume={55},
  number={3},
  pages={1--38},
  year={2022},
  publisher={ACM New York, NY}
}

@article{zheng2019finbrain,
  title={FinBrain: when finance meets AI 2.0},
  author={Zheng, Xiao-lin and Zhu, Meng-ying and Li, Qi-bing and Chen, Chao-chao and Tan, Yan-chao},
  journal={Frontiers of Information Technology \& Electronic Engineering},
  volume={20},
  number={7},
  pages={914--924},
  year={2019},
  publisher={Springer}
}

@article{milana2021artificial,
  title={Artificial intelligence techniques in finance and financial markets: a survey of the literature},
  author={Milana, Carlo and Ashta, Arvind},
  journal={Strategic Change},
  volume={30},
  number={3},
  pages={189--209},
  year={2021},
  publisher={Wiley Online Library}
}

@article{parekh2022review,
  title={A review on autonomous vehicles: Progress, methods and challenges},
  author={Parekh, Darsh and Poddar, Nishi and Rajpurkar, Aakash and Chahal, Manisha and Kumar, Neeraj and Joshi, Gyanendra Prasad and Cho, Woong},
  journal={Electronics},
  volume={11},
  number={14},
  pages={2162},
  year={2022},
  publisher={MDPI}
}

@article{bathla2022autonomous,
  title={Autonomous vehicles and intelligent automation: Applications, challenges, and opportunities},
  author={Bathla, Gourav and Bhadane, Kishor and Singh, Rahul Kumar and Kumar, Rajneesh and Aluvalu, Rajanikanth and Krishnamurthi, Rajalakshmi and Kumar, Adarsh and Thakur, RN and Basheer, Shakila},
  journal={Mobile Information Systems},
  volume={2022},
  number={1},
  pages={7632892},
  year={2022},
  publisher={Wiley Online Library}
}

@article{wu2023brief,
  title={A brief overview of ChatGPT: The history, status quo and potential future development},
  author={Wu, Tianyu and He, Shizhu and Liu, Jingping and Sun, Siqi and Liu, Kang and Han, Qing-Long and Tang, Yang},
  journal={IEEE/CAA Journal of Automatica Sinica},
  volume={10},
  number={5},
  pages={1122--1136},
  year={2023},
  publisher={IEEE}
}

@article{liu2024deepseek,
  title={Deepseek-v3 technical report},
  author={Liu, Aixin and Feng, Bei and Xue, Bing and Wang, Bingxuan and Wu, Bochao and Lu, Chengda and Zhao, Chenggang and Deng, Chengqi and Zhang, Chenyu and Ruan, Chong and others},
  journal={arXiv preprint arXiv:2412.19437},
  year={2024}
}

@article{madakam2015internet,
  title={Internet of Things (IoT): A literature review},
  author={Madakam, Somayya and Ramaswamy, Ramya and Tripathi, Siddharth},
  journal={Journal of Computer and Communications},
  volume={3},
  number={5},
  pages={164--173},
  year={2015},
  publisher={Scientific Research Publishing}
}

@inproceedings{tetzlaff2010intelligent,
  title={Intelligent task mapping using machine learning},
  author={Tetzlaff, Dirk and Glesner, Sabine},
  booktitle={2010 International Conference on Computational Intelligence and Software Engineering},
  pages={1--4},
  year={2010},
  organization={IEEE}
}

@inproceedings{subedi2019leveraging,
  title={Leveraging machine learning for anticipatory data delivery in extreme scale in-situ workflows},
  author={Subedi, Pradeep and Davis, Philip E and Parashar, Manish},
  booktitle={2019 IEEE International Conference on Cluster Computing (CLUSTER)},
  pages={1--11},
  year={2019},
  organization={IEEE}
}

@book{augustine2021applying,
  title={Applying machine learning on linux interprocess communication graphs for intrusion detection},
  author={Augustine, William Anthony},
  year={2021},
  publisher={State University of New York at Albany}
}

@article{filelis2018framework,
  title={A framework for simulating large scale cloud infrastructures},
  author={Filelis-Papadopoulos, Christos K and Gravvanis, George A and Kyziropoulos, Panagiotis E},
  journal={Future Generation Computer Systems},
  volume={79},
  pages={703--714},
  year={2018},
  publisher={Elsevier}
}

@article{lin2025byos,
  title={BYOS: Knowledge-driven Large Language Models Bring Your Own Operating System More Excellent},
  author={Lin, Hongyu and Li, Yuchen and Luo, Haoran and Yao, Kaichun and Zhang, Libo and Xing, Mingjie and Wu, Yanjun},
  journal={arXiv preprint arXiv:2503.09663},
  year={2025}
}

@article{polze2012trends,
  title={Trends and challenges in operating systems—from parallel computing to cloud computing},
  author={Polze, Andreas and Tr{\"o}ger, Peter},
  journal={Concurrency and Computation: Practice and Experience},
  volume={24},
  number={7},
  pages={676--686},
  year={2012},
  publisher={Wiley Online Library}
}



\end{document}